\documentclass[10pt,journal]{IEEEtran}
\usepackage{subfigure}
\usepackage{graphicx}
\usepackage{psfrag}
\usepackage{latexsym}
\usepackage{color}
\usepackage[cmex10]{amsmath}
\usepackage{amstext,amssymb, amsfonts}
\usepackage{subfigure}
\usepackage{longtable}
\usepackage{chngcntr}
\def\sinc{\mbox{sinc}}

\def\bmo{\mathcal{B}_\Omega}
\def\nom{\mathcal{N}_\Omega}
\def\nomk{\mathcal{N}_\Omega^\kappa}

\def\amo{\mathcal{A}_\Omega}
\def\emoh{\mathcal{E}_\Omega^\mathcal{H}}
\def\nomh{\mathcal{N}_\Omega^{d-1}}
\def\emo{\mathcal{E}_\Omega}
\def\fmo{\mathcal{F}_\Omega}
\def\pd{path density }

\def\sfN{\mathsf{N}}
\def\sfM{\mathsf{M}}

\DeclareMathOperator{\BigO}{O}
\DeclareMathOperator{\littleo}{o}

\def\mindex#1{\index{#1}}

\def\sq{\hbox{\rlap{$\sqcap$}$\sqcup$}}
\def\qed{\ifmmode\sq\else{\unskip\nobreak\hfil
\penalty50\hskip1em\null\nobreak\hfil\sq
\parfillskip=0pt\finalhyphendemerits=0\endgraf}\fi\medskip}

\long\def\defbox#1{\framebox[.9\hsize][c]{\parbox{.85\hsize}{%
\parindent=0pt
\baselineskip=12pt plus .1pt      
\parskip=6pt plus 1.5pt minus 1pt 
 #1}}}

\long\def\beginbox#1\endbox{\subsection*{}%
\hbox{\hspace{.05\hsize}\defbox{\medskip#1\bigskip}}%
\subsection*{}}

\def\endbox{}

\newsavebox{\junk}
\savebox{\junk}[1.6mm]{\hbox{$|\!|\!|$}}

\def\det{{\mathop{\rm det}}}
\def\limsup{\mathop{\rm lim\ sup}}

\def\argmin{\mathop{\rm arg\, min}}

\def\argmax{\mathop{\rm arg\, max}}

\newcommand{\field}[1]{\mathbb{#1}}

\def\Re{\field{R}}

\def\bC{{\mathbb C}}

\def\bI{{\mathbb I}}

\def\bN{{\mathbb N}}

\def\bZ{{\mathbb Z}}

\def\bfmath#1{{\mathchoice{\mbox{\boldmath$#1$}}%
{\mbox{\boldmath$#1$}}%
{\mbox{\boldmath$\scriptstyle#1$}}%
{\mbox{\boldmath$\scriptscriptstyle#1$}}}}

\def\bfmY{\bfmath{Y}}

\def\bfmhhaY{\bfmath{\hhaY}} 
\def\bfmhhaY{\hbox to 0pt{$\widehat{\bfmY}$\hss}\widehat{\phantom{\raise 1.25pt\hbox{$\bfmY$}}}}

\def\til={{\widetilde =}}

\def\clD{{\cal D}}

\def\clH{{\cal H}}

\def\clK{{\cal K}}

\def\clN{{\cal N}}

\def\clQ{{\cal Q}}

\def\clS{{\cal S}}

\def\clU{{\cal U}}
\def\clV{{\cal V}}
\def\clW{{\cal W}}

 \def\FRAC#1#2#3{\genfrac{}{}{}{#1}{#2}{#3}}

\def\ddtp{{\mathchoice{\FRAC{1}{d^{\hbox to 2pt{\rm\tiny +\hss}}}{dt}}%
{\FRAC{1}{d^{\hbox to 2pt{\rm\tiny +\hss}}}{dt}}%
{\FRAC{3}{d^{\hbox to 2pt{\rm\tiny +\hss}}}{dt}}%
{\FRAC{3}{d^{\hbox to 2pt{\rm\tiny +\hss}}}{dt}}}}

\def\half{{\mathchoice{\FRAC{1}{1}{2}}%
{\FRAC{1}{1}{2}}%
{\FRAC{3}{1}{2}}%
{\FRAC{3}{1}{2}}}}

\def\average#1,#2,{{1\over #2} \sum_{#1}^{#2}}

\def\eye(#1){{\bf(#1)}\quad}

\newtheorem{theorem}{Theorem}[section]
\newtheorem{corollary}{Corollary}[section]
\newtheorem{proposition}[theorem]{Proposition}
\newtheorem{lemma}[theorem]{Lemma}

\def\eq#1/{(\ref{e:#1})}

\newcommand{\beqn}[1]{\notes{#1}%
\begin{eqnarray} \elabel{#1}}

\newcommand{\eeqn}{\end{eqnarray} }

\newcommand{\beq}[1]{\notes{#1}%
\begin{equation}\elabel{#1}}

\newcommand{\eeq}{\end{equation}}

\def\bdes{\begin{description}}
\def\edes{\end{description}}

\newcounter{rmnum}
\newenvironment{romannum}{\begin{list}{{\upshape (\roman{rmnum})}}{\usecounter{rmnum}
\setlength{\leftmargin}{14pt}
\setlength{\rightmargin}{8pt}
\setlength{\itemindent}{-1pt}
}}{\end{list}}

\newcounter{anum}

{\end{list}}

\def\ass(#1:#2){(#1\ref{#1:#2})}

\def\ritem#1{
\item[{\sf \ass(\current_model:#1)}]
}

\newenvironment{recall-ass}[1]{%
\begin{description}
\def\current_model{#1}}{
\end{description}
}

\newcommand{\bd}{\begin{description}}
\newcommand{\ed}{\end{description}}
\newcommand{\bt}{\begin{theorem}}
\newcommand{\et}{\end{theorem}}
\newcommand{\ba}{\begin{array}{rcl}}
\newcommand{\ea}{\end{array}}

\newtheorem{example}{Example}[section]

\counterwithin{corollary}{theorem}
\newcounter{con}
\newenvironment{condition}{\begin{list}{{ \bf (C\arabic{con}) \ }}{\usecounter{con}
\setlength{\leftmargin}{14pt}
\setlength{\rightmargin}{12pt}
\setlength{\itemindent}{-25 pt}
\setlength{\labelsep}{-3 pt}
\setlength{\itemsep}{5 pt}
}}{\end{list}}

\def\Con#1{\textbf{(C\ref{con:#1})}}

\newcounter{mon}
\newenvironment{mondition}{\begin{list}{{\upshape \bf (M\arabic{mon}) \ }}{\usecounter{mon}
\setlength{\leftmargin}{14pt}
\setlength{\rightmargin}{12pt}
\setlength{\itemindent}{-25 pt}
\setlength{\labelsep}{-3 pt}
\setlength{\itemsep}{5 pt}
}}{\end{list}}

\def\Mon#1{\textbf{(M\ref{mon:#1})}}

\newcounter{kon}
\newenvironment{kondition}{\begin{list}{{\upshape \bf (K\arabic{kon}) \ }}{\usecounter{kon}
\setlength{\leftmargin}{14pt}
\setlength{\rightmargin}{12pt}
\setlength{\itemindent}{-25 pt}
\setlength{\labelsep}{-3 pt}
\setlength{\itemsep}{5 pt}
}}{\end{list}}

\def\Kon#1{\textbf{(K\ref{kon:#1})}}

\newtheorem{definition}{Definition}[section]

\newlength{\noteWidth}
\long\def\notes#1{\ifinner
             {\tiny {#1}}
             \else
              \marginpar{\parbox[t]{\noteWidth}{\raggedright\scriptsize {#1}}}
               \fi}

\begin{document}
\title{Sampling High-Dimensional Bandlimited Fields on Low-Dimensional Manifolds}
\author{\authorblockN{Jayakrishnan Unnikrishnan and Martin Vetterli\\}
\authorblockA{Audiovisual Communications Laboratory\\
School of Computer and Communication Sciences\\
Ecole Polytechnique F\'{e}d\'{e}rale de Lausanne (EPFL)\\
Switzerland\thanks{Some of the results presented here were published in an abridged form in \cite{unnvet11Allerton} and \cite{unnvet12ISIT}.}\\
Email:  \{jay.unnikrishnan, martin.vetterli\}@epfl.ch}}
\maketitle
\begin{abstract}
Consider the task of sampling and reconstructing a bandlimited spatial field in $\Re^2$ using moving sensors that take measurements along their path. It is inexpensive to increase the sampling rate along the paths of the sensors but more expensive to increase the total distance traveled by the sensors per unit area, which we call the \emph{path density}. In this paper we introduce the problem of designing sensor trajectories that are minimal in path density subject to the condition that the measurements of the field on these trajectories admit perfect reconstruction of bandlimited fields. We study various possible designs of sampling trajectories. Generalizing some ideas from the classical theory of sampling on lattices, we obtain necessary and sufficient conditions on the trajectories for perfect reconstruction. We show that a single set of equispaced parallel lines has the lowest path density from certain restricted classes of trajectories that admit perfect reconstruction.

We then generalize some of our results to higher dimensions. We first obtain results on designing sampling trajectories in higher dimensional fields. Further, interpreting trajectories as $1$-dimensional manifolds, we extend some of our ideas to higher dimensional sampling manifolds. We formulate the problem of designing $\kappa$-dimensional sampling manifolds for $d$-dimensional spatial fields that are minimal in \emph{manifold density}, a natural generalization of the path density. We show that our results on sampling trajectories for fields in $\Re^2$ can be generalized to analogous results on $d-1$-dimensional sampling manifolds for $d$-dimensional spatial fields.
\end{abstract}



\section{Introduction}
\subsection{Problem description}
\begin{figure}
\centering
\subfigure[Static sampling on points]{
\includegraphics[width=1.6in]
{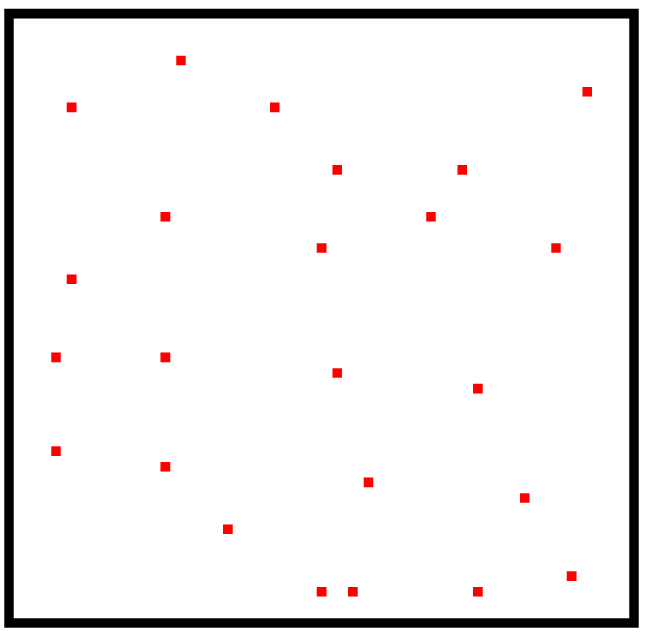}
\label{fig:sampR2a}
}
\subfigure[Mobile sampling on a curve]{
\includegraphics[width=1.6in]
{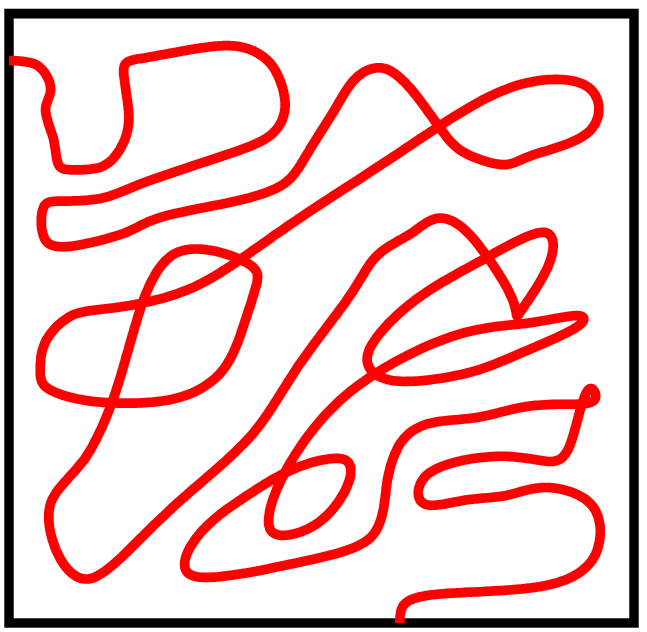}
\label{fig:sampR2b}
}
\caption[Two approaches for sampling a field in $\Re^2$]{Two approaches for sampling a field in $\Re^2$}
\end{figure}

Consider the problem of sampling a $d$-dimensional time-invariant spatial field $f(r): r \in \Re^d$, where $r$ represents a $d$-dimensional spatial location. If $f$ is bandlimited, results from classical sampling theory (see, e.g., \cite{mar91, mar93}) provide schemes for sampling and reconstructing the field based on measurements of the field at a countable number of spatial locations, e.g., points on a lattice or a non-uniform collection of points like the one depicted in Figure \ref{fig:sampR2a}. The performance metric used in designing such sampling schemes is the sampling density - i.e., the number of sampling locations per unit spatial volume. Such a metric is motivated by the fact that one typically employs static sensors to measure the field at their locations and hence the sampling density is equal to the spatial density of the sensor deployment.

The scenario is different in some practical cases. Consider for instance the problem of sampling a $d$-dimensional spatial field (where $d = 2 \mbox{ or } 3$) using a mobile sensor that moves along a continuous path through space and takes measurements along its path. An example of the path of a sensor moving in $\Re^2$ is shown in Figure \ref{fig:sampR2b}. In such cases it is often inexpensive to increase the spatial sampling rate along the sensor's path. Hence it is reasonable to assume that the sensor can record the field values at an arbitrarily high but finite resolution on its path. The objective now is to reconstruct the $d$-dimensional field using only the values of the field at closely spaced points on the path of the sensor through $\Re^d$. For such a sampling scheme, the density of the sampling points in $\Re^d$ is no longer a relevant performance metric. Instead, a more relevant metric is the average distance that needs to be traveled by the sensor per unit spatial volume (or area, for $d =2$). We call this metric the \textit{path density}. Such a metric is relevant in applications like environmental monitoring using moving sensors \cite{keh07, sinnowram06}, where the path density directly measures the distance moved by the sensor per unit area. This metric is also useful in designing $k$-space trajectories for Magnetic Resonance Imaging (MRI) \cite{myrcha96}, where the path density captures the total length of the trajectories per unit area in $k$-space which can be used as a proxy for the total scanning time per unit area in $k$-space.

Now consider a different related problem. Suppose we want to reconstruct a $3$-dimensional bandlimited spatial field using measurements of the field along $2$-dimensional surfaces. Such a scheme is employed in applications like Transmission Electron Microscopy (TEM) \cite{harperboufeiosthur06} and MRI \cite{vlaboe03}. In these cases it may be relatively inexpensive to increase the sampling resolution on the $2$-dimensional measurement surfaces but it may be more expensive to increase the total area of the measurement surfaces. Thus it may be reasonable to assume that the measurements reveal the value of the field at arbitrarily high resolutions on the surfaces. The objective now is to design measurement surfaces that admit perfect reconstruction of bandlimited fields and are simultaneously minimal in area.
%

\begin{figure}
\centering
\subfigure[Sampling lattice]{
\includegraphics[width=1in]
{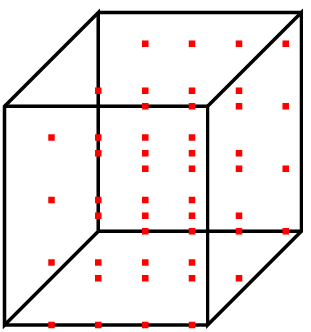}
\label{fig:samanifsR3a}
}
\subfigure[Sampling lines]{
\includegraphics[width=1in]
{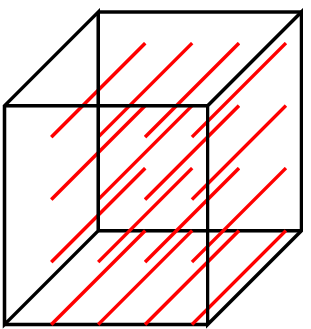}
\label{fig:samanifsR3b}
}
\subfigure[Sampling planes]{
\includegraphics[width=1in]
{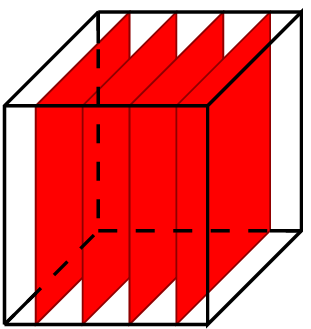}
\label{fig:samanifsR3c}
}
\label{fig:samanifsR3}
\caption[Three approaches for sampling a field in $\Re^3$]{Three approaches for sampling a field in $\Re^3$: sampling on a lattice, along lines, and along planes.}
\end{figure}

Motivated by such problems we introduce a generalization of the classical theory of sampling of $d$-dimensional fields on countable sets of points to a theory of sampling on countable sets of $\kappa$-dimensional manifolds on $\Re^d$ where $\kappa < d$. Some examples of sampling schemes for sampling a field in $\Re^3$ are illustrated in Figure \ref{fig:samanifsR3}. Figure \ref{fig:samanifsR3a} depicts sampling on a lattice, Figure \ref{fig:samanifsR3b} sampling on a set of equispaced parallel lines, and Figure \ref{fig:samanifsR3c} sampling on a set of equispaced parallel planes, corresponding to $\kappa = 0$, $\kappa= 1$, and $\kappa= 2$ respectively.

Initially we focus on the $\kappa = 1$ case since our primary motivation comes from the problem of mobile sensing. The trajectory of a mobile sensor can be interpreted as a $1$-dimensional manifold, or in other words, a curve, through space. Mobile sensing has an advantage over classical static sensing in that a single sensor can be used to take measurements at several positions within an area of interest \cite{ajdsbavet07}. Moreover, in some applications \cite{ciglurripvan08} moving sensors can sample the fields along their paths at high spatial frequencies thereby reducing the amount of spatial aliasing introduced in the samples. Furthermore, as we point out in \cite{unnvet12ICASSP} and \cite{unnvet12TSP}, a moving sensor admits filtering over space in the direction of motion of the sensor whereas no such spatial filtering is possible in the case of static sampling. Such spatial filtering helps in reducing the amount of aliasing and the contribution of out-of-band noise in the reconstructed field.

\subsection{Main results}
Our results for $\kappa =1$ provide guidance for designing trajectories for mobile sensors moving through space. Focusing initially on straight line trajectories we identify some configurations of straight line trajectories that admit perfect reconstruction of bandlimited fields. This can be interpreted as a generalization of the classical Nyquist sampling criterion for sampling on a lattice to sampling on lines. We also formulate the problem of designing sampling trajectories with minimal path density that admit perfect reconstruction of bandlimited fields and obtain partial solutions to this problem, restricting ourselves to specific classes of straight line trajectories. Our main results for straight line trajectories are the following:
\begin{itemize}
\item Necessary and sufficient conditions on a union of $N \leq 2$ sets of equispaced parallel lines that admit perfect reconstructions of bandlimited fields in $\Re^2$ (see Theorem \ref{thm:gentrajsetnecsuff}).
\item Optimality of a single set of equispaced parallel lines from among unions of sets of equispaced parallel lines for sampling bandlimited fields in $\Re^2$ (see Theorem \ref{thm:regparopt}).
\item Optimum configuration of a uniform set of parallel lines for sampling bandlimited fields in $\Re^d$ (see Corollary \ref{cor:cor1}).
\item Optimality of a uniform set of parallel lines from among all trajectory sets that visit all points in a sampling lattice (see Propositions \ref{prop:bestlatticetraj} and \ref{prop:bestlatticetrajdD}).
\end{itemize}
We also consider non-affine trajectories such as concentric circles and interleaved spirals and discuss some known results and some new results on sufficient conditions for perfect reconstruction. These results are based on an application of Beurling's theorem on sufficient conditions for non-uniform sampling.

We then consider higher dimensional sampling manifolds corresponding to $\kappa > 1$. We introduce a \textit{manifold density} metric for sampling manifolds that generalizes the path density metric for sampling trajectories. We generalize our results on sampling on lines in $\Re^2$ to sampling on hyperplanes in $\Re^d$ which are $\kappa$-dimensional manifolds where $\kappa = d-1$. Our main results for hyperplanes are:
\begin{itemize}
\item Necessary and sufficient conditions on a union of $N \leq d$ sets of equispaced parallel hyperplanes that admit perfect reconstructions of bandlimited fields in $\Re^d$ (see Theorem \ref{thm:ddimsgentrajsetnecsuffmanifolds}).
\item Optimality of  a single set of equispaced parallel hyperplanes from among unions of sets of equispaced parallel hyperplanes for sampling bandlimited fields in $\Re^d$ (see Theorem \ref{thm:ddimsregparopt}).
\end{itemize}
The dependencies between various results is summarized in the form of a table in Table \ref{tbl:results} in the conclusion section of the paper.

\subsection{Related work} Although there has not been any past work specifically on the problem of designing optimal sampling manifolds or trajectories, some reconstruction schemes based on measurements taken along concentric circular trajectories have been proposed by Tewfik et al. \cite{tewwil88} and Myridis et al. \cite{myrcha98}. Various sampling trajectories have also been studied in the context of Magnetic Resonance Imaging (see, e.g., \cite{vlaboe03}, \cite{benpowwu02}, \cite{pip99}, \cite{benwu00}). However, the literature on trajectory design for MRI is of a different flavor from ours, since the primary focus in these works is to suppress noise in the reconstruction and they do not aim for exact reconstruction. To the best of our knowledge this paper is the first to introduce the notion of path density and manifold density and the notions of optimal sampling trajectories and optimal sampling manifolds for exact reconstruction of bandlimited fields.
%
%

\subsection{Notations and conventions}
Most of the notations and conventions we use are described where they first arise. In Appendix \ref{sec:symbols} we provide a detailed list of symbols. Here we present some commonly used notations.

We denote a field in $d$-dimensional space by a complex-valued mapping $f: \Re^d \mapsto \bC$. For a field $f(.)$, we define its Fourier transform $F$ as
\begin{equation}
F(\omega) = \int_{\Re^d} f(r) \exp(-{\sf i}\langle \omega, r\rangle ) dr, \qquad \omega \in \Re^d \label{eqn:fourtran}
\end{equation}
where ${\sf i}$ denotes the imaginary unit, and $\langle u ,v \rangle$ denotes the scalar product between vectors $u$ and $v$ in $\Re^d$. We use $\bmo$ to denote the collection of fields with finite energy such that the Fourier transform $F$ of $f$ is supported on a set $\Omega \subset \Re^d$, i.e.,
\begin{equation}
\bmo: = \{f \in L^2(\Re^d): F(\omega) = 0 \mbox{ for } \omega \notin \Omega\}.\label{eqn:bmodefn}
\end{equation}

For $\Omega \subset \Re^d$ we use $\overset{\circ}{\Omega}$ to denote its interior in $\Re^d$. Also for any $s\in \Re^d$ we use $\Omega(s)$ to denote the set obtained by shifting $\Omega$ by $s$, defined as
\begin{equation}
\Omega(s) := \{x \in \Re^d: x-s \in \Omega\}.\label{eqn:omegashiftdefn}
\end{equation}

We use the following notations for order statistics. When $a \to \infty$ we say that a real-valued function $h(a) \in \BigO(a)$ if there exists $k, A \in \Re$ such that for all $a \geq A$, we have,
\[
|h(a)| \leq k a.
\]
Similarly, we say $h(a) \in \littleo(a)$ if for every $\epsilon >0$ there exists $A \in \Re$ such that for all $a \geq A$, we have,
\[
|h(a)| \leq \epsilon a.
\]

We use $\delta(.)$ to denote the Dirac-delta function in $d$-dimensions. The value of $d$ is understood from context. We use $\delta_{ij}$ to denote the Kronecker delta function. For vectors $v,w \in \Re^d$, we use $\|v\|$ to denote the Euclidean norm of $v$ and $\langle v, w\rangle$ to denote the Euclidean inner product.

\subsection{Outline}
The rest of this paper is organized as follows. In Section \ref{sec:samtraj} we formally define the notion of sampling trajectories and introduce the Nyquist criterion for sampling trajectories. We describe various designs of sampling trajectories and provide conditions on these trajectories for perfect reconstruction of bandlimited fields. We provide results on trajectory sets composed of straight lines, circles, and spirals. In Section \ref{sec:opttraj} we study the problem of optimizing sampling trajectories in terms of the path density metric and present some optimality results from certain restricted configurations of sampling trajectories. In Section \ref{sec:manifs} we discuss sampling on higher-dimensional manifolds. We generalize our results on sampling on straight line trajectories to sampling on hyperplanes in high-dimensional spaces. We discuss reconstruction schemes in Section \ref{sec:recon} and conclude in Section \ref{sec:conc}.

\section{Sampling trajectories}\label{sec:samtraj}
We now discuss the problem of designing trajectories for sampling. We  consider natural generalizations of sampling lattices and unions of sampling lattices to trajectories.

\subsection{Preliminaries}\label{sec:defterm}




We use the following terminology in this section. A \textit{trajectory} $p_i$ in $\Re^d$ refers to a curve in $\Re^d$. We represent a trajectory by a continuous function $p(.)$ of a real variable taking values on $\Re^d$:
\[
p: \Re \mapsto \Re^d.
\]
A simple example of a trajectory is a straight line defined by $p(t) = w + vt$ for some $w, v\in \Re^d$. A \textit{trajectory set} $P$ is defined as a countable collection of trajectories:
\begin{equation}
P = \{p_i: i\in \bI\} \label{eqn:trajset}
\end{equation}
where $\bI$ is a countable set of indices and for each $i \in \bI$, $p_i$ is a trajectory in the trajectory set $P$. A simple example of a trajectory set is a countable collection of parallel lines through $\Re^d$.

We introduce a natural generalization of the sampling density metric that characterizes the sampling efficiency of sampling lattices. In a sampling scheme using mobile sensors it is much more difficult to increase the spatial density of trajectories than to increase the sampling rate along the trajectories. Hence, unlike in classical sampling theory, the density of sampling points in space is no longer an appropriate metric for quantifying the efficiency of a mobile sampling scheme. A more reasonable metric is the total length of the trajectories required to span a field of given spatial volume.
Let $B_a^d$ and $B_a^d(x)$ denote $d$-dimensional spherical balls of radii $a$ centered respectively at the origin and at a point $x \in \Re^d$. For any given trajectory set $P$ we denote its \textit{\pd} by $\ell(P)$ defined as follows:
\begin{equation}
\ell(P) := \limsup_{a \to \infty} \frac{\sup_{x\in\Re^d} \clD^P(a,x)}{\mbox{Vol}_d(a)} \label{eqn:pddefn}
\end{equation}
where $\clD^P(a,x)$ represents the total arc-length of trajectories from $P$ located within the ball $B_a^d(x)$ and $\mbox{Vol}_d(a)$ represents the volume of the ball. Clearly, $\mbox{Vol}_2(a) = \pi a^2$ and $\mbox{Vol}_3(a) = \frac{4}{3}\pi a^3$. For a trajectory set composed of differentiable functions $p_i$ we note that $\clD^P(a,x)$ can be explicitly calculated as
\begin{equation}
\clD^P(a,x) = \sum_{i\in\bI} \int_{t\in T_i(a,x)} \left\| \frac{d p_i(t)}{dt}\right\| dt \label{eqn:clLdefn}
\end{equation}
where $\|x\|$ represents the Euclidean norm of $x \in \Re^d$ and
\[
T_i(a,x) := \{t \in\Re: \|p_i(t) -x \| \leq  a\}
\]
represents the portion of trajectory $p_i(.)$ that lies within $B_a^d(x)$.
%


We say that a set of points $\Lambda \subset \Re^d$ is \emph{uniformly discrete} if we have $\inf\{\|x-y\|: x,y \in \Lambda, x \neq y\} > 0$, i.e., there exists $r > 0$ such that for any two distinct points $x,y \in \Lambda$ we have $\|x-y\| > r$.\footnote{For example lattices in $\Re^d$ are uniformly discrete, but a sequence in $\Re^d$ converging to a point in $\Re^d$ is not.}
For a set $\Omega \subset \Re^d$, let $\bmo$ denote the class of fields bandlimited to $\Omega$ as defined in (\ref{eqn:bmodefn}). Let $\amo$ denote the collection of all uniformly discrete sets $\Lambda \subset \Re^d$ which have the property that any field $f \in \bmo$ can be reconstructed exactly from its values on $\Lambda$, i.e. $f \in \bmo$ is uniquely determined from $\{f(x): x \in \Lambda\}$. Classical sampling theory is primarily concerned with the elements of $\amo$, e.g., Nyquist sampling lattices \cite{petmid62}. We now introduce the desirable properties of sampling trajectory sets.
\begin{definition}
A trajectory set $P$ of the form (\ref{eqn:trajset}) is called a \emph{Nyquist trajectory set} for $\Omega \subset \Re^d$ if it satisfies the following conditions:
\begin{quote}
\nobreak%
\begin{condition}

\item \label{con:recon2} [\emph{Nyquist}] There exists a uniformly discrete collection $\Lambda$ of points on the trajectories in $P$  such that $\Lambda$ admits perfect reconstruction of fields in $\bmo$, i.e., $\Lambda \subset \{p_i(t) : i \in \bI, t \in \Re \}$ and $\Lambda \in \amo$.

\item \label{con:path3} [\emph{Non-degeneracy}] For any $x \in \Re^d$, there is a continuous curve of length no more than $\clD^P(a,x) + \littleo(a^d)$ that contains the portion of the trajectory set $P$ that is located within $B_a^d(x)$.
\end{condition}
\end{quote}
\end{definition}
We also introduce a special notation for the collection of all Nyquist trajectory sets:
\begin{definition}
We use $\nom$ to denote the collection of all Nyquist trajectory sets for $\Omega$, i.e., $\nom$ is the collection of all trajectory sets $P$ of the form (\ref{eqn:trajset}) that satisfy conditions~\Con{recon2} and~\Con{path3}.
\end{definition}

The condition~\Con{recon2} ensures that the entire field can be reconstructed exactly from samples taken on the trajectories. This is the direct analogue of the Nyquist condition for sampling on points. The restriction that only uniformly discrete collections of samples are allowed is a standard one in non-uniform sampling theory (see, e.g., \cite{groraz96}, \cite{benwu00}). There is a subtle reason for adopting this restriction in this work. We know that a bandlimited signal is an entire function and hence can be completely recovered (see, e.g., \cite{gia84, pap75}) using its values on an interval or using its values on a convergent sequence of points. Such results can be used to construct a trajectory in $\Re^d$ with finite total length such that bandlimited fields can be perfectly reconstructed using the field values on the trajectory. However, in reality it is impossible to accurately measure the field values on an interval or on a convergent sequence of points. The condition~\Con{path3} ensures that the \pd metric does indeed capture the total length that needs to be traversed by a single moving sensor using the trajectories in $P$ for sampling. This avoids degenerate static cases like the situation in which every trajectory $p_i$ corresponds to a single point in a sampling lattice for the field $f$. Such a degenerate trajectory set has \pd equal to zero, but it is not possible to visit all of these points by using a single sensor. In fact, we know from Nyquist sampling theory that if we had $\BigO(a^d)$ sensors available for sampling spherical regions of radius $a$, it may be possible to sample bandlimited fields without any movement at all.

\begin{figure}
\centering
\psfrag{v}{$v$}
\psfrag{Om}{$\Omega$}
\psfrag{h1}{$\{\omega: \langle \omega, v \rangle = \overline{\rho}\}$}
\psfrag{h2}{$\{\omega: \langle \omega, v \rangle = \underline{\rho}\}$}
\includegraphics[width=1.9 in]{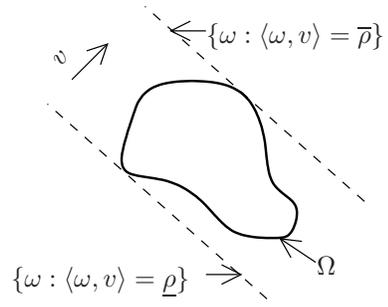}
\caption{Illustration of the result of Lemma \ref{lem:lintrajBW} for fields in $\Re^2$. The vector $v$ represents the velocity vector in the spatial domain and $\underline{\rho}$ and $\overline{\rho}$ represents the limits of the spectrum of $f(p(.))$ in the frequency domain.}  \label{fig:BWaffine}
\end{figure}

In a practical deployment, it is not possible to take measurements of the field at all points along a continuous path because a continuous path has an infinite number of points. However, if the sensor moving along a trajectory is exposed to a bandlimited function of time it is possible to reconstruct the entire field along its path from uniformly spaced samples.
This motivates the following additional desirable condition~\Kon{BL1} of a trajectory set $P$ :
\begin{quote}
\nobreak%
\begin{kondition}
\item
\label{kon:BL1}
 For any field $f \in \bmo$ the $1$-dimensional signal $f(p_i(.))$ is bandlimited\footnote{i.e. has compact support in the Fourier domain} for all $p_i \in P$.
\end{kondition}
\end{quote}
The following lemma shows that trajectory sets composed of straight lines satisfy condition~\Kon{BL1}.
%
%
%
\begin{lemma} \label{lem:lintrajBW}
For $w, v \in \Re^d$ let $p(t) = w + vt$ denote a straight line trajectory parameterized by $t$. Then for $f \in \bmo$ the function $f(p(.))$ is bandlimited to the set $[\underline{\rho}, \overline{\rho}]$ where
\[
\underline{\rho} = \min \{\langle \omega, v \rangle: \omega \in \Omega\} \mbox{ and } \overline{\rho} = \max \{\langle \omega, v \rangle: \omega \in \Omega\}.
\]
\qed
\end{lemma}
We do not prove this lemma since it follows easily from elementary properties of the Fourier transform (see, e.g., \cite{bra99}). The limits of the spectral support $\underline{\rho}$ and $\overline{\rho}$ are illustrated in Figure \ref{fig:BWaffine}. Because of this desirable bandlimitedness property of $f(p(.))$ for straight line trajectories, almost all the trajectory sets that we study in this paper are collections of straight line trajectories.

\subsection{Sampling trajectories for $\Re^2$}\label{sec:samtraj2d}
We now present some simple examples of trajectory sets on $\Re^2$ and compute their path densities. We discuss conditions required to ensure that these trajectory sets are Nyquist trajectory sets for specific choices of $\Omega \subset \Re^2$. These results can be interpreted as a generalization of known results on conditions on sampling lattices \cite{petmid62} and unions of sampling lattices \cite{behfar02} for perfect reconstruction of bandlimited fields.


\begin{figure}
\centering
\subfigure[Uniform set]{
\includegraphics[width=1in]
{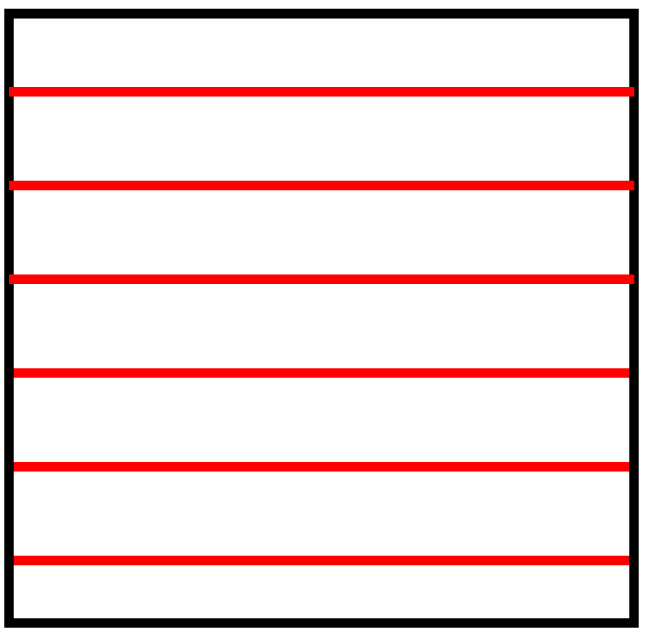}
\label{fig:samtrajsR2a}
}
\subfigure[Union of uniform sets]{
\includegraphics[width=1in]
{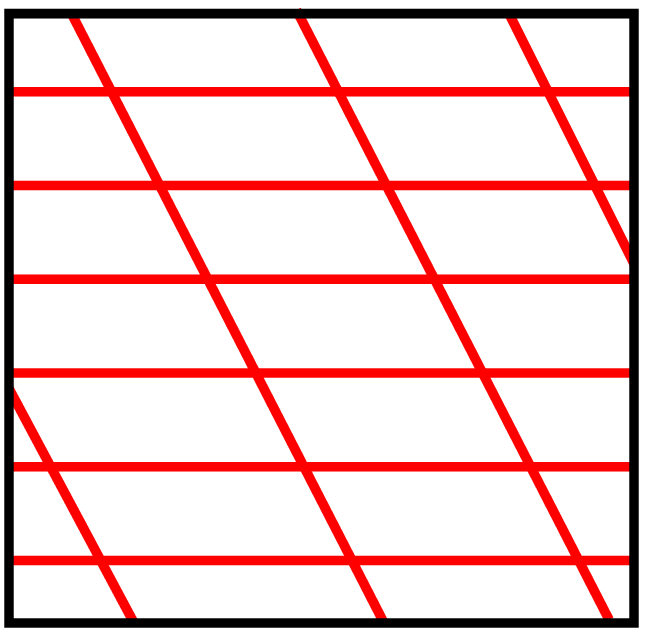}
\label{fig:samtrajsR2b}
}
\subfigure[Concentric circles]{
\includegraphics[width=1in]
{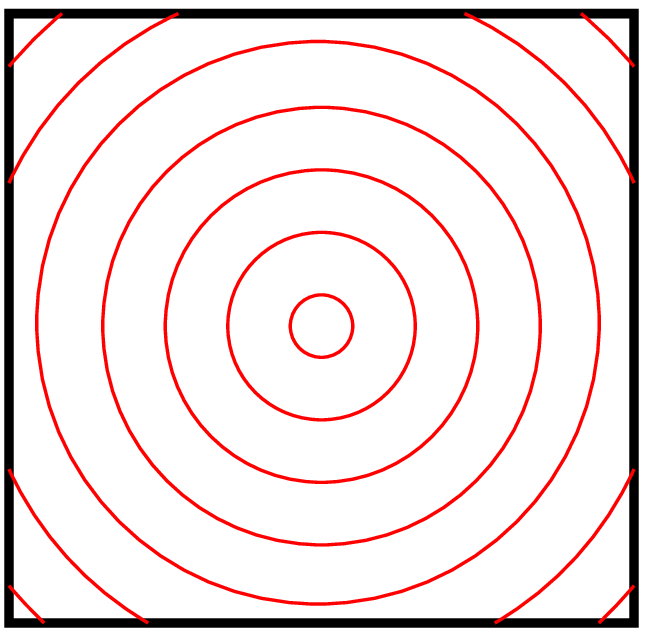}
\label{fig:samtrajsR2c}
}
\label{fig:samtrajsR2}
\caption[Three ]{Three choices of sampling trajectory sets for $\Re^2$: uniform sets, union of uniform sets, and concentric equispaced circular trajectories.}
\end{figure}

%

\subsubsection{Uniform set}\label{sec:regpartraj}
The most natural choice of a trajectory set in $\Re^2$ is a collection of equispaced straight line trajectories of the form
\begin{equation}
P = \{p_i: i \in \bZ\} \quad \mbox{ with } \quad p_i(t) = w + i \Delta v^\perp + vt\label{eqn:unifsetdefn}
\end{equation}
where $\Delta \in \Re_+$, and $w, v$ are fixed vectors in $\Re^2$ with $v^\perp$ denoting the unit vector orthogonal to $v$. These trajectories are lines oriented parallel to $v$ and are spaced $\Delta$ units apart. We refer to such a set of equispaced parallel line trajectories as \emph{uniform sets} in $\Re^2$. Uniform sets form a periodic configuration of straight lines and hence can be interpreted as a natural generalization of sampling lattices to sampling trajectory sets. It is also immediate from Lemma \ref{lem:lintrajBW} that uniform sets satisfy the desirable condition~\Kon{BL1}. 

\begin{figure}
\centering
\subfigure[Sampled spectrum]{
\psfrag{O}{$\Omega$}
\psfrag{Dx}{$\frac{2\pi}{\epsilon}$}
\psfrag{Dy}{$\frac{2\pi}{\Delta}$}
\psfrag{ox}{$\omega_x$}
\psfrag{oy}{$\omega_y$}
\includegraphics[height=2.3in]
{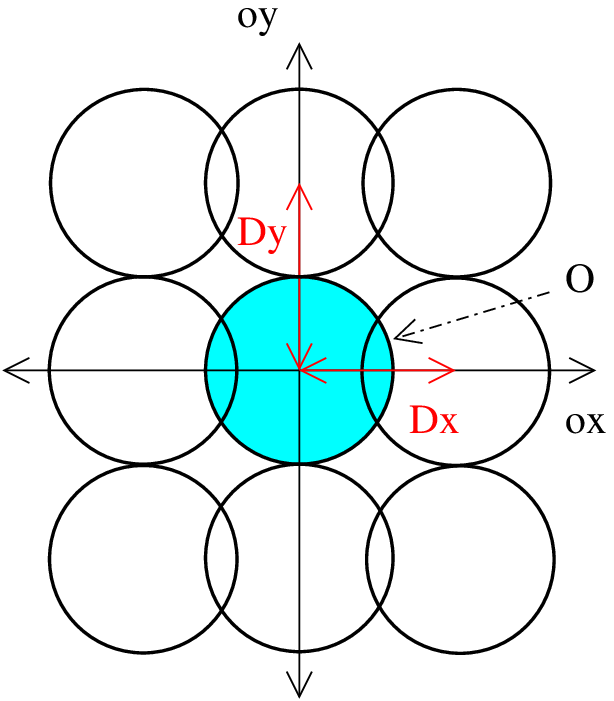}
\label{fig:specreps2da}
}
\subfigure[Limit as $\epsilon \to 0$]{
\psfrag{O}{$\Omega$}
\psfrag{Dx}{$\frac{2\pi}{\epsilon}$}
\psfrag{Dy}{$\frac{2\pi}{\Delta}$}
\psfrag{ox}{$\omega_x$}
\psfrag{oy}{$\omega_y$}
\includegraphics[height=2.3in]
{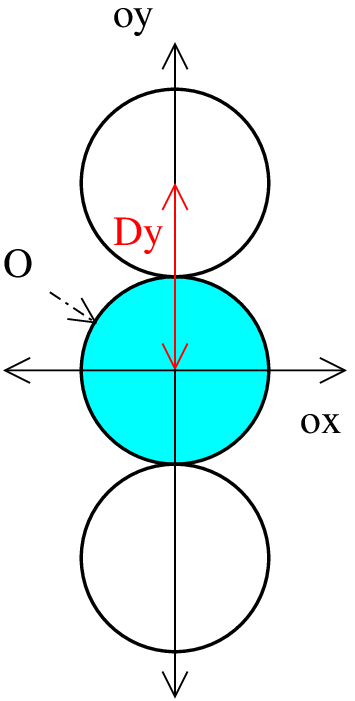}
\label{fig:specreps2db}
}
\label{fig:specreps2d}
\caption[Three ]{Sampled spectrum of a two-dimensional field bandlimited to a circle $\Omega$. The support of the sampled spectrum is composed of periodic repetitions of $\Omega$ on the plane.}
\end{figure}

As an example let $P$ be a uniform set parallel to the x-axis with the spacing between adjacent lines equal to $\Delta$, defined by $P = \{p_i: i \in \bZ\}$ where
\begin{equation}
p_i(t) = (t, \Delta i)^T, i \in \bZ, t\in \Re. \label{eqn:regpartraj}
\end{equation}
Such a uniform set is shown in Figure \ref{fig:samtrajsR2a}. Suppose $\Omega$ is a compact convex subset of $\Re^2$. Let
\[
\Delta^*:= \sup \{\Delta: \Omega \cap \Omega\left(\left(0, \frac{2\pi}{\Delta}\right)^T \right) \mbox{ is empty}\}
\]
where we use the convention of (\ref{eqn:omegashiftdefn}) for a shifted version of a set. It can be easily shown via classical sampling theory \cite{petmid62} that $P$ forms a Nyquist trajectory set for $\Omega$ if $\Delta < \Delta^*$. More precisely, it can be shown that
\begin{equation}
P \in \nom \mbox{ if }\Delta < \Delta^* \label{eqn:unifsetsuff}
\end{equation}
and
\begin{equation}
P \notin \nom \mbox{ if }\Delta > \Delta^*.\label{eqn:unifsetnec}
\end{equation}
These results, as well as analogous results for general uniform sets, follow as special cases of Theorem \ref{thm:gentrajsetnecsuff} which we prove later in the paper. We now illustrate the main idea behind the result via a simple visual proof. Since the sensors can measure the field at high resolutions, we can assume that for any $\epsilon > 0$ the sensors can measure the field value at points of the form $p_i(j \epsilon), j \in \bZ$. In this case, we have access to field samples $f((j \epsilon, i \Delta)^T), i,j \in \bZ$, which correspond to samples on a rectangular lattice. We know from classical sampling theory that the sampled impulse stream,
\[
f_s(r) = \sum_{i,j \in \bZ} f((j \epsilon, i \Delta)^T) \delta(r - (j \epsilon, i \Delta)^T)
\]
has a Fourier spectrum $F_s$ composed of spectral repetitions of $F$ over a reciprocal lattice of points in the Fourier domain:
\[
F_s(\omega) = \sum_{i,j \in \bZ} F\left(\omega - \left(j \frac{2 \pi}{\epsilon}, i \frac{2 \pi}{\Delta}\right)^T\right).
\]
Such a spectrum with repetitions is illustrated in Figure \ref{fig:specreps2da} where $\Omega$ is chosen to be a circle. Now, $\epsilon$ can be made arbitrarily small. In the limit as $\epsilon \to 0$, the sampled spectrum satisfies
\begin{equation}
F_s(\omega) = \sum_{i \in \bZ} F\left(\omega - \left(0, i \frac{2 \pi}{\Delta}\right)^T\right), \omega \in \Omega.\label{eqn:specrepsregpar}
\end{equation}
This means that the spectral repetitions along the $x$-axis no longer overlap, leading to Figure \ref{fig:specreps2db}. Now perfect recovery is possible whenever the spectral repetitions along the $y$-direction do not overlap, or equivalently when $\Delta < \Delta^*$.

It is straightforward to generalize the results of (\ref{eqn:unifsetsuff}) and (\ref{eqn:unifsetnec}) to general uniform sets of the form (\ref{eqn:unifsetdefn}) by interpreting these sets as shifted and rotated versions of the simple example we considered in (\ref{eqn:regpartraj}). The exact conditions can be explicitly obtained from Theorem \ref{thm:gentrajsetnecsuff} which is stated in the next section. In the following lemma we characterize the \pd of a uniform set.
\begin{lemma}\label{lem:pdforregular}
The path density of a uniform set $P$ of the form in (\ref{eqn:unifsetdefn}) is given by
\begin{equation}
\ell(P) = \frac{1}{\Delta}.\label{eqn:regpartrajratio}
\end{equation}
Furthermore, $P$ satisfies condition~\Con{path3}.
\qed
\end{lemma}
We provide a proof in Appendix \ref{lemproof:pdforregular}.

The analyses for uniform sets can be extended to a more general configuration of straight line trajectories composed of unions of uniform sets which we discuss next.

\subsubsection{Union of uniform sets}\label{sec:unionunif}
For vectors $w_i, v_i\in \Re^2$ let $P_i$ denote the uniform set defined by $P_i = \{p_{i,j}: j \in \bZ\}$ where
\begin{equation}
p_{i,j}(t) = w_i + j  \Delta_i v_i^\perp + t v_i, j \in \bZ, t\in \Re, \label{eqn:trajthetai2}
\end{equation}
where $v^\perp$ is a unit vector in $\Re^2$ orthogonal to $v$ and $\Delta_i > 0$. Thus $P_i$ is a uniform set oriented parallel to $v_i$ with the spacing between adjacent lines equal to $ \Delta_i$ apart. The vector $w_i$ is just an offset from the origin. We define the vector $u_i \in \Re^2$ as
\begin{equation}
u_i = \frac{2\pi v_i^\perp}{\Delta_i}.\label{eqn:uicondndefn}
\end{equation}
If we assume that we are given samples of the field $f$ at points of the form $p_{i,j}(k \epsilon), j,k \in \bZ$, then the sampled spectrum from these samples corresponds to repetitions of the field spectrum $F$. As in the example we considered in Section \ref{sec:regpartraj} if we choose a small enough value of $\epsilon$, then it can be shown via classical sampling results \cite{petmid62} that the sampled spectrum comprises spectral repetitions in one direction. We obtain the following spectrum analogous to (\ref{eqn:specrepsregpar}):
\begin{equation}
F_s^i(\omega) = \sum_{j \in \bZ} \exp({\sf i} \langle j u_i,w_i\rangle) F(\omega + j u_i), \omega \in \Omega.\label{eqn:specrepsregparunion}
\end{equation}

Now suppose we have $N$ such uniform sets $\{P_i: 1\leq i \leq N\}$ of the form (\ref{eqn:trajthetai2}). Let $ P$ denote the union of all the sets:
\begin{equation}
P := \bigcup_{i=1}^N P_i. \label{eqn:unionphat}
\end{equation}
An example of such a trajectory set for $N=2$ is depicted in Figure \ref{fig:samtrajsR2b}. It follows from Lemma \ref{lem:pdforregular} that the \pd of this trajectory set is given by
\begin{equation}
\ell( P) = \sum_{i=1}^N\frac{1}{ \Delta_i}. \label{eqn:pdunionregpar}
\end{equation}
We now seek the conditions under which $P$ forms a Nyquist trajectory set for $\Omega$. Since we now have samples from $N$ different uniform sets, it may be possible to reconstruct the field although the sampled spectrum $F_s^i$ from each individual uniform set given by (\ref{eqn:specrepsregparunion}) is aliased. Let $\clQ \subset \Re^2$ denote the set of points
\begin{eqnarray}
\clQ := \left\{\sum_{i=1}^N (-1)^{k_i}\frac{u_i}{2}: k_i \in \{0,1\}, 1 \leq i \leq N\right\}. \label{eqn:qalphadefn}
\end{eqnarray}
The following proposition gives a necessary condition that must be satisfied by any Nyquist trajectory set for $\Omega$ that is a union of uniform sets represented in the form (\ref{eqn:unionphat}).
\begin{proposition}\label{prop:neccondunion}
Let $\Omega \subset \Re^2$ be a compact convex set. Suppose $\clQ \subset \overset{\circ}{\Omega}(s)$ for some $s \in \Re^2$. Then $P \notin \nom$. \qed
\end{proposition}
Here $\overset{\circ}{\Omega}$ denotes the interior of $\Omega$ and $\overset{\circ}{\Omega}(s)$ denotes the shifted version of the interior as defined in (\ref{eqn:omegashiftdefn}). The above result can be proved by constructing a sinusoidal field that vanishes on all the lines in $P$, and has a Fourier transform supported on $\clQ(-s) \subset \overset{\circ}{\Omega}$. We provide a proof in Appendix \ref{propproof:neccondunion}. The result is equivalent to the fact that every $P \in \nom$ must satisfy
\[
\clQ \nsubseteq \overset{\circ}{\Omega}(s) \mbox{ for all } s \in \Re^2.
\]
This necessary condition must be satisfied by all unions of uniform sets $P$ that form Nyquist trajectory sets for $\Omega$. When $N \leq 2$, it can be shown that the tightest necessary condition given by Proposition \ref{prop:neccondunion} is also sufficient. Our main result in two dimensions is the following theorem which provides necessary and sufficient conditions to ensure that $P$ forms a Nyquist trajectory set for $\Omega$ when $N \leq 2$.
\begin{theorem}\label{thm:gentrajsetnecsuff}
Let $\Omega \subset \Re^2$ be a compact convex set. Let ${P}$ denote a union of uniform sets expressed in the form of (\ref{eqn:unionphat}) with $N=1 \mbox{ or }2$. For $N=2$ assume that $v_1$ and $v_2$ are non-collinear. Then we have
\begin{equation}
P \in \nom \mbox{ if } \clQ \nsubseteq \Omega(s) \mbox{ for all }s \in \Re^2\label{eqn:gentrajssufcondn}
\end{equation}
and,
\begin{equation}
P \notin \nom \mbox{ if } \clQ  \subset \overset{\circ}{\Omega}(s) \mbox{ for some } s \in \Re^2 \label{eqn:gentrajsneccondn}
\end{equation}
where $\clQ$ is defined in (\ref{eqn:qalphadefn}).
\qed
\end{theorem}
The result of (\ref{eqn:gentrajsneccondn}) clearly follows from Proposition \ref{prop:neccondunion}. The rest of the theorem is proved in Appendix \ref{thmproof:gentrajsetnecsuff}.

Theorem \ref{thm:gentrajsetnecsuff} gives us necessary and sufficient conditions on unions of uniform sets that admit perfect reconstruction. The conclusion shows that it is possible to recover the field completely using readings on multiple uniform sets even when the measurements on each individual uniform set may be aliased. The result of Theorem \ref{thm:gentrajsetnecsuff} can be generalized to higher dimensions where sampling on lines is replaced with sampling on hyperplanes. This generalization is contained in Theorem \ref{thm:ddimsgentrajsetnecsuffmanifolds}.
%
%

The results of Theorem \ref{thm:gentrajsetnecsuff} can be interpreted as special cases of classical sampling on a discrete collection of points. Since uniform sets are composed of straight line trajectories we know from Lemma \ref{lem:lintrajBW} that the restriction $f(p_{i,j}(.))$ of the bandlimited field $f$ to any line $p_{i,j}$ in the uniform set of (\ref{eqn:trajthetai2}) is bandlimited. Clearly, for a fixed $i$, the bandwidths of the $f(p_{i,j}(.))$ are identical for all $j$. Now suppose that the sensors moving along each of the lines in the set $P_i$ take uniform spatial samples that are $\epsilon_i$ apart. It follows via the bandlimitedness of $f(p_{i,j}(.))$ that the field values $\{f(p_{i,j}(t)): t\in \Re\}$ can be recovered exactly from the samples $\{f(p_{i,j}(m\epsilon_i)): m, j \in \bZ\}$ provided $\epsilon_i$ is small enough. It is also clear that the points $\{p_{i,j}(m\epsilon_i): m, j \in \bZ\}$ lie on a shifted version of a rectangular lattice. Thus for $N =1$, the conditions given in Theorem \ref{thm:gentrajsetnecsuff} can be interpreted as the condition for perfect recovery under a special case of sampling on a shifted rectangular lattice, viz., when the sampling interval along one direction of the lattice is arbitrarily small. We have already seen this interpretation in Section \ref{sec:regpartraj} where we illustrated the idea using Figure \ref{fig:specreps2d}. We can extend this interpretation to the $N=2$ case. For $N = 2$, the collection of sample locations from the two uniform sets, viz., the collection $\{p_{i,j}(m\epsilon_i): m, j \in \bZ, 1\leq i \leq 2\}$ defines a union of two shifted rectangular lattices. Hence the result of Theorem \ref{thm:gentrajsetnecsuff} can be interpreted  \cite{unnvet12ISIT} as the condition for perfect recovery under a special case of sampling on a union of two shifted lattices \cite{behfar02}, when the sampling intervals $\epsilon_i$ are arbitrarily small. We note that in this example, the samples taken on each individual shifted lattice are aliased but given all sets of samples, perfect recovery is possible. The value of $\epsilon_i$ required can be determined from the bandwidth of $f(p_{i,j}(.))$ via Lemma \ref{lem:lintrajBW}.

We also note that the result of Theorem \ref{thm:gentrajsetnecsuff} on unions of uniform sets is similar in spirit to known works on the related problem of sampling on the union of shifted versions of a lattice (see, e.g., \cite{koh53}, \cite{yen56}, \cite{pap77}). These works (e.g., \cite[Example 2]{pap77}) provide necessary and sufficient conditions for sampling on unions of shifted versions of a lattice, whereas, as we argued in the previous paragraph, our results provide conditions for sampling on certain unions of shifted non-identical lattices. Furthermore, as can be seen from the proof of Theorem \ref{thm:gentrajsetnecsuff}, the condition
\[
\clQ \nsubseteq \Omega(s), \mbox{ for all }s\in\Re^2
\]
is equivalent to the condition for invertibility of a linear system of equations relating the values of the Fourier transform of the field. This is analogous to the invertibility condition of Papoulis \cite{pap77}.

We conclude this section on unions of uniform sets with an example that illustrates the result of Theorem \ref{thm:gentrajsetnecsuff} for mutually orthogonal uniform sets under two different choices of $\Omega$.

\begin{figure*}[ht]
\centering
\subfigure[Orthogonal sets of trajectories.]{
\psfrag{D}{$\Delta$}
\psfrag{x}{$x$}
\psfrag{y}{$y$}
\includegraphics[width=1.9in]{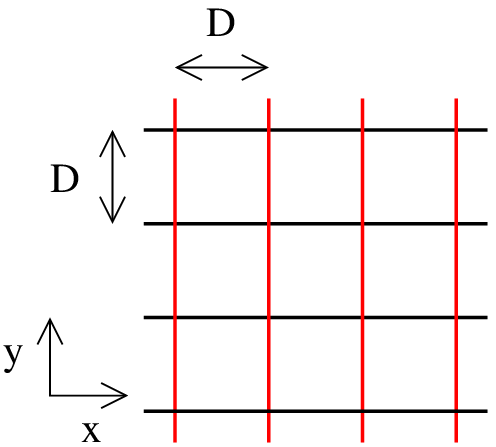}
\label{fig:orthotrajs}
}
\subfigure[Sampled spectra from the two sets.]{
\psfrag{Dx}{$\frac{2\pi}{\Delta}$}
\psfrag{Dy}{$\frac{2\pi}{\Delta}$}
\psfrag{ox}{$\omega_x$}
\psfrag{oy}{$\omega_y$}
\psfrag{O}{$\Omega$}
\includegraphics[width=2.9in]{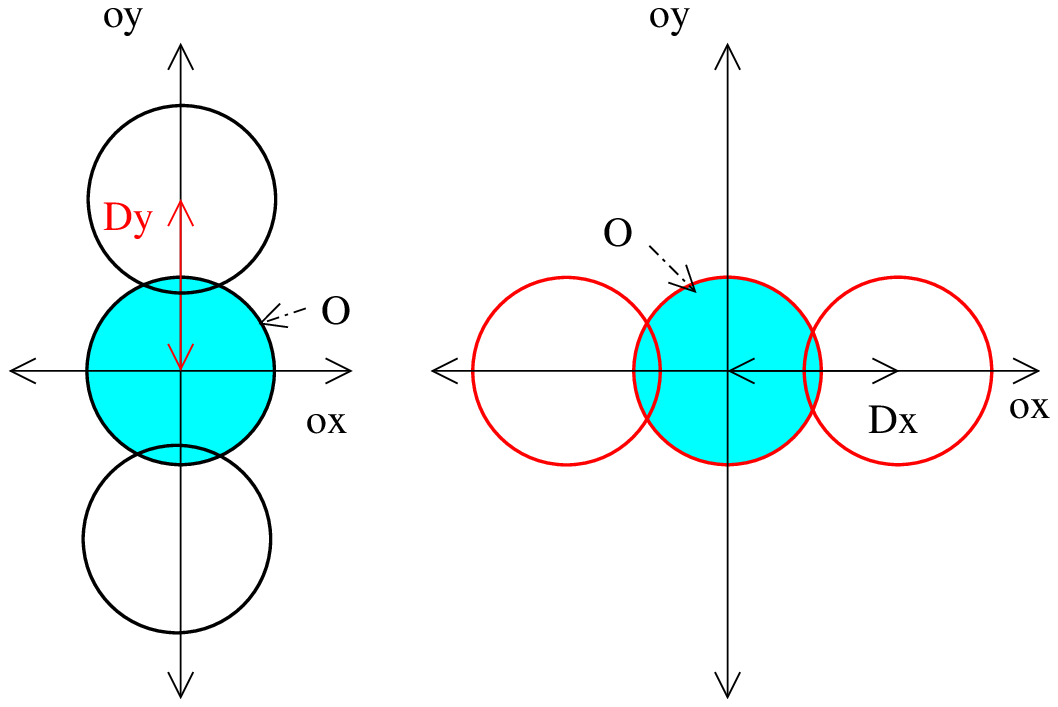}
\label{fig:orthospec}
}
\subfigure[Critical sampling.]{
\psfrag{Dx}{$\frac{2\pi}{\Delta^*}$}
\psfrag{Dy}{$\frac{2\pi}{\Delta^*}$}
\psfrag{ox}{$\omega_x$}
\psfrag{oy}{$\omega_y$}
\psfrag{O}{$\Omega$}
\includegraphics[width=1.9in]{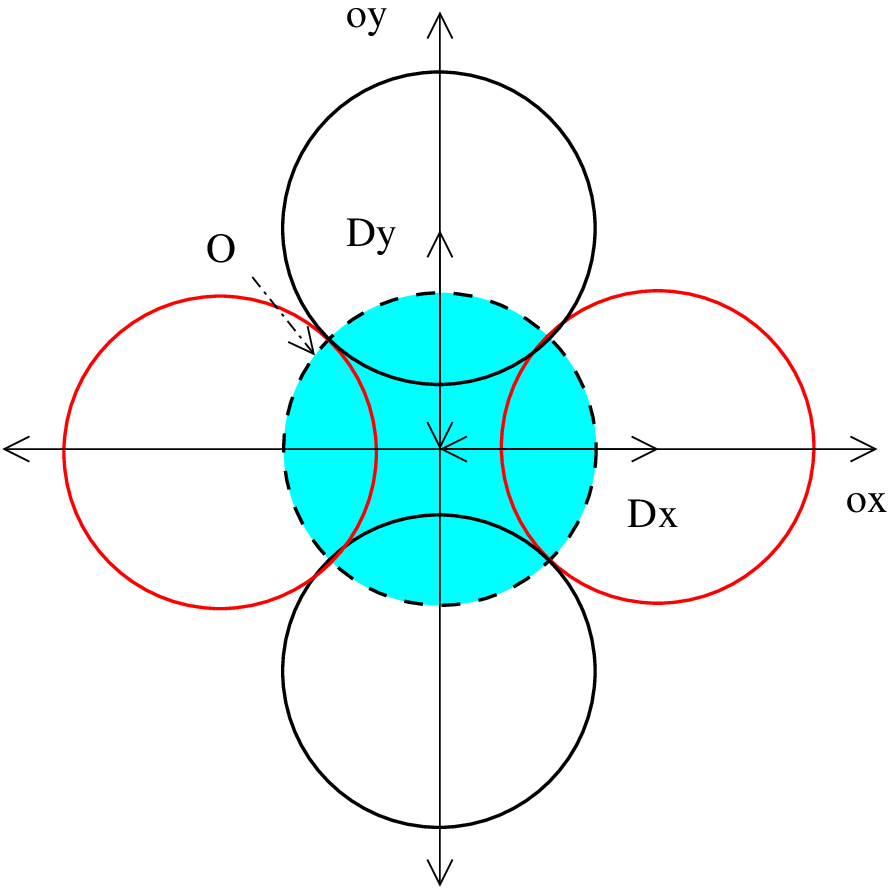}
\label{fig:orthospectouch}
}
\label{fig:subfigureExample}
\caption[Optional caption for list of figures]{Two sets of mutually orthogonal uniform sets and the sampled spectra from samples taken on these trajectories discussed in the first part of Example \ref{eg:orthounif}.}
\end{figure*}

\begin{figure*}[ht]
\centering
\subfigure[Orthogonal sets of trajectories.]{
\psfrag{D}{$\Delta$}
\psfrag{D2}{$2\Delta$}
\psfrag{x}{$x$}
\psfrag{y}{$y$}
\includegraphics[width=2.0in]{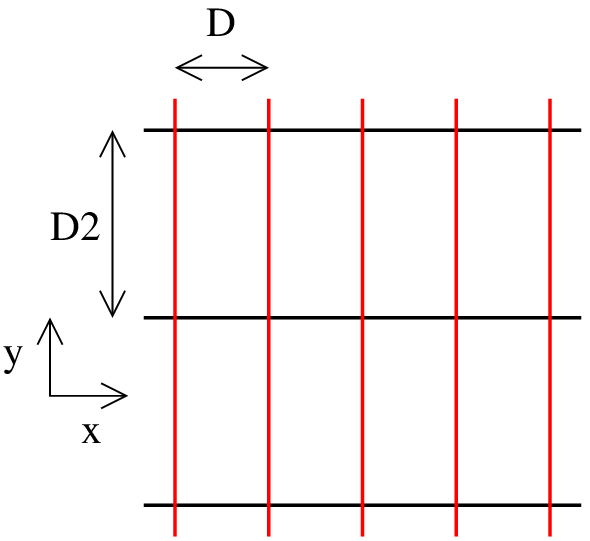}
\label{fig:orthotrajsnew}
}
\subfigure[Sampled spectra from the two sets.]{
\psfrag{d2}{$\frac{2\pi}{\Delta}$}
\psfrag{d}{$\frac{\pi}{\Delta}$}
\psfrag{ox}{$\omega_x$}
\psfrag{oy}{$\omega_y$}
\psfrag{O}{$\Omega$}
\includegraphics[width=4.9in]{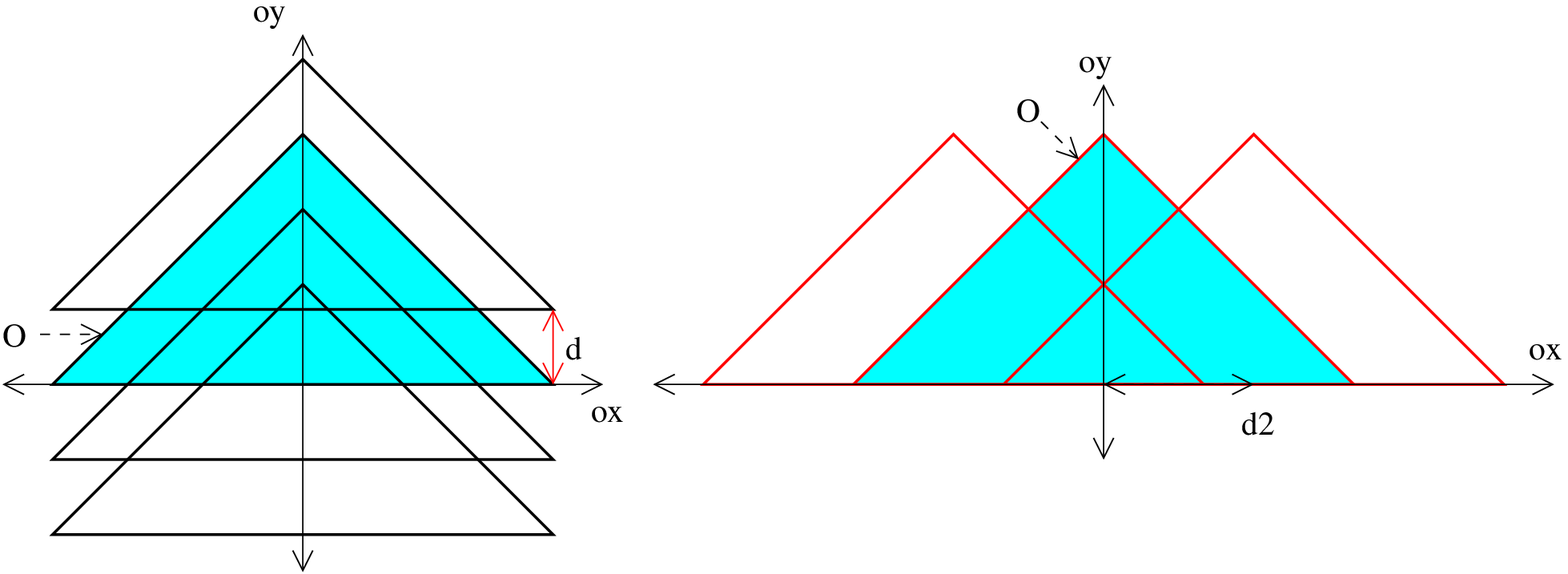}
\label{fig:orthospecnew}
}
\subfigure[Critical sampling.]{
\psfrag{d}{$\frac{\pi}{\Delta^*}$}
\psfrag{d2}{$\frac{2\pi}{\Delta^*}$}
\psfrag{ox}{$\omega_x$}
\psfrag{oy}{$\omega_y$}
\psfrag{O}{$\Omega$}
\psfrag{O1}{$\scriptstyle\Omega_1$}
\psfrag{O2}{$\Omega_2$}
\psfrag{O3}{$\Omega_3$}
\psfrag{O4}{$\Omega_4$}
\psfrag{O5}{$\Omega_5$}
\psfrag{O6}{$\Omega_6$}
\psfrag{O7}{$\Omega_7$}
\includegraphics[width=4.9in]{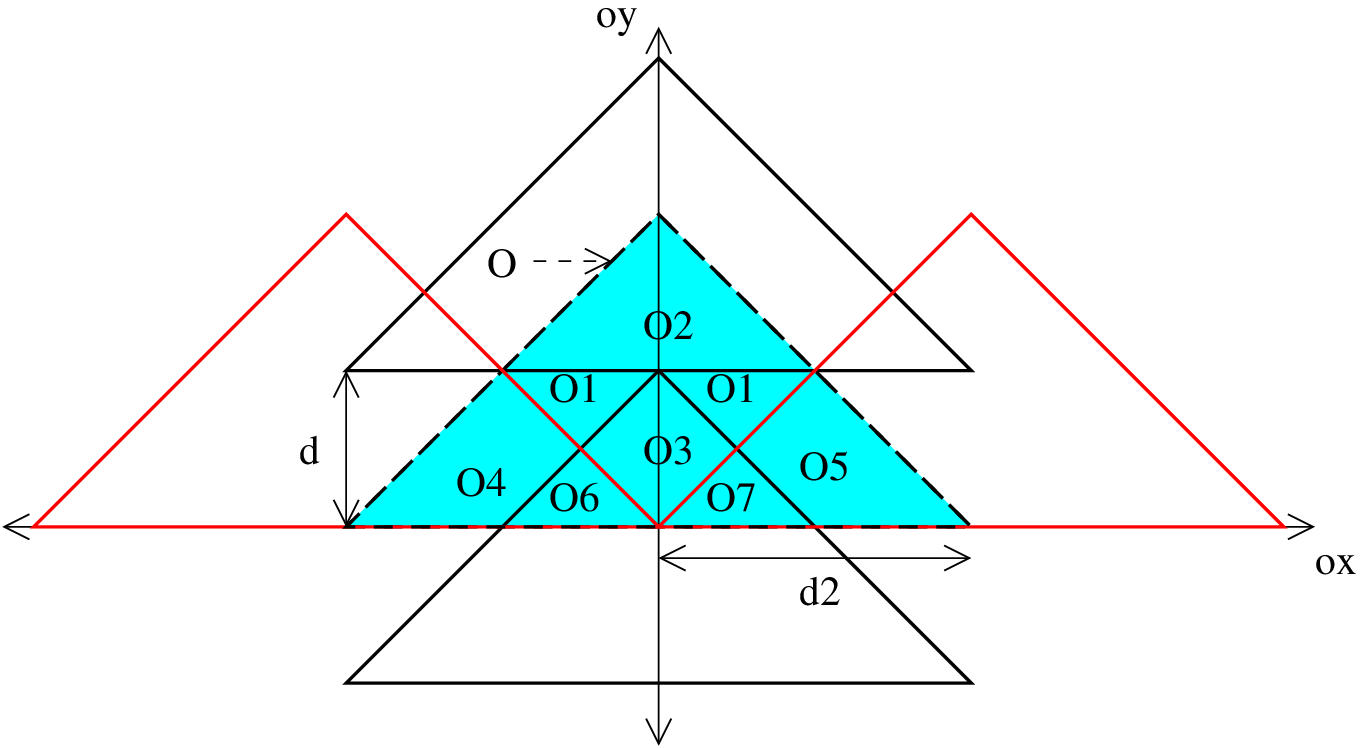}
\label{fig:orthospectouchnew}
}
\caption[Optional caption for list of figures]{Two sets of mutually orthogonal uniform sets and the sampled spectra from samples taken on these trajectories discussed in the second part of Example \ref{eg:orthounif}.}
\end{figure*}

\begin{example}[Mutually orthogonal uniform sets]\label{eg:orthounif}
Consider a trajectory set $P$ composed of the union of two uniform sets, one oriented parallel to the $x$-axis spaced $\Delta_1$ units apart and another parallel to the $y$-axis spaced $\Delta_2$ units apart.
\begin{romannum}
\item Suppose $\Delta_1 = \Delta_2 = \Delta$. Such a set of trajectories is illustrated in Figure \ref{fig:orthotrajs} with the trajectories parallel to $x$-axis ($y$-axis) colored in black (red). Suppose $\Omega$ is a circular disc of radius $\rho$ centered at the origin. The sampled spectra corresponding to the samples from each set of trajectories is illustrated in Figure \ref{fig:orthospec}. From Theorem \ref{thm:gentrajsetnecsuff} it follows that in order to guarantee perfect recovery we must have at least one point of the form $(\pm \frac{\pi}{\Delta}, \pm \frac{\pi}{\Delta})^T$ must lie outside $\Omega$. Hence the maximum value of $\Delta$ required to ensure that the field can be perfectly reconstructed is $\Delta^* = \frac{\sqrt 2 \pi}{\rho}$. From simple geometry it can be seen that when $\Delta = \Delta^*$, the first spectral repetitions along the $x$-axis touch the first spectral repetitions along the $y$-axis as illustrated in Figure \ref{fig:orthospectouch}. Thus when $\Delta < \Delta^*$, these first repetitions along the two axes do not overlap and thus every point $\omega \in \Omega$ satisfies either $F(\omega) = F_s^1(\omega)$ or $F(\omega) = F_s^2(\omega)$ and thus the whole spectrum $F(\omega), \omega \in \Omega$ can be recovered using both sampled spectra.

\item If $\Omega$ is not a circular disc, the conditions are a bit more complex to state. Suppose $\Delta_1 = 2\Delta_2 = 2\Delta$. Such a set of trajectories is illustrated in Figure \ref{fig:orthotrajsnew} with the trajectories parallel to $x$-axis ($y$-axis) colored in black (red).
    Consider this example from \cite[Example 3.2]{unnvet12ISIT} when $\Omega$ has the shape of a right-triangle given by $\Omega:= \{\omega \in \Re^2: \omega_y \geq 0, |\omega_x| +|\omega_y| \leq \rho\}$. The sampled spectra corresponding to the samples from each set of trajectories is illustrated in Figure \ref{fig:orthospecnew}. In this case Theorem \ref{thm:gentrajsetnecsuff} predicts that perfect reconstruction is possible whenever $\Delta < \Delta^* := \frac{2 \pi}{\rho}$. When $\Delta = \Delta^*$ the sampled spectra are as shown in Figure \ref{fig:orthospectouchnew}. We have partitioned $\Omega$ into seven distinct portions, each portion being characterized by the nature of spectral overlap in the two sampled spectra. We see that the portions of $\Omega$ such as $\Omega_6$ and $\Omega_7$ in Figure \ref{fig:orthospectouchnew} are aliased in the spectra from both the uniform sets. Nevertheless, Theorem \ref{thm:gentrajsetnecsuff} shows that whenever $\Delta < \Delta^* := \frac{2 \pi}{\rho}$ perfect reconstruction is possible. In this case, it can be verified that the sampled spectra at the different portions of $\Omega$  satisfy the following relations
\begin{eqnarray*}
F_s^1(\omega) &=& F(\omega), \quad \omega \in \Omega_4\cup\Omega_5\\
F_s^1(\omega) &=& F(\omega) + F(\omega+u_1), \quad \omega \in \Omega_6\cup\Omega_7\\
F_s^2(\omega) &=& F(\omega), \quad \omega \in \Omega_1 \cup \Omega_2\cup\Omega_3
\end{eqnarray*}
where $F_s^1$ and $F_s^2$ denotes the sampled spectra from the trajectories parallel to the $x$-axis and $y$-axis respectively.

Thus the original spectrum can be recovered as
\begin{eqnarray*}
F(\omega) &=& F_s^2(\omega)\chi_{\Omega_1 \cup \Omega_2\cup\Omega_3}(\omega) +F_s^1(\omega) \chi_{\Omega_4\cup\Omega_5}(\omega) \\
&& \qquad +(F_s^1(\omega) - F_s^2(\omega+u_1)) \chi_{\Omega_6\cup\Omega_7}(\omega)
\end{eqnarray*}
where $\chi_A(.)$ denotes the indicator function of set $A$. Hence the original field can be recovered by inverting the Fourier spectrum.
    \qed
\end{romannum}
\end{example}
The two cases considered in Example \ref{eg:orthounif} are distinctly different. In the first case, we saw that the conditions for perfect recovery is the same as the requirement that every part of $\Omega$ is unaliased in at least one of the two spectra. Whereas, in the second case, we saw that perfect recovery is possible even when some portions of the spectrum are aliased in both sampled spectra.

\subsubsection{Concentric equispaced circular trajectories}\label{sec:circtraj}
We now consider an example of a set of non-affine trajectories. Suppose $\Omega$ is a circular disc of radius $\rho$ centered at the origin. Let $C = \{c_i: i \in \bN \cup \{0\}\}$ where $c_i$ denotes a circular trajectory of radius $i \Delta$ centered at the origin. Such a trajectory set is shown in Figure \ref{fig:samtrajsR2c}. It is easily verified that condition~\Con{path3} is satisfied by this trajectory set. It is also known from \cite{myrcha98} that any field bandlimited to $\Omega$ is reconstructible from its values on these trajectories whenever $\Delta < \frac{\pi}{\rho}$. However, this does not verify whether condition~\Con{recon2} is satisfied because exact reconstruction of the field using the scheme in \cite{myrcha98} requires the field values at all points on the continuous circular curve. Nevertheless, one can use the Beurling frame theorem to show that indeed condition~\Con{recon2} is also satisfied when $\Delta < \frac{\pi}{\rho}$. Below we give a version of Beurling's theorem for isotropic bandlimited fields in $\Re^2$, i.e., fields bandlimited to a circular disc in the Fourier transform. This version of the theorem is taken from \cite{benwu00}. The slight difference from the statement in \cite{benwu00} is due to the different definition of the Fourier transform.
\begin{theorem}[Beurling's Covering Theorem]\label{thm:beur}
Let $\Omega \subset \Re^2$ be a circular disc of radius $\rho$ centered at the origin. For any uniformly discrete set $\Lambda \subset \Re^d$ let
\[
\varrho(\Lambda) := \sup_{x \in \Re^d} \inf_{y\in\Lambda}\|x - y\|.
\]
If $\varrho(\Lambda) < \frac{\pi}{2 \rho}$ then $\Lambda \in \amo$ , i.e., any $f \in \bmo$ is uniquely determined from $\{f(x): x \in \Lambda\}$. \qed
\end{theorem}
We can now use this theorem to verify that condition~\Con{recon2} is satisfied by $C$. For $\epsilon > 0$ small enough we note that on each circle $c_i$ we can choose a discrete collection of equispaced points such that the arc-length between nearest neighbors on the circle lies in the interval $[\half \epsilon, \epsilon]$. Let $\Lambda$ denote the set of all such points together with the origin. It follows that $\Lambda$ is uniformly discrete and satisfies $\varrho(\Lambda) < \half (\Delta + \epsilon)$. Now if $\Delta < \frac{\pi}{\rho}$ we can choose $\epsilon$ small enough to ensure that $\varrho(\Lambda) < \frac{\pi}{2 \rho}$. Thus by Beurling's covering theorem it follows that $C \in \nom$  whenever $\Delta < \frac{\pi}{\rho}$.

We now compute the \pd for this trajectory set. Clearly, the maximum density of trajectories is at the origin and hence the supremum in (\ref{eqn:pddefn}) is achieved at the origin. A disc of radius $a$ with center at the origin contains a total of $K \approx \frac{a}{\Delta}$ concentric circles separated by a radial distance of $\Delta$. This leads to a total trajectory length of $\pi \Delta K(K+1)$ whence we get $\ell(C) = \frac{1}{\Delta}$. Since we require $\Delta < \frac{\pi}{\rho}$ to ensure that $C \in \nom$ it follows that the lowest possible path density among trajectory sets that satisfy this condition is achieved when $\Delta$ meets the upper bound and is given by $\frac{\rho}{\pi}$.

We also note that this trajectory set does not satisfy condition~\Kon{BL1} in Section \ref{sec:defterm} because the signals along the circles are not bandlimited. Nevertheless, these signals have a finite essential bandwidth as established in Lemma \ref{lem:bwlincirc} below.
\begin{lemma} \label{lem:bwlincirc}
Let $f \in \bmo$ where $\Omega \subset \Re^2$. Consider a sensor moving at a constant angular velocity of $\nu$ along a circle of radius $a$ centered at the origin. The time-domain signal $s(t) = f((a \cos(\nu t), a\sin(\nu t)))$ is essentially bandlimited to $[-\rho_s, \rho_s]$ where $\rho_s = \nu(1+ a \sup_{\omega \in \Omega} \|\omega\|)$.
\qed
\end{lemma}
We provide a proof in Appendix \ref{lemproof:bwlincirc}.

We note that the signal $s(t)$ in Lemma \ref{lem:bwlincirc} is periodic in time. Hence it can be represented by a Fourier series expansion $s(t) = \sum_{k=-\infty}^\infty s_k \exp({\sf i} k \nu t)$. The coefficients $s_k$ can be computed explicitly using Bessel functions as:
\begin{eqnarray}
s_k &=& \frac{1}{\nu} \int_{\Re^2} F(\omega) \exp({\sf i} k \beta)  J_k(\alpha)d \omega \label{eqn:fourseriescoeff}
\end{eqnarray}
where $\alpha = a \|\omega\|$ and $\beta = \tan^{-1}\frac{\omega_y}{ \omega_x}$. The Fourier series coefficients are negligible for $|k| > \overline k := \lceil \frac{\rho_s}{\nu}\rceil$. Thus the signal $s(t)$ can be approximated as
\begin{equation}
s(t) \approx \sum_{k=-\overline k}^{\overline k} s_k \exp({\sf i} k \nu t). \label{eqn:fourseriesapprox}
\end{equation}
Hence if we have measurements of $s(t)$ at a finitely many points of time, then we can use the approximation of (\ref{eqn:fourseriesapprox}) to estimate the periodic $s(t)$ from these samples.

\subsubsection{Union of spiral trajectories}\label{sec:spirtraj}
\begin{figure}
\centering
\includegraphics[width=3.5in]
{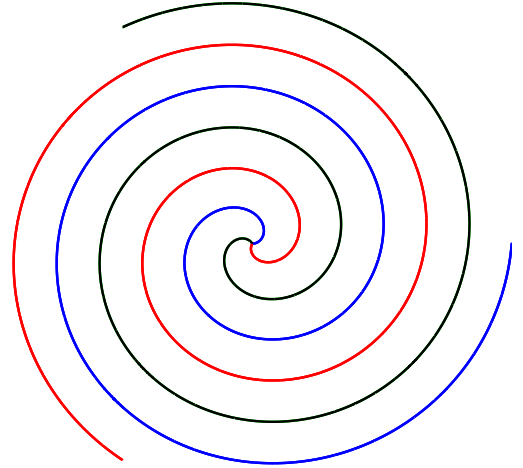}
\caption{A trajectory set in $\Re^2$ composed of a union of three interleaved Archimedean spirals.}
\label{fig:spir}
\end{figure}

Another type of non-affine trajectories are spirals. The geometry of spiral trajectories are particularly convenient to use in MRI applications and hence they have been studied extensively in the field of MRI (see, e.g, \cite{yudsta88}, \cite{benwu00}). For example the work \cite{benwu00} contains some results on perfect reconstruction of bandlimited fields from their measurements taken on spiral trajectories and unions on spiral trajectories. Suppose $\Omega$ is the circular disc of radius $\rho$ centered at the origin. Consider a trajectory set composed of $N$ interleaved spirals of the form \begin{equation}
SP = \{sp_i: i \in \{0,1,\ldots, N-1\}\} \label{eqn:spirunion}
\end{equation}
where $sp_i$ are Archimedean spirals of the form
\begin{eqnarray*}
sp_i(t) = \left(\begin{array}{c} c t \cos (2 \pi (t - i/N))\\ c t \sin (2 \pi (t - i/N))\end{array}\right), t\in \Re^+,
\end{eqnarray*}
for $i \in \{0,1,\ldots, N-1\}$. An example of such a trajectory set for $N =3$ is shown in Figure \ref{fig:spir}. It follows via \cite[Example 3]{benwu00} that $SP$ satisfies condition~\Con{recon2} whenever $\frac{c}{N} <\frac{\pi}{\rho}$. This result is established by applying Beurling's covering theorem (Theorem \ref{thm:beur}) to identify a uniformly discrete set $\Lambda$ of points on the spirals satisfying $\Lambda \in \amo$. Moreover, since $SP$ is composed of a finite number of continuous trajectories it follows trivially that the condition~\Con{path3} is satisfied by this trajectory set. Hence it follows that $SP \in \nom$ whenever $\frac{c}{N} < \frac{\pi}{\rho}$.

In the following lemma we compute the path density of the trajectory set of interleaved spirals.

\begin{lemma}\label{lem:pdspirals}
The path density of the union of spiral trajectories of (\ref{eqn:spirunion}) satisfies $\ell(SP) = \frac{N}{c}$. \qed
\end{lemma}
We provide a proof in Appendix \ref{lemproof:pdspirals}. Since we require $\frac{c}{N} < \frac{\pi}{\rho}$ to ensure that $SP \in \nom$ it follows that the lowest possible path density among trajectory sets that satisfy this condition is achieved when $c$ meets the upper bound and is given by $\frac{\rho}{\pi}$ for all values of $N$.

Before we conclude this section, we mention a caveat on the use of Beurling's covering theorem (Theorem \ref{thm:beur}) for identifying optimal spacings in trajectory sets. This theorem gives only a sufficient condition on sampling sets. Although we used the conditions for this theorem to find conditions on the trajectory sets of Sections \ref{sec:circtraj} and \ref{sec:spirtraj} required to ensure that they are Nyquist trajectory sets, these are only sufficient conditions. Some trajectory sets that do not satisfy these conditions may also be Nyquist trajectory sets. For instance, in Example \ref{eg:orthounif} (i) of Section \ref{sec:unionunif}, we saw that the maximum possible spacing possible for a union of equispaced orthogonal uniform sets is $\Delta^* = \frac{\sqrt 2 \pi}{\rho}$. However, for a union of equispaced orthogonal uniform sets satisfying the conditions of Beurling's theorem we need to have $\Delta < \frac{\pi}{\rho}$. This proves the existence of Nyquist trajectory sets that do not satisfy the conditions of Beurling's theorem.

\subsection{Sampling trajectories for $\Re^d$ where $d \geq 3$}\label{sec:samtrajrd}
\begin{figure}
\centering
\psfrag{x}{$x$}
\psfrag{y}{$y$}
\psfrag{z}{$z$}
\psfrag{a}{$\Delta_1$}
\psfrag{b}{$\Delta_2$}
\psfrag{p}{$P$}
\includegraphics[width=3in]
{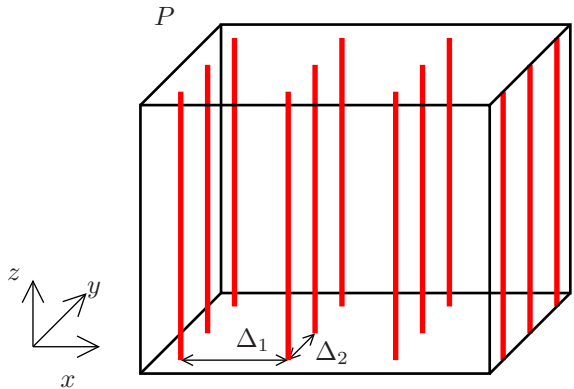}
\caption{A simple example of a uniform set in $\Re^3$.}
\label{fig:recttraj3d}
\end{figure}

For $d \geq 3$, we consider only trajectory sets composed of periodically spaced parallel straight lines analogous to the uniform sets we considered in Section \ref{sec:regpartraj} for $d=2$. Let $\{v_1, v_2,\ldots,v_d\}$ denote a basis for $\Re^d$ such that $v_d$ is a unit vector orthogonal to the hyperplane spanned by $\{v_1, v_2,\ldots,v_{d-1}\}$. In other words $\langle v_i, v_d \rangle = \delta_{i d}$ for all $i \leq d$. Consider trajectories $p_m$ of the form
\begin{equation}
p_m(t) = \sum_{i=1}^{d-1}m_i v_i + t v_d, \qquad t \in \Re \label{eqn:trajpm}
\end{equation}
where $m = (m_1,m_2,\ldots, m_{d-1})^T \in \bZ^{d-1}$. Let $P$ denote the trajectory set
\begin{equation}
P = \{p_m: m \in \bZ^{d-1}\}. \label{eqn:linetrajsdD}
\end{equation}
We refer to such a set of periodically spaced parallel line trajectories as \emph{uniform sets}\footnote{For an exact generalization of uniform sets in $\Re^2$, we have to include an additive shift of some vector $w \in \Re^d$ to $p_m(t)$. We avoid it here to keep the presentation simple. Generalization of the results to non-zero shifts is trivial.} in $\Re^d$. A simple example of a uniform set in $\Re^3$ is illustrated in Figure \ref{fig:recttraj3d} corresponding to
\begin{equation}
v_1 = \left(\begin{array}{c}\Delta_1\\0\\0\end{array}\right), \, v_2 = \left(\begin{array}{c}0\\\Delta_2\\0\end{array}\right), \, v_3 = \left(\begin{array}{c}0\\0\\1\end{array}\right).\label{eqn:3dbasesexample}
\end{equation}

We use $\fmo \subset \nom$ to denote the collection of all uniform sets that form Nyquist trajectory sets for $\Omega$. In other words, $\fmo$ is the collection of uniform sets in $\Re^d$ that satisfy conditions~\Con{recon2} and~\Con{path3}. The following theorem provides conditions on the vectors $\{v_1, v_2,\ldots,v_d\}$ and on the set $\Omega$ so that $P \in \nom$.
\begin{theorem} \label{thm:suffcondsnD}
Let $P$ denote the uniform set defined in (\ref{eqn:linetrajsdD}). Let $\{u_1, \ldots, u_{d-1}\}$ denote vectors in $\Re^d$ satisfying $\langle u_i, v_j\rangle  = 2\pi \delta_{ij}$ for $1\leq i\leq d-1$ and $1\leq j\leq d$ and let $\Omega \subset \Re^d$ denotes a compact convex set with a point of symmetry at the origin. Then we have $P \in \nom$ if
\begin{equation}
\half \sum_{i=1}^{d-1}m_i u_i \notin \Omega, \mbox{ for all } m \in\bZ^{d-1} \setminus \{0\}. \label{eqn:suffcondndD}  
\end{equation}
The condition (\ref{eqn:suffcondndD}) is also necessary in the sense that if there exists $m \in\bZ^{d-1} \setminus \{0\}$ such that $\half \sum_{i=1}^{d-1}m_i u_i \in \overset{\circ}{\Omega}$, where $\overset{\circ}{\Omega}$ denotes the interior of $\Omega$ in $\Re^d$, then $P \notin \nom$.
\qed
\end{theorem}
We provide a proof in Appendix \ref{thmproof:suffcondsnD}. The crux of the proof lies in the fact that the sufficient condition to ensure that $P$ satisfies condition~\Con{recon2} is that for $\epsilon$ small enough the lattice of points defined by $\{\sum_{i=1}^{d-1}m_i v_i + m_d \epsilon v_d: m \in \bZ^d\}$ forms a sampling lattice for $\Omega$, i.e., admits perfect reconstruction \cite{petmid62} of fields in $\bmo$ from its samples on the lattice. Since $\Omega$ is convex and symmetric about the origin this condition is equivalent to (\ref{eqn:suffcondndD}). We illustrate the result for a simple example below.

\begin{example}[Rectangular uniform sets]
Consider the uniform set shown in Figure \ref{fig:recttraj3d} corresponding to the basis vectors of (\ref{eqn:3dbasesexample}). In this case it is easily verified that
\begin{equation}
u_1 = \left(\begin{array}{c}\frac{2\pi}{\Delta_1}\\0\\0\end{array}\right), \, u_2 = \left(\begin{array}{c}0\\\frac{2\pi}{\Delta_2}\\0\end{array}\right).
\end{equation}
Hence, for this trajectory set, the condition in (\ref{eqn:suffcondndD}) is equivalent to the condition that the vectors
\begin{equation*}
\left(\begin{array}{c}\Delta_1\\0\end{array}\right) \mbox{ and } \left(\begin{array}{c}0\\\Delta_2\end{array}\right)
\end{equation*}
generate a sampling lattice for the set $\Omega_{xy}$, the intercept of $\Omega$ with the $xy$ plane defined as $\Omega_{xy} = \{(x,y): (x,y,0)\in \Omega\}$. It is easily verified from classical sampling results \cite{petmid62} that for compact convex sets $\Omega$ with a point of symmetry this is exactly the condition required to ensure that there is no aliasing in the samples of the field measured at points of the form $\{m_1 v_1+ m_2 v_2 + \epsilon m_3 v_3: m \in \bZ^3\}$ for $\epsilon$ small enough. Now suppose that $\Omega$ is a spherical ball of radius $\rho$ centered at the origin. Then $\Omega_{xy}$ is just a circular disc so that the condition to ensure perfect reconstruction becomes
\[
\max\{\Delta_1, \Delta_2\} \leq \frac{\pi}{\rho}.
\]
\qed
\end{example}

The \pd of uniform sets in $\Re^d$ is characterized in the following lemma.
\begin{lemma}\label{lem:pdDdimns}
The \pd of the trajectory set $P$ defined in (\ref{eqn:linetrajsdD}) is given by
\[
\ell(P) = \frac{1}{|\mbox{det}(G)|^\half}
\]
where $\mbox{det}(G)$ denotes the determinant of the $d-1 \times d-1$ Gram matrix defined by
$G_{ij} = \langle v_i, v_j\rangle, 1\leq i,j \leq d-1$. \qed
\end{lemma}
We provide a proof outline in Appendix \ref{lemproof:pdDdimns}.

As we did with Theorem \ref{thm:gentrajsetnecsuff}, the results of Theorem \ref{thm:suffcondsnD} can also be interpreted as a special case of sampling on lattices. Since uniform sets in $\Re^d$ are composed of straight line trajectories we know from Lemma \ref{lem:lintrajBW} that the restriction  $f(p_m(.))$ of the bandlimited field $f$ to any line in the uniform set is bandlimited. It follows that the field along each line can be perfectly reconstructed using uniform sampling along the line, provided the sampling interval $\epsilon$ is small enough. If we further assume that the sampling locations on the lines are aligned with each other, then the collection of the sampling locations on all lines can be expressed as $\{p_m(n \epsilon): n \in \bZ, m \in \bZ^{d-1}\}$ where $\epsilon$ is small enough. Clearly, this collection of points forms a sampling lattice in $\Re^d$. Thus the results of Theorem \ref{thm:suffcondsnD} can be interpreted as the conditions for perfect reconstruction under a special case of sampling on a lattice in $\Re^d$ when the sampling interval along one direction is sufficiently small. The value of $\epsilon$ required can be determined from the bandwidth of $f(p_m(.))$ via Lemma \ref{lem:lintrajBW}.


\section{Optimal sampling trajectories}\label{sec:opttraj}
As we argued earlier, the \pd of a trajectory set captures the total distance required to be traveled per unit area for sampling spatial fields using a mobile sensor moving on the trajectory set. Hence it is of interest to characterize the optimal trajectory set for sampling fields that are bandlimited to a given set $\Omega \subset \Re^d$. We seek a solution to the following problem:
\begin{equation}
\min_{P \in \nom} \ell(P).\label{eqn:optmzntraj}
\end{equation}
In this section we identify partial solutions to the problem, solving it exactly for trajectory sets restricted to some subsets of $\nom$.

We introduce some notation. Suppose $\{b_1, b_2,\ldots, b_d\}$ forms a basis for $\Re^d$. A lattice $b$ generated by the vectors $\{b_1, b_2,\ldots, b_d\}$ is a collection of points in $\Re^d$ of the form
\begin{equation}
b =\left\{\sum_{i=1}^d m_i b_i: m \in \bZ^d\right\}. \label{eqn:samplatexample}
\end{equation}
For any lattice $b$, it is known that (see, e.g., \cite[p. 276]{vaz04}) it is always possible to define a basis $\{c_1, c_2,\ldots, c_d\}$ of $\Re^d$ such that $b$ is generated by $\{c_1, c_2,\ldots, c_d\}$ and the basis vector $c_1$ is a vector with shortest length in $b$, i.e.,
\[
c_1 \in \argmin_{x \in b \setminus \{0\}} \|x\|.
\]
A lattice $b$ is called a \emph{sampling lattice} for a set $\Omega \subset \Re^d$ if every bandlimited field $g: \Re^d \mapsto \Re$ with Fourier transform supported on $\Omega$ can be recovered perfectly using only the values of the field $g$ at points in the lattice. The sampling density of a sampling lattice of the form (\ref{eqn:samplatexample}) is defined as the average number of points per unit volume in $\Re^d$.

In the following lemma we present a simple result that identifies the shortest strategy for visiting all points in a sampling lattice for $\Omega \subset \Re^d$.
\begin{lemma}\label{lem:bestrajforlat}
Let $b$ be a lattice of points in $\Re^d$ of the form (\ref{eqn:samplatexample}) such that $b_1 \in \argmin_{x \in b \setminus \{0\}} \|x\|$ is a shortest vector of the lattice. Let $\clN^b$ denote the set of trajectory sets $P$ of the form (\ref{eqn:trajset}) that visits all points in $b$:
\begin{eqnarray*}
\clN^b &:=& \{P: b \subset \{p_i(t): i \in \bI, t\in\Re\} \mbox{ and } \\
&& \quad P \mbox{ satisfies condition}~\Con{path3}\}.
\end{eqnarray*}
Then the minimum in the following problem
\begin{equation}
\min_{P \in \clN^b} \ell(P)
\end{equation}
is achieved by the uniform set given by $\hat P = \{\hat p_m: m\in \bZ^{d-1}\}$ where
\[
\hat p_m(t) = \sum_{i=1}^{d-1}m_i b_{i+1} + t b_1, t \in \Re.
\]
\end{lemma}
\begin{IEEEproof}
Let $P \in \clN^b$ and $x\in\Re^d$ be arbitrary. Since $P$ satisfies condition~\Con{path3} it follows that there is a continuous curve $q$ of length $\clD^P(a,x) + \littleo(a^d)$ that contains all the lattice points in $b \cap B_a^d(x)$. If one follows the curve $q$ starting at one of its end-points, one will eventually visit all points in $b \cap B_a^d(x)$ before reaching the end of $q$. By construction the shortest path connecting any two points in $b \cap B_a^d(x)$ has a length no less than $\|b_1\|$. Hence the total length of $q$ should satisfy
\[
\text{Length}(q) \geq (\#(b \cap B_a^d(x))-1) \|b_1\|
\]
where $\#$ denotes cardinality. This further implies that
\[
\ell(P) \geq \lim_{a \to \infty} \sup_{x \in \Re^d} \frac{(\#(b \cap B_a^d(x))-1) \|b_1\|}{\mbox{Vol}_d(a)}.
\]
By the definition of $\hat P$ it is easily verified that choosing $P = \hat P$ achieves equality in the above relation. The desired result follows.
\end{IEEEproof}
An immediate corollary of this lemma is the fact that for any set $\Omega$ a uniform set has the shortest path density among all trajectory sets containing a sampling lattice for $\Omega$. In this section we use this lemma together with results from Section \ref{sec:samtraj} to establish some optimality properties of uniform sets.

\subsection{Optimality for $\Re^2$}\label{sec:opt2D}
It is difficult to solve (\ref{eqn:optmzntraj}) exactly because it is difficult to characterize all the trajectory sets that satisfy conditions~\Con{recon2} and~\Con{path3}. However, as we show below, it is possible to identify the optimal trajectory set among those that can be written as a finite union of uniform sets like in (\ref{eqn:unionphat}). Such trajectory sets have the added advantage that they satisfy the desirable property of~\Kon{BL1}, as proved in Lemma \ref{lem:lintrajBW}.

\begin{figure}
\centering
\subfigure[Width $\clW(\Omega)$ of a convex set $\Omega$ and the direction $\hat u$ along which $\Omega$ is narrowest. The orientation $\hat u$ is the minimizer in (\ref{eqn:widthdefn}).]{
\psfrag{u}{$\hat u$}
\psfrag{Om}{$\Omega$}
\psfrag{W(O)}{$\clW(\Omega)$}
\includegraphics[width=1.6in]
{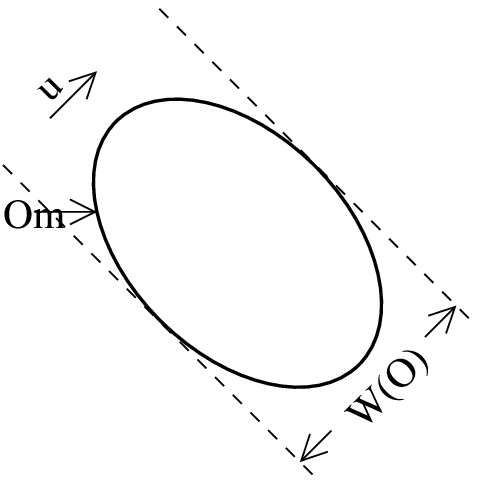}
\label{fig:widthconvset}
}
\subfigure[Optimal uniform set for $\Omega$ corresponding to critical sampling obtained by plugging in $\epsilon  = 0$ in the solution of Theorem \ref{thm:regparopt}.]{
\psfrag{T}{$\frac{2\pi}{\clW(\Omega)}$}
\includegraphics[width=1.6in]
{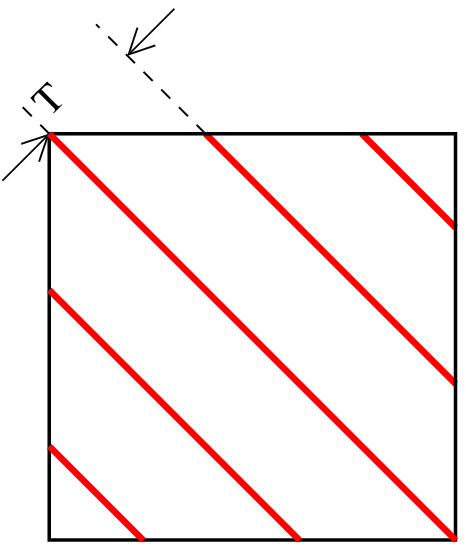}
\label{fig:optunifilled}
}
\caption[]{The width $\clW(\Omega)$ of a convex set $\Omega$, and the limiting optimal uniform set corresponding to critical sampling given by Theorem \ref{thm:regparopt}. The optimal uniform set is oriented orthogonal to the direction $\hat u$ in which $\Omega$ is narrowest.}
\end{figure}

Let $\emo \subset \nom$ denote the collection of trajectory sets $Q$ in $\nom$ such that $Q$ is a finite union of uniform sets of the form
\[
Q = \bigcup_{i=1}^n Q_i
\]
where for each $i$, $Q_i$ is a uniform set. We need the following definitions. For a nonempty compact convex set $\Omega \subset \Re^d$ and any $u \in \Re^d$ let $B^u(\Omega)$ denote the distance between the two parallel supporting hyperplanes of $\Omega$ that are perpendicular to the vector $u$. We refer to $B^u(\Omega)$ as the breadth of $\Omega$ in the direction $u$. The width of $\Omega$ is defined by the relation
\begin{equation}
\clW(\Omega) := \min_{u \in \Re^d} B^u(\Omega). \label{eqn:widthdefn}
\end{equation}
A chord of $\Omega$ is defined as the nonempty intersection of $\Omega$ with a line in $\Re^d$. For $u \in \Re^d$, $\clW^u(\Omega)$ is defined as the maximum length of a chord of $\Omega$ parallel to $u$. The width $\clW(\Omega)$ can alternately also be interpreted as (see, e.g., \cite{val64})
\begin{equation}
\clW(\Omega) = \min_{u \in \Re^d} \clW^u(\Omega). \label{eqn:widthdefn2}
\end{equation}
In the following theorem, we identify the union of uniform sets with minimal path density such that the trajectory set is a Nyquist trajectory set for $\Omega$.
\begin{theorem} \label{thm:regparopt}
For any nonempty compact convex set $\Omega \subset \Re^2$, let $\hat u$ be the minimizer in (\ref{eqn:widthdefn}), and for $\epsilon >0$ let $P^\epsilon$ denote a uniform set given by $P^\epsilon = \{p^\epsilon_j:j\in\bZ\}$ where
\[
p^\epsilon_j(t) = j \left[\left(\frac{2\pi}{\clW(\Omega)}- \epsilon\right) \frac{\hat u}{\| \hat u\|} \right] + \hat u^\perp t, t\in \Re, j\in \bZ
\]
where $\hat u^\perp \in \Re^2$ is orthogonal to $\hat u$. Then $P^\epsilon \in \emo$ for all $\epsilon \in (0, \frac{2\pi}{\clW(\Omega)})$ and is optimal in \pd as $\epsilon \to 0$, i.e.,
\[
\lim_{\epsilon \to 0}\ell(P^\epsilon) = \inf_{Q \in \emo}  \ell(Q) = \frac{\clW(\Omega)}{2\pi}.
\]
\qed
\end{theorem}
The width $\clW(\Omega)$ and the optimizer $\hat u$ appearing in the statement of the theorem are illustrated in Figure \ref{fig:widthconvset}.
This result is established by identifying the union of uniform sets with minimal path density that satisfies the conditions of Theorem \ref{thm:gentrajsetnecsuff}. We provide a proof for the theorem in Appendix \ref{thmproof:regparopt}. This optimality result can be generalized to the case of sampling $d$-dimensional fields on unions of uniform sets of hyperplanes. This result is the topic of Theorem \ref{thm:ddimsregparopt}.

In short, Theorem \ref{thm:regparopt} establishes the optimality of a uniform set $P$ from all trajectory sets in $\emo$. In particular, we have identified a sequence of uniform sets of trajectories indexed by $\epsilon$ with path densities converging to the infimum path density from $\emo$ as $\epsilon$ tends to zero. The limiting uniform set corresponding to critical sampling is obtained by plugging in $\epsilon = 0$. This critically sampled uniform set is a set of parallel lines oriented parallel to $\hat u$ and spaced $\frac{2\pi}{\clW(\Omega)}$ apart, as shown in Figure \ref{fig:optunifilled}.

In addition to the optimality property of Theorem \ref{thm:regparopt}, uniform sets satisfy a different optimality property. In particular, uniform sets are optimal among all trajectory sets in $\nom$ that contain all the points on a sampling lattice for $\Omega$, as the following result shows.
\begin{proposition}\label{prop:bestlatticetraj}
For any set $\Omega \subset \Re^2$, let $Q \in \nom$ be a trajectory set that visits all points in a sampling lattice for $\Omega$. Then there exists a uniform set $P \in \nom$ such that $\ell(P) \leq \ell(Q)$.
\end{proposition}
\begin{IEEEproof}
Let $b$ denote the sampling lattice for $\Omega$ that is visited by $Q$. It is immediate from Lemma \ref{lem:bestrajforlat} that the minimum path density in the collection of all trajectory sets that visit $b$, is achieved by a uniform set. This uniform set automatically satisfies condition~\Con{recon2} since it contains a sampling set for $\Omega$. Since uniform sets further satisfy condition~\Con{path3} by lemma \ref{lem:pdforregular}, the result follows.
\end{IEEEproof}

We now revisit the original problem (\ref{eqn:optmzntraj}) that we wished to solve. The results of Theorem \ref{thm:regparopt} and Proposition \ref{prop:bestlatticetraj} suggest that the uniform set given in Theorem \ref{thm:regparopt} is optimal from a wide class of trajectory sets. It is tempting to consider the possibility that this optimality extends to a wider class so that this uniform set also solves (\ref{eqn:optmzntraj}). Some further evidence for this is obtained by identifying the trajectory sets with minimal path densities among those studied in Sections \ref{sec:circtraj} and \ref{sec:spirtraj}. Assume that $\Omega$ a circular disc $\Omega$ of radius $\rho$. In this case, we know from Theorem \ref{thm:regparopt} that the best path density possible with a uniform set is $\frac{\rho}{\pi}$. We also see that among all the configurations of equispaced concentric circular trajectories in $\bmo$ discussed in Section \ref{sec:circtraj}, the lowest path density is given by $\frac{\rho}{\pi}$ achieved when $\Delta$ is equal to the upper limit of $\frac{\pi}{\rho}$. A similar conclusion also holds for the union of interleaved spiral trajectories in Section \ref{sec:spirtraj}. The lowest path density among all configurations of interleaved spiral trajectories in $\bmo$ discussed in Section \ref{sec:spirtraj} is also equal to $\frac{\rho}{\pi}$ achieved when the parameter $c$ is equal to the upper limit of $\frac{N\pi}{\rho}$. Thus the best known path densities from the examples in Sections \ref{sec:circtraj} and \ref{sec:spirtraj} also match the best path density of a uniform set when $\Omega$ is a circular disc. However, it is not easy to verify whether this is indeed the optimal value of (\ref{eqn:optmzntraj}). We recall that in the classical case of sampling fields in $\bmo$ on a uniformly discrete collection of points, the 
best known lower bound on the sampling density is obtained from Landau's necessary conditions \cite{lan67} (see also \cite[Corollary 1]{groraz96}). It is also known that one can identify a collection of sampling points with a sampling density that is arbitrarily close to this bound \cite{che93}, \cite[Cor. 4.5]{mar08}. It may be possible to extend Landau's conditions and the constructions of \cite{che93} and \cite[Cor. 4.5]{mar08} to Nyquist trajectory sets and thus obtain a solution to the problem (\ref{eqn:optmzntraj}) of identifying the Nyquist trajectory set with minimal path density. 

In the following section we generalize some of the above optimality results for sampling trajectories in two dimensions to higher dimensions.

\subsection{Optimality for $\Re^d$ where $d \geq 3$}\label{sec:optdD}
For fields in $\Re^d$ with $d \geq 3$, we consider only unifom sets of the form (\ref{eqn:linetrajsdD}). We also restrict ourselves to fields bandlimited to sets $\Omega$ that are compact convex subsets of $\Re^d$ and have a point of symmetry at the origin. As in Section \ref{sec:samtrajrd}, let $\fmo \subset \nom$ denote the collection of all uniform sets in $\Re^d$ that form Nyquist trajectory sets for $\Omega$. From Theorem \ref{thm:suffcondsnD} we know the necessary and sufficient conditions on the vectors $\{v_1, v_2, \ldots, v_d\}$ required for $P \in \fmo$. We now seek the solution to the problem
\begin{equation}
\min_{P \in \fmo} \ell(P)\label{eqn:optmznemo}
\end{equation}
where $\Omega \subset \Re^d$ is a compact convex set with a point of symmetry at the origin. In this section we outline a procedure for solving the above problem. In our approach we relate this problem to the problem of designing optimal sampling lattices for static sampling in $\Re^{d-1}$.

Let $P$ be a uniform set as defined in (\ref{eqn:linetrajsdD}) where $p_m$ is defined in (\ref{eqn:trajpm}) with the vectors $\{v_1, v_2,\ldots,v_d\}$ forming a basis for $\Re^d$ with $\langle v_i, v_d\rangle = \delta_{id}$ for all $i \in \{1,\ldots, d\}$. Let $U$ denote a $d \times d$ unitary matrix such that $U v_d = e^d$, the unit vector along the $d$-th principal axis. Define
\begin{equation}
\hat \Omega^U := \{s \in \Re^{d-1}: U^{-1}\left( \begin{array}{c}s\\0\end{array}\right) \in \Omega\}\label{eqn:mhoU}
\end{equation}
Also let $\tilde v_i := U v_i$ and let $\breve v_i \in \Re^{d-1}$ be the vector obtained from $\tilde v_i$ by omitting its last component.
We have the following result that relates the problem of designing optimal uniform sets to the well-studied problem of designing optimal sampling lattices.
\begin{proposition}\label{prop:nesufdD}
Let $\Omega \subset \Re^d$ be a compact convex set with a point of symmetry at the origin and let $P$ be a uniform set in $\Re^d$ as described above. Let $U$, $\hat \Omega^U$ and $\breve v_i$ be as defined above. Then the uniform set $P$ satisfies $P \in \nom$ if and only if the lattice of points defined by $b^U := \{\sum_{i=1}^{d-1} m_i \breve v_i: m \in \bZ^{d-1}\}$ forms a sampling lattice for $\hat \Omega^U$. Furthermore the path density $\ell(P)$ is equal to the sampling density of $b^U$.
\end{proposition}
\begin{IEEEproof}
Consider the field $\tilde f$ defined by
\[
\tilde f(r) = f(U^{-1} r), r \in \Re^d.
\]
Clearly the Fourier transform of $\tilde f$ is supported on the set
\[
\Omega^U := \{Us :s  \in \Omega\}.
\]
Now the problem of sampling the field $f$ along the line $p_m(t)$ is equivalent to sampling the field $\tilde f$ along the line $\tilde p_m(t)$ defined by
\begin{equation}
\tilde p_m(t) = \sum_{i=1}^{d-1}m_i \tilde v_i + t \tilde v_d, \qquad t \in \Re \label{eqn:trajpmtilde}
\end{equation}
where $\tilde v_i = U v_i$. Hence if $\tilde P := \{\tilde p_m : m \in \bZ^{d-1}\}$ then it follows that
\[
P \in \nom \quad \Leftrightarrow \quad \tilde P \in \mathcal{N}_{\Omega^U}
\]
and that the \pd of $\tilde P$ is identical to that of $P$. We know from Theorem \ref{thm:suffcondsnD} that the necessary and sufficient condition for $\tilde P \in \mathcal{N}_{\Omega^U}$ is given by
\begin{equation}
\half \sum_{i=1}^{d-1}m_i \tilde u_i \notin \Omega^U, \mbox{ for all } m \in\bZ^{d-1} \setminus \{0\} \label{eqn:cond1}
\end{equation}
where $\tilde u_i$ are defined as vectors in $\Re^d$ that satisfy $\langle \tilde u_i, \tilde  v_j\rangle  = 2\pi \delta_{ij}$ for $1\leq i\leq d-1$ and $1\leq j\leq d$. Now since $\tilde v_d = e^d$, it follows that $\langle \tilde u_i, e^d\rangle = 0$ for $i \leq d-1$. This fact together with the fact that $\Omega$ is convex and symmetric about the origin implies that the condition (\ref{eqn:cond1}) is equivalent to
\begin{equation}
\half \sum_{i=1}^{d-1}m_i \breve u_i \notin \hat \Omega^U, \mbox{ for all } m \in\bZ^{d-1} \setminus \{0\} \label{eqn:cond2}
\end{equation}
where $\breve u_i \in \Re^{d-1}$ is the vector obtained from $\tilde u_i$ by omitting its last component.
By construction it is clear that $\langle \breve u_i, \breve v_j\rangle  = \langle \tilde u_i, \tilde  v_j\rangle  = 2\pi \delta_{ij}$ for $1\leq i,j\leq d-1$. Thus it follows from \cite{petmid62} that condition (\ref{eqn:cond2}) is exactly the necessary and sufficient condition to ensure that $b^U$ forms a sampling lattice for $\hat \Omega^U$.

We now consider the \pd $\ell(\tilde P)$. Since $\tilde v_d = e^d$ it is clear that the collection of points
\[
\{\sum_{i=1}^{d-1} m_i \tilde v_i + t \tilde v_d: m \in \bZ^d, t\in \Re\}
\]
remains unaltered if we replace $\tilde v_i$ by
\[
\hat v_i:= \left(\begin{array}{c}\breve v_i\\ 0\end{array} \right)
\]
for $1 \leq i \leq d-1$. Thus it follows via Lemma \ref{lem:pdDdimns} that the \pd $\ell(\tilde P)$ satisfies
\[
\ell(\tilde P) = |\det(G)|^{-\half}
\]
where $G$ is a $d-1 \times d-1$ matrix with entries $G_{i,j} = \langle \hat v_i, \hat v_j \rangle = \langle \breve v_i, \breve v_j \rangle$. Since the sampling density of $b^U$ is equal to $|\det (G)|^{-\half}$ (see \cite{petmid62}) and $\ell(P) = \ell(\tilde P)$ the result follows from Lemma \ref{lem:pdDdimns}.
\end{IEEEproof}

As an immediate consequence of the above result we have the following corollary on optimal sampling trajectory sets from $\fmo$.
\begin{corollary}\label{cor:cor1}
Let $\Omega \subset \Re^d$ denote a compact convex set with a point of symmetry at the origin. Among all possible choices of $d \times d$ unitary matrices let $\widehat U$ be the one such that the set $\hat \Omega^{\widehat U}$ defined as in (\ref{eqn:mhoU}) admits a sampling lattice with minimal sampling density in $\Re^{d-1}$. Also suppose that the vectors $\{w_1,w_2,\ldots,w_{d-1}\} \subset \Re^{d-1}$ generate an optimal sampling lattice for fields bandlimited to $\hat \Omega^{\widehat U}$. Let $\hat P$ be a uniform set as defined in (\ref{eqn:linetrajsdD}) where the vectors $v_i$ are given by
\begin{equation}
v_i = \widehat U^{-1}\left(\begin{array}{c} w_i\\0\end{array} \right), 1 \leq i \leq d-1\mbox{ and }v_d = \widehat U^{-1} e^d. \label{eqn:soltnvectorsdD}
\end{equation}
Then $\hat P \in \fmo$ and solves\footnote{We are being imprecise here to keep the presentation simple. In reality the optimal orientation $\hat U$ can be found. But the optimal uniform set $\hat P$ can only be approached since an optimal sampling lattice for $\hat \Omega^{\widehat U}$ can only approached. To be precise, one would have to consider a series of uniform sets $\hat P^\epsilon$ satisfying $\lim_{\epsilon \to 0} \ell(\hat P^\epsilon) = \inf_{Q \in \fmo}\ell(Q)$ akin to the statement of Theorem \ref{thm:regparopt}. The $\hat P$ mentioned here is the limit of $\hat P^\epsilon$.} the optimization problem (\ref{eqn:optmznemo}).\qed
\end{corollary}
The problem of identifying sampling lattices with minimal density is well studied in literature (see, e.g., \cite{petmid62}, \cite{ludolau09}, \cite{mer79}). Such results can be used in conjunction with the above corollary to design optimal uniform sets in $\Re^d$. Below we present some examples of $\Omega \subset \Re^d$ for which we can explicitly solve for the optimal uniform set.

\medskip

\begin{figure}
\centering
\psfrag{x}{$x$}
\psfrag{y}{$y$}
\psfrag{z}{$z$}
\psfrag{1byr}{$\Delta$}
\psfrag{d}{$\Delta=\frac{2\pi}{\sqrt 3\rho}$}
\psfrag{Latp}{}
\psfrag{p}{$P$}
\includegraphics[width=3.5in]
{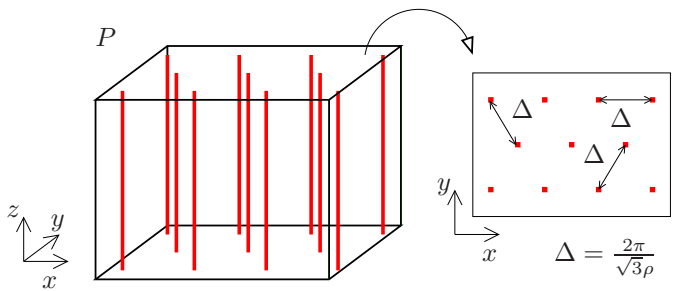}
\caption{Optimal configuration of a uniform set for sampling an isotropic field in $\Re^3$. The cross-section forms a hexagonal sampling lattice as shown.}
\label{fig:opttraj3d}
\end{figure}

\begin{example}
Suppose $\Omega$ is the spherical ball $B_\rho^d \subset \Re^d$. In this case it is easy to see that $\hat \Omega^U$ defined in (\ref{eqn:mhoU}) is a $d-1$-dimensional spherical ball of radius $\rho$ in $\Re^{d-1}$ for all choices of the unitary matrix $U$. Hence without loss of optimality we choose $\widehat U$ in Corollary \ref{cor:cor1} to be the identity matrix. Now suppose that $\{w_1,w_2,\ldots,w_{d-1}\} \subset \Re^{d-1}$ generate a sampling lattice with minimal sampling density for fields in $\Re^{d-1}$ bandlimited to $B_\rho^{d-1}$. Define
\[
v_i := \left(\begin{array}{c}
w_i\\
0
\end{array} \right), 1\leq i \leq {d-1} \qquad v_d = e^d.
\]
Then it follows via Corollary \ref{cor:cor1} that the uniform set $P$ defined in (\ref{eqn:linetrajsdD}) with the above choices for vectors $v_i$ achieves the minimum in (\ref{eqn:optmznemo}). For $d=3$ the vectors $v_i$ can be chosen as
\begin{equation*}
v_1 = \frac{\pi}{\sqrt 3\rho}\left(\begin{array}{c}1\\\sqrt 3\\0\end{array}\right), \, v_2 = \frac{2\pi}{\sqrt 3\rho} \left(\begin{array}{c}1\\0\\0\end{array}\right), \, v_3 = \left(\begin{array}{c}0\\0\\1\end{array}\right).
\end{equation*}
This follows from the fact that a sampling grid based on a hexagonal
lattice is the optimal lattice \cite{petmid62} for two-dimensional isotropic fields. This optimal configuration of a uniform set and the associated hexagonal sampling lattice is illustrated in Figure \ref{fig:opttraj3d}. The exact choices of vectors $v_i$ for fields bandlimited to spherical balls in $\Re^d$ for all $d < 8$ can be obtained from the above result using the results on optimal lattices for isotropic fields presented in \cite[Table C.I]{petmid62}. As mentioned in \cite{petmid62}, these results are based on results on closest packing of spheres in $\Re^d$. \qed
\end{example}
\medskip
\begin{example}
Now suppose $\Omega$ is a cuboidal region in $\Re^3$ given by $\Omega =\{(x,y,z): |x| \leq \rho_x, |y| \leq \rho_y,|z| \leq \rho_z\}$ where $\rho_x \leq \rho_y \leq \rho_z$. We are interested in choosing the vectors $v_1, v_2, v_3$ in a uniform set $P$ of the form (\ref{eqn:linetrajsdD}) such that it solves (\ref{eqn:optmznemo}). In this case it can be seen that the $\hat \Omega^{\hat U}$ mentioned in Corollary \ref{cor:cor1} admits a lattice with minimal sampling density when we have $v_3 = e^3$. In this case $\hat U$ is the identity matrix and the set $\hat \Omega^{\hat U} = \{(x,y): |x| \leq \rho_x, |y| \leq \rho_y\}$. Therefore, by the corollary it follows that (\ref{eqn:optmznemo}) is solved by
\begin{equation*}
v_1 = \left(\begin{array}{c}\frac{\pi}{\rho_x}\\0\\0\end{array}\right), \, v_2 = \left(\begin{array}{c}0\\\frac{\pi}{\rho_y}\\0\end{array}\right), \, v_3 = \left(\begin{array}{c}0\\0\\1\end{array}\right).
\end{equation*}
Such a trajectory set is illustrated in Figure \ref{fig:recttraj3d} where $\Delta_1 = \frac{\pi}{\rho_x}$ and $\Delta_2 = \frac{\pi}{\rho_y}$.\qed
\end{example}

In the following proposition we present a different optimality property of uniform sets. We show that uniform sets are optimal among all sets in $\nom$ that contain all points in a sampling lattice for $\Omega$. This result is the $\Re^d$ analogue of Proposition \ref{prop:bestlatticetraj} for $\Re^2$.
\begin{proposition}\label{prop:bestlatticetrajdD}
Let $\Omega \subset \Re^d$ denote a compact convex set with a point of symmetry at the origin, and $Q \in \nom$ be any trajectory set that visits all points in a sampling lattice for $\Omega$. Then there exists a uniform set $P \in \nom$ such that $\ell(P) \leq \ell(Q)$.
\end{proposition}
\begin{IEEEproof}[Sketch of proof]
This follows from Lemma \ref{lem:bestrajforlat} and Lemma \ref{lem:pdforregular} via the same argument used in proving Proposition \ref{prop:bestlatticetraj}.
\end{IEEEproof}
The above result implies that the optimal uniform set given by Corollary \ref{cor:cor1} is optimal among all trajectory sets that visit all points in a sampling lattice for $\Omega$. However, we have not solved (\ref{eqn:optmzntraj}). Nevertheless, as we mentioned in the case of trajectories in $\Re^2$, it may be possible to extend Landau's necessary conditions \cite{lan67} and the results on optimal configurations of points \cite{che93}, \cite[Cor. 4.5]{mar08} for sampling in $\Re^d$ to Nyquist trajectory sets in $\Re^d$ and thus obtain a solution to (\ref{eqn:optmzntraj}).

%


\section{Generalization to higher dimensional sampling manifolds}\label{sec:manifs}
In Sections \ref{sec:samtraj} and \ref{sec:opttraj} we studied the problem of sampling bandlimited fields in $\Re^d$ on trajectories. Trajectories can be regarded as one-dimensional manifolds in $\Re^d$. We now consider a generalization of this problem in which we replace trajectories with general $\kappa$-dimensional manifolds in $\Re^d$ for $\kappa < d$. Such a setting is relevant in applications where the process of measuring a $d$-dimensional field involves the process of recording the field values on several $\kappa$-dimensional manifolds at a fine resolution. For example consider the problem of sampling and reconstructing a $3$-dimensional object from its views along several $2$-dimensional sections. Such a sampling and reconstruction scheme is often employed in applications like Transmission Electron
Microscopy (TEM) \cite{harperboufeiosthur06} and Magnetic Resonance Imaging (MRI) \cite{vlaboe03} where one tries to recreate the $3$-dimensional structure of an object after imaging or scanning the surfaces of several cross-sections of the object. In such cases, it may be of interest to identify the optimal orientations along which the images should be taken so as to minimize the total area of the images taken while still being able to reconstruct the $3$-dimensional structure of the object.

We use the following terminology. A \emph{$\kappa$-manifold in $\Re^d$} will mean a topological manifold $M$ embedded in $\Re^d$ such that every point in $M$ has a neighborhood in $M$ that is homeomorphic to $\Re^\kappa$. E.g., a $1$-manifold in $\Re^3$ is a curve in $\Re^3$ and a $2$-manifold in $\Re^3$ is a surface in $\Re^3$. Generalizing the concept of a trajectory set, we define a $\kappa$-manifold set in $\Re^d$ as a countable collection of $\kappa$-manifolds in $\Re^d$ of the form $P = \{p_i: i\in\bI\}$ where $p_i$ denotes a $\kappa$-manifold in $\Re^d$ and $\bI \subseteq \bZ$. We also define a generalization of the \pd metric to manifolds. If one assumes that it is inexpensive to increase the sampling density on the sampling manifolds, the metric of interest is the total volume of the all the sampling manifolds per unit spatial volume. We call this metric the \textit{manifold density} of the manifold set. For a set $P$ of $\kappa$-manifolds in $\Re^d$, the manifold density is defined by
\begin{equation}
\mu_\kappa(P) := \limsup_{a \to \infty} \frac{ \sup_{x\in\Re^d} \clV^P_\kappa(a,x)}{\mbox{Vol}_d(a)} \label{eqn:manifdendefn}
\end{equation}
where $\clV^P_\kappa(a,x)$ represents the total $\kappa$-dimensional volume of the portions of the manifolds from $P$ that are located within the ball $B_a^d(x)$ and $\mbox{Vol}_d(a)$ represents the volume of the ball. For example, when $\kappa=1$ the manifold density is exactly the same as the path density defined in (\ref{eqn:pddefn}), and when $\kappa = 2$ and $d=3$, the manifold density $\mu_2(P)$ represents the total area of the sampling surfaces per unit volume. Similarly, we generalize Nyquist trajectory sets to manifolds.

\begin{definition}
A $\kappa$-manifold set in $\Re^d$ of the form $P = \{p_i: i \in \bI\}$ is called a Nyquist $\kappa$-manifold set for $\Omega \subset \Re^d$ if it satisfies the following conditions:

\begin{quote}
\nobreak%
\begin{mondition}
\item \label{mon:recon1} [\emph{Nyquist}] There exists a uniformly discrete collection $\Lambda$ of points on the manifolds in $P$  such that $\Lambda$ admits perfect reconstruction of fields in $\bmo$, i.e., $\Lambda \subset \{r : r \in p_i, i\in \bI \}$ and $\Lambda \in \amo$.

\end{mondition}
\end{quote}
\end{definition}
When the value of $\kappa$ is clear from context, we just say Nyquist manifold set. We use a special notation for collections of Nyquist $\kappa$-manifold sets:

\begin{definition}
For $\Omega \subset \Re^d$ we define $\nomk$ as the collection of all Nyquist $\kappa$-manifold sets for $\Omega$.
\end{definition}
The examples of Nyquist trajectory sets in $\Re^d$ that we considered in Section \ref{sec:samtraj} are Nyquist $1$-manifold sets, or equivalently, elements of $\nomk$ for $\kappa = 1$. We now seek generalizations of these examples to $\kappa > 1$. The other problem of interest is a generalization of (\ref{eqn:optmzntraj}). However, we note that the problem $\min_{P \in \nomk} \mu_\kappa(P)$ is in general ill-posed under the current definition of $\nomk$ since it is possible to construct a set of manifolds with vanishing manifold density, e.g., the degenerate case in which each manifold $p_i$ is confined to a small neighborhood of some point in a sampling lattice for $\Omega$. This problem can be made more meaningful if further restrictions are placed on the elements of $\nomk$ by imposing a condition analogous to condition~\Con{path3} in the definition of $\nom$ in Section \ref{sec:defterm}. However, we do not consider such a definition in this paper to keep the presentation simple. Instead, restricting ourselves to hyperplanes, we establish a generalization of Theorem \ref{thm:regparopt} for sets of hyperplanes, i.e., affine manifolds of dimension $\kappa = d-1$.



For some $h \in \Re^d$ with $\|h\| = 1$,  let $H$ denote the hyperplane $H := \{r \in \Re^d: \langle r,  h\rangle = 0\}$. Generalizing uniform sets of straight line trajectories we define a \emph{uniform hyperplane set} in $\Re^d$ to be a collection of equispaced parallel hyperplanes of the form $P = \{p_i: i \in \bZ\}$ where
\[
p_i = w + i  \Delta h + H
\]
is a shifted version\footnote{We avoid using the notational convention of (\ref{eqn:omegashiftdefn}) here which we reserve for sets in the frequency space.} of the hyperplane $H$. Thus $P$ is a collection of hyperplanes parallel to $H$ located at intervals of $\Delta$ units. The manifold density of the uniform hyperplane set $P$ is given in the following result.
\begin{lemma}\label{lem:ddimsmdunion}
The manifold density of the uniform hyperplane set $P$ defined above is given by
\[
\mu_\kappa(P) = \frac{1}{ \Delta}
\]
where ${\Delta}$ is the spacing between adjacent parallel hyperplanes.
\end{lemma}
\begin{IEEEproof}[Sketch of proof]
This lemma can be proved in exactly the same way as Lemma \ref{lem:pdforregular} by considering $d-1$-dimensional rectangular regions located within a $d$-dimensional spherical ball. We skip the details.
\end{IEEEproof}

We first obtain generalizations of Proposition \ref{prop:neccondunion} and Theorem \ref{thm:gentrajsetnecsuff} to unions of uniform hyperplane sets. Given finite sets of vectors $\{w_i\}_{i=1}^N$, unit vectors $\{h_i\}_{i=1}^N$, and scalars $\{\Delta_i\}_{i=1}^N$, let $P_i$ denote the uniform hyperplane set
\begin{equation}
P_i = \{p_{i,j}: j \in \bZ\} \label{eqn:ddimsmanifset}
\end{equation}
where
\begin{equation}
p_{i,j} := w_i + j  \Delta_i h_i + H_i \label{eqn:hyperplane}
\end{equation}
represents a shifted version of hyperplane $H_i := \{r \in \Re^d: \langle r,  h_i\rangle = 0\}$. The union of the $N$ uniform hyperplane sets $P_i:1 \leq i \leq N$ is defined as
\begin{equation}
P := \bigcup_{i=1}^N P_i. \label{eqn:ddimsunionphat}
\end{equation}
We are interested in finding the conditions on $P$ so that $P \in \nomh$. Since we assume that the field can be sampled at a fine resolution on each of the sampling manifolds $P_i$, we can construct a sufficiently fine lattice in $\Re^d$ composed of points from $P_i$ such that every point on $P_i$ is arbitrarily close to some point from the lattice. For such a lattice, it follows via classical sampling theory \cite{petmid62} that the spectral repetitions in the sampled spectrum obtained from the samples taken on such a lattice are spaced arbitrarily far apart in all directions except one. In this case if $\Omega$ is a compact set, the sampled spectrum from observations on $P_i$ satisfies:
\begin{equation}
F_s^i(\omega) = \sum_{j \in \bZ} \exp({\sf i} \langle j u_i,w_i\rangle) F(\omega + j u_i), \omega \in \Omega\label{eqn:specrepsregparunionddimns}
\end{equation}
where $u_i:= \frac{2 \pi h_i}{\Delta_i}$. This is the analogue to (\ref{eqn:specrepsregparunion}) and is derived explicitly in Lemma \ref{lem:sampspec1set} in Appendix \ref{thmproof:ddimsgentrajsetnecsuffmanifolds}.

Let $\clQ \subset \Re^d$ denote the set of points
\begin{eqnarray*}
\clQ := \left\{\sum_{i=1}^N (-1)^{k_i}\frac{u_i}{2 }: k_i \in \{0,1\}, 1 \leq i \leq N\right\}.
\end{eqnarray*}
Then we have the following results that generalize Proposition \ref{prop:neccondunion} and Theorem \ref{thm:gentrajsetnecsuff}. The following proposition gives a necessary condition that must be satisfied by any union of uniform hyperplane sets that forms a Nyquist manifold set for $\Omega$.
\begin{proposition}\label{prop:neccondunionnD}
Let $\Omega \subset \Re^d$ be a compact convex set and let $P$ denote a union of uniform hyperplane sets of the form (\ref{eqn:ddimsunionphat}). Suppose $\clQ \subset \overset{\circ}{\Omega}(s)$ for some $s \in \Re^d$. Then $P \notin \nomh$. \qed
\end{proposition}
As in the case of Proposition \ref{prop:neccondunion} the above proposition can be proved by constructing a sinusoidal field that vanishes on all the hyperplanes in $P$, and has a Fourier transform supported on $\clQ(-s) \subset \overset{\circ}{\Omega}$. We provide a proof in Appendix \ref{propproof:neccondunionnD}. In other words, this proposition implies that every $P \in \nomh$ must necessarily satisfy
\[
\clQ \nsubseteq \overset{\circ}{\Omega}(s) \mbox{ for all } s \in \Re^d.
\]
This necessary condition must be satisfied by all uniform hyperplane sets $P$ that form Nyquist manifold sets for $\Omega$. When the vectors $h_i$ are linearly independent (which automatically forces the constraint $N \leq d$), it can be shown that the tightest necessary condition given by Proposition \ref{prop:neccondunionnD} is also sufficient. In that scenario, the following theorem provides necessary and sufficient conditions to ensure that $P$ forms a Nyquist manifold set for $\Omega$.
\begin{theorem}\label{thm:ddimsgentrajsetnecsuffmanifolds}
Let $\Omega \subset \Re^d$ be a compact convex set. Let ${P}$ denote the union of uniform hyperplane sets defined in (\ref{eqn:ddimsunionphat}). Suppose that the vectors $\{h_1, h_2, \ldots, h_N\}$ are linearly independent. Then we have the following generalization of Theorem \ref{thm:gentrajsetnecsuff}.
\begin{equation}
P \in \nomh \mbox{ if } \clQ \nsubseteq \Omega(s) \mbox{ for all }s\in\Re^d, \label{eqn:ddimsgentrajssufcondn}
\end{equation}
and moreover,
\begin{equation}
P \notin \nomh \mbox{ if } \clQ \subset \overset{\circ}{\Omega}(s) \mbox{ for some }s\in\Re^d. \label{eqn:ddimsgentrajsneccondn}
\end{equation}
\qed
\end{theorem}
We provide a proof in Appendix \ref{thmproof:ddimsgentrajsetnecsuffmanifolds}. The proof of the first result (\ref{eqn:ddimsgentrajssufcondn}) is very similar to that of the analogous result (\ref{eqn:gentrajssufcondn}) for $\Re^2$. We also note that the second result (\ref{eqn:ddimsgentrajsneccondn}) is a restatement of Proposition \ref{prop:neccondunionnD}. As we saw in the discussion following Theorem \ref{thm:gentrajsetnecsuff}, these results can also be interpreted as the conditions for sampling on a union of shifted lattices \cite{behfar02} under special case in which each of the lattices is finely sampled in all directions except one. We do not repeat the discussion here since the argument is very similar to that for uniform sets in $\Re^2$.

Below, we illustrate the theorem using a simple example of sampling a field in $\Re^3$. In such a case, the hyperplanes in $\Re^3$ are just planes.
\begin{figure*}[ht]
\centering
\subfigure[Uniform set]{
\psfrag{D1}{$\Delta_1$}
\psfrag{x}{$x$}
\psfrag{y}{$y$}
\psfrag{z}{$z$}
\includegraphics[height=2in]{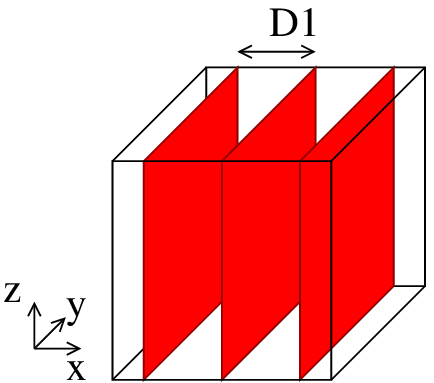}
\label{fig:hyp1}
}
\subfigure[Two orthogonal uniform sets]{
\psfrag{D2}{$\Delta_2$}
\psfrag{Dx}{$\frac{2\pi}{\Delta}$}
\psfrag{Dy}{$\frac{2\pi}{\Delta}$}
\psfrag{ox}{$\omega_x$}
\psfrag{oy}{$\omega_y$}
\psfrag{O}{$\Omega$}
\includegraphics[height=2in]{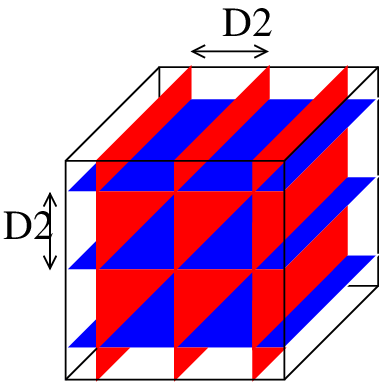}
\label{fig:hyp2}
}
\subfigure[Three orthogonal uniform sets]{
\psfrag{D3}{$\Delta_3$}
\psfrag{Dx}{$\frac{2\pi}{\Delta^*}$}
\psfrag{Dy}{$\frac{2\pi}{\Delta^*}$}
\psfrag{ox}{$\omega_x$}
\psfrag{oy}{$\omega_y$}
\psfrag{O}{$\Omega$}
\includegraphics[height=2in]{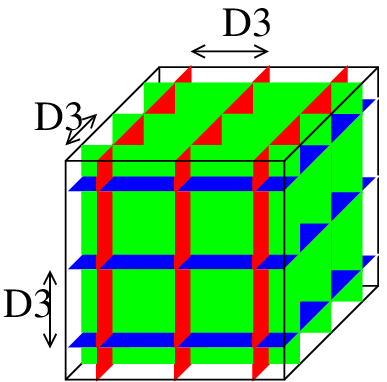}
\label{fig:hyp3}
}
\label{fig:hypExample}
\caption[Optional caption for list of figures]{Three configurations of sampling on planes discussed in Example \ref{eg:unifhypsets}}
\end{figure*}
\begin{example}\label{eg:unifhypsets}
Suppose $\Omega \subset \Re^3$ is a spherical ball of radius $\rho$ centered at the origin. For $\Delta_i > 0$ let $s_{x,i} = (\frac{2\pi}{\Delta_i},0,0)$, $s_{y,i} = (0,\frac{2\pi}{\Delta_i},0)$, and $s_{z,i} = (0,0,\frac{2\pi}{\Delta_i})$. We consider three examples of unions of uniform hyperplane sets.
\begin{enumerate}
\item \textit{Uniform set of planes} Consider a uniform set of planes parallel to the $yz$ plane spaced at intervals of $\Delta_1$ units given by $P = \{p_i: i \in \bZ\}$ where
    \[
    p_i = \{(i\Delta_1,y,z): y,z \in \Re\}.
    \]
    Such a configuration is illustrated in Figure \ref{fig:hyp1}. In this case, we have spectral repetitions along $\omega_x$ axis. Hence it follows from Theorem \ref{thm:ddimsgentrajsetnecsuffmanifolds} that the maximum value of $\Delta_1$ to ensure that $P \in \nomh$ is $\Delta^*_1 = \frac{\pi}{\rho}$. It is easily verified via elementary geometry that $\Delta_1 < \Delta^*_1$ is the condition to ensure that the intersection of two spheres,
    \[
    \Omega \cap \Omega(s_{x,1})  \mbox{ is empty}.
    \]
    It follows that this is exactly the condition to ensure that there is no aliasing in (\ref{eqn:specrepsregparunionddimns}).

\item \textit{Union of two orthogonal uniform sets of planes} Consider a union of two uniform sets of planes one parallel to the $yz$ plane spaced at intervals of $\Delta_2$ units and another parallel to the $xy$ plane spaced at intervals of $\Delta_2$ units given by $P = \cup_{i=1}^2 P_i$ where $P_i = \{p_{i,j}: j \in \bZ\}$ where
    \begin{eqnarray*}
    p_{1,j} &=& \{(j\Delta_2,y,z): y,z \in \Re\} \mbox{ and }\\
    p_{2,j} &=& \{(x,y,j\Delta_2): x,y \in \Re\}.
    \end{eqnarray*}
    Such a configuration is illustrated in Figure \ref{fig:hyp2}. In this case, we obtain spectral repetitions along the $\omega_x$ axis from $P_1$ and along $\omega_z$ axis from $P_2$. Hence it follows from Theorem \ref{thm:ddimsgentrajsetnecsuffmanifolds} that the maximum value of $\Delta_2$ to ensure that $P \in \nomh$ is $\Delta^*_2 = \frac{\sqrt 2 \pi}{\rho}$. It is easily verified that $\Delta_2 < \Delta^*_2$ is the condition to ensure that the intersection of three spheres,
    \[
    \Omega \cap \Omega(s_{x,2}) \cap \Omega(s_{z,2}) \mbox{ is empty}
    \]
    In other words, for each $\omega \in \Omega$ there is at least one set among $\Omega(s_{x,2})$ and $\Omega(s_{z,2})$ that does not contain $\omega$. It follows via (\ref{eqn:specrepsregparunionddimns}) that this is exactly the condition required to ensure that for every $\omega \in \Omega$, there is at least one $i \in \{1,2\}$ such that the sampled spectrum $F_s^i(\omega)$ from the samples on $P_i$ matches the true spectrum $F(\omega)$.

\item \textit{Union of three mutually orthogonal uniform sets of planes} Consider a union of three uniform sets of planes one parallel to the $yz$ plane, another parallel to the $xy$ plane and a third parallel to the $xz$ plane, with each spaced at intervals of $\Delta_3$ units given by $P = \cup_{i=1}^3 P_i$ where $P_i = \{p_{i,j}: j \in \bZ\}$ where
    \begin{eqnarray*}
    p_{1,j} &=& \{(j\Delta_3,y,z): y,z \in \Re\} \\
    p_{2,j} &=& \{(x,y,j\Delta_3): x,y \in \Re\}\mbox{ and }\\
    p_{3,j} &=& \{(x,j\Delta_3,z): x,z \in \Re\}.
    \end{eqnarray*}
    Such a configuration is illustrated in Figure \ref{fig:hyp3}. In this case, we obtain spectral repetitions along the $\omega_x$ axis from $P_1$, along $\omega_z$ axis from $P_2$ and along $\omega_y$ from $P_3$. Hence it follows from the theorem that the maximum value of $\Delta_3$ to ensure that $P \in \nomh$ is $\Delta^*_3 = \frac{\sqrt 3 \pi}{\rho}$. It is easily verified that $\Delta_3 < \Delta^*_3$ is the condition to ensure that the intersection of four spheres,
    \[
    \Omega \cap \Omega(s_{x,3}) \cap \Omega(s_{y,3}) \cap \Omega(s_{z,3}) \mbox{ is empty}.
    \]
    In other words, for each $\omega \in \Omega$ there is at least one set among $\Omega(s_{x,3})$, $\Omega(s_{z,3})$, and $\Omega(s_{y,3})$ that does not contain $\omega$. It follows via (\ref{eqn:specrepsregparunionddimns}) that this is exactly the condition required to ensure that for every $\omega \in \Omega$, there is at least one $i \in \{1,2,3\}$ such that the sampled spectrum $F_s^i(\omega) = F(\omega)$, the true spectrum.

\end{enumerate}
\qed
\end{example}

We now present a generalization of the optimality result of Theorem \ref{thm:regparopt}. We denote by $\emoh$ the collection of manifold sets in $\nomh$ that are unions of uniform hyperplane sets. We have the following optimality result for a uniform hyperplane set.
\begin{theorem} \label{thm:ddimsregparopt}
For any nonempty compact convex set $\Omega \subset \Re^d$, let $\hat u$ be the minimizer in (\ref{eqn:widthdefn}), and for $\epsilon >0$ let $P^\epsilon$ denote a uniform hyperplane set given by $P^\epsilon = \{p^\epsilon_j:j\in\bZ\}$ where
\[
p^\epsilon_j = j \left[\left(\frac{2\pi}{\clW(\Omega)}- \epsilon\right) \frac{\hat u}{\| \hat u\|} \right] + H_{\hat u}
\]
with $H_{\hat u} = \{r \in \Re^d: \langle r, \hat u \rangle = 0\}$. Then $P^\epsilon \in \emoh$ for all $\epsilon > 0$ and is optimal in manifold density as $\epsilon \to 0$, i.e.,
\[
\lim_{\epsilon \to 0}\mu_\kappa(P^\epsilon) = \inf_{Q \in \emoh}  \mu_\kappa(Q) = \frac{\clW(\Omega)}{2\pi}.
\]
\qed
\end{theorem}
This result is established by identifying the union of uniform hyperplane sets with minimal manifold density that satisfies the conditions of Theorem \ref{thm:ddimsgentrajsetnecsuffmanifolds}. We provide a proof in Appendix \ref{thmproof:ddimsregparopt}. In short, Theorem \ref{thm:ddimsregparopt} establishes the optimality of a uniform set $P$ from all trajectory sets in $\emoh$. In particular, we have identified a sequence of uniform sets of hyperplanes indexed by $\epsilon$ with manifold densities converging to the infimum manifold density from $\emoh$ as $\epsilon$ tends to zero. The limiting uniform set corresponding to critical sampling is obtained by plugging in $\epsilon = 0$. This critically sampled uniform hyperplane set is a set of parallel hyperplanes oriented parallel to $\hat u$ and spaced $\frac{2\pi}{\clW(\Omega)}$ apart.

For fields in $\Re^3$, this result implies that for sampling a $3$-dimensional field it is more efficient to sample along one set of equispaced parallel planes rather than along several different sets of equispaced parallel planes. In particular, in Example \ref{eg:unifhypsets}, this result implies that the first configuration of planes has a lower manifold density than the other two examples. This conclusion can be verified by calculating the optimal manifold densities in the three cases using Lemma \ref{lem:ddimsmdunion}. The minimum manifold densities in the first, second and third cases are respectively, $\frac{\rho}{\pi}$, $\frac{\sqrt 2 \rho}{\pi}$, and $\frac{\sqrt 3\rho}{\pi}$.

\section{Reconstruction schemes}\label{sec:recon}
We now consider schemes for reconstructing bandlimited fields using measurements of the field taken along the various sampling trajectories and manifolds proposed above. The reconstruction schemes we propose can be broadly classified into two types. For trajectory sets and manifold sets composed of affine sets such as lines and hyperplanes, we use results from sampling on lattices and unions of lattices. For non-affine trajectories and manifolds, we use schemes for non-uniform sampling based on the Beurling frame theorem \cite{benwu00}.

We first consider sampling trajectories. We know from Lemma \ref{lem:lintrajBW} that uniform sets of the form (\ref{eqn:unifsetdefn}) for fields in $\Re^2$ and (\ref{eqn:linetrajsdD}) for fields in $\Re^d$ satisfy condition~\Kon{BL1}. Hence the field values at all points on these straight line trajectories can be reconstructed from samples taken at uniform spatial intervals. The maximum spacing allowed between adjacent samples on the lines can be calculated from Lemma \ref{lem:lintrajBW}. Suppose that the sensors on each of the parallel lines take samples at uniform intervals and that the sample locations on the various parallel lines are aligned with each other. In such a case the collection of all samples obtained on all these lines effectively corresponds to a collection of samples of the field taken over a periodic lattice of points. Hence, for uniform sets of trajectories, any reconstruction algorithm used for reconstructing bandlimited fields on lattices is sufficient (see, e.g., \cite{petmid62}). For unions of uniform sets like in (\ref{eqn:unionphat}) the resulting set of points forms a union of sampling lattices. In this case reconstruction schemes for sampling on unions of lattices are applicable (see, e.g., \cite{behfar02}). Alternatively, more general schemes for non-uniform sampling such as the projection on convex sets (POCS) algorithm \cite{gro92} can be used.

Reconstruction schemes for the circular trajectories considered in Section \ref{sec:circtraj} are provided in \cite{tewwil88} and \cite{myrcha98}. However, since these trajectory sets do not satisfy condition~\Kon{BL1}, these reconstruction schemes require the exact field values at all points on the circular trajectories. For an approximate reconstruction from samples taken by sensors moving at constant angular velocity one can use the approximation provided in (\ref{eqn:fourseriesapprox}) in conjunction with the reconstruction scheme of \cite{myrcha98}. In addition, if one is interested in reconstructing the field accurately only over a finite number of lines through the origin, then it is sufficient to sample the field at all points of intersection between these lines and the circles and use the reconstruction scheme of \cite{myrcha98}. Alternatively, as we showed in Section \ref{sec:circtraj}, it is possible to reconstruct the field from a uniformly discrete collection of points on the concentric circles chosen to satisfy the conditions of Beurling's covering theorem. The reconstruction algorithm is based on a frame analysis and is described in \cite{benwu00}. The same reconstruction algorithm is also applicable for the union of spiral trajectories of Section \ref{sec:spirtraj}. Similarly, even for arbitrary non-affine trajectories like the one shown in Figure \ref{fig:sampR2b} it is possible to identify a set of sampling points and a corresponding exact reconstruction scheme via the Beurling theorem.

Now consider the case of sampling on uniform hyperplane sets. We can make an argument analogous to that of sampling on uniform sets of trajectories. The restriction of a bandlimited field to the hyperplane is also bandlimited. Like in (\ref{eqn:hyperplane}) any hyperplane $H$ in $\Re^d$ can be expressed in the 
form
\[
H = \{r+b: \langle r, n\rangle = 0, r  \in \Re^d\}
\]
for some vectors $n, b \in \Re^d$. Now if we let $A$ denote a $d \times d-1$ matrix with linearly independent columns that are also orthogonal to $n$, then the hyperplane $H$ can be expressed as $H = \{Ax + b: x \in \Re^{d-1}\}$. In the following lemma we identify the bandwidth of the field $f$ restricted to $H$.
\begin{lemma}\label{lem:BLhypBL}
Let $\{f(r): r\in \Re^d\}$ denote a field bandlimited to some set $\Omega \subset \Re^d$. Consider any hyperplane $H \subset \Re^d$ expressed in the form $H = \{Ax + b: x\in\Re^{d-1}\}$ where $A$ is a $d \times d-1$ matrix. Then the restricted field $g$ defined by
\[
g(x) = f(Ax + b), \quad x \in \Re^{d-1}
\]
is bandlimited to $\Omega_G:=\{A^T \omega: \omega \in \Omega\} \subset \Re^{d-1}$.
\end{lemma}
\begin{IEEEproof}
Using $F$ and $G$ to denote the Fourier transforms of $f$ and $g$ respectively, we have
\[
g(x) = \int_{\omega \in \Omega} F(\omega) \exp({\sf i} \langle \omega, Ax + b \rangle) d\omega
\]
whence
\begin{eqnarray*}
G(s) &=& \int_{\omega \in \Omega} \int_{x} F(\omega) \exp({\sf i} (\langle \omega, Ax + b \rangle - \langle s, x \rangle)) d x d\omega \\
&=&\int_{\omega \in \Omega}  F(\omega) \exp({\sf i} \langle \omega, b\rangle) (2\pi)^{d-1} \delta(A^T \omega - s)  d\omega \\
&=& 0 \mbox{ for } s\notin \Omega_G.
\end{eqnarray*}
\end{IEEEproof}
The conclusion of the lemma above shows that the restriction of a bandlimited field to a hyperplane is also bandlimited. Hence the field values on all points on the hyperplane can be reconstructed exactly using only the field values measured on a sufficiently fine lattice of points on the hyperplane. Thus a sampling scheme using a uniform hyperplane set of the form in (\ref{eqn:ddimsmanifset}) can be practically implemented by using measurements on a $d-1$ dimensional lattice of points for each of the hyperplanes. If the sampling lattices on each parallel hyperplane are aligned with each other then the collection of all sampling points from all the hyperplanes forms a lattice in $\Re^d$. Thus reconstruction schemes for reconstructing bandlimited fields on lattices can be used in these cases (see, e.g., \cite{petmid62}, \cite{gro92}). Similarly, if we consider sampling manifold sets like in (\ref{eqn:ddimsunionphat}) composed of unions of uniform hyperplane sets, reconstruction schemes for sampling on unions of lattices are applicable (see, e.g., \cite{behfar02}). Alternatively, the POCS algorithm \cite{gro92} can also be used for non-uniform sampling.

\section{Conclusion and future work}\label{sec:conc}
In this paper we have introduced the problem of sampling bandlimited fields in $\Re^d$ on trajectories and other lower dimensional manifolds. We have presented various examples of Nyquist trajectory sets and Nyquist manifold sets that admit perfect reconstruction of bandlimited fields. We obtained necessary and sufficient conditions on a union of uniform hyperplane sets to be able to reconstruct bandlimited fields. We also introduced the \pd and manifold density metrics and illustrated some optimality properties of select trajectory sets and manifold sets in terms of these metrics. In particular, we established the fact that a uniform hyperplane set achieves the minimum manifold density among unions of uniform hyperplane sets that admit perfect reconstruction of bandlimited fields. Some of our main results are summarized in Table \ref{tbl:results}. Besides these results, we also have Propositions \ref{prop:bestlatticetraj} and \ref{prop:bestlatticetrajdD} which show that uniform sets in $\Re^2$ and $\Re^d$ are optimal from among all Nyquist trajectory sets that visit all points in a sampling lattice for a given $\Omega$. In addition, we have considered non-affine trajectories such as concentric circles and interleaved spirals and discussed some known results and some new results on sufficient conditions for perfect reconstruction.

\begin{table}
    \begin{tabular}{ | p{1in} | p{1in} | p{1in} | }
    \hline
\multicolumn{2}{|c|}{Lines} & Hyperplanes in $\Re^d$ \\ \hline
Unions of uniform sets in $\Re^2$ & Uniform sets in $\Re^d$ & Unions of uniform hyperplane sets \\ \hline
Prop \ref{prop:neccondunion} (Nec) & Thm \ref{thm:suffcondsnD} (Nec \& Suf) & Prop \ref{prop:neccondunionnD} (Nec) \\ \hline
Thm \ref{thm:gentrajsetnecsuff} (Nec \& Suf) & Prop \ref{prop:nesufdD} (Nec \& Suf) & Thm \ref{thm:ddimsgentrajsetnecsuffmanifolds} (Nec \& Suf)\\ \hline
Thm \ref{thm:regparopt} (Opt)& Cor \ref{cor:cor1} (Opt) & Thm \ref{thm:ddimsregparopt} (Opt)\\ \hline
    \end{tabular}
        \caption{Summary of some results. Each result uses the one above it. ``Nec'' stands for necessary conditions and ``Suf'' for sufficient conditions for perfect reconstruction and ``Opt'' stands for optimality conditions under the sampling scheme specified by the column.}\label{tbl:results}
\end{table}

This paper opens numerous avenues for future work in terms of extensions and generalizations. We have not completely solved the general problem that we posed in (\ref{eqn:optmzntraj}) which seeks the trajectory set in $\nom$ with minimal path density. The best results we have so far are Theorem \ref{thm:regparopt}, Corollary \ref{cor:cor1}, and Proposition \ref{prop:bestlatticetraj} which establish different sub-optimality properties of uniform sets. We also made the interesting observation that for isotropic fields bandlimited to a circular disc $\Omega$ of radius $\rho$ in $\Re^2$, the minimum path density for uniform sets in $\nom$ matches the minimum path density for the sets of concentric circles that we considered as well as the minimum path density for the set of interleaved spirals we considered. This minimum value is given by $\frac{\rho}{\pi}$. It remains to be seen whether this can be bettered by employing some other trajectory sets, perhaps involving a non-uniform collection of sampling trajectories. As mentioned earlier, a potential approach would be to try to generalize Landau's lower bound on the minimum sampling density required for sampling bandlimited fields on points \cite{lan67} (see also \cite[Corollary 1]{groraz96}) to analogous bounds on the minimum path density and minimum manifold density required for sampling on trajectories and manifolds. It may then be possible to extend known results on sampling configurations with minimal sampling density \cite{che93}, \cite[Cor. 4.5]{mar08} to Nyquist trajectory sets to solve (\ref{eqn:optmzntraj}). Similar extensions to the problem of designing manifolds with minimal manifold density are also of interest. However, before proceeding we need to first generalize condition~\Con{path3} to manifolds in such a way that the class of allowed sampling manifolds can be restricted to a meaningful set that does not admit trivial solutions composed of degenerate manifolds.

Many results in this paper can potentially be extended to more general settings. For instance, most of our results are under the assumption that the field is bandlimited to a convex subset of $\Re^d$. Ideas from the proofs of these results can be used to extend these results to more general classes of fields, e.g., bandpass fields bandlimited to non-convex regions, and non-bandlimited fields that form shift-invariant spaces. The results on sampling on circles and spirals in $\Re^2$ can be extended to analogous results on sampling on concentric spherical shells, concentric cylindrical shells, and concentric helixes in $\Re^d$ for $d \geq 2$. The Beurling theorem can be applied to design sampling and reconstruction schemes for such non-affine manifolds.

An important practical aspect that we have ignored in this work is ambient noise which affects the measurement process. In the presence of noise, perfect reconstruction is not possible. In such a case, it would be of interest to study the tradeoff between path (manifold) density and SNR in the reconstructed field and to identify optimal sampling trajectories (manifolds) that optimizes the tradeoff at a given noise level. This tradeoff is also relevant in sampling non-bandlimited parametric fields such as diffusion fields using mobile sensors.

It is also of interest to extend our analysis to the design of $k$-space trajectories for MRI. The bandlimited field model we use in this paper is analogous to using a spatially limited field model in MRI. However, in MRI literature one typically assumes that in addition to the object being spatially limited, the object has more energy in the lower frequencies in $k$-space. It would be interesting to generalize our approach to such a setting wherein one makes assumptions on the field both in the spatial and frequency domains.

\section*{Acknowledgements}
We thank the anonymous reviewers for several helpful suggestions including the connections to Beurling's covering theorem, Bruce Hajek for his insightful comments on a previous version of this work, and Yue Lu for helpful discussions. This research was supported by ERC Advanced Investigators Grant: Sparse Sampling: Theory, Algorithms and Applications –- SPARSAM –- no. 247006.

\appendix
\subsection{List of symbols and definitions}\label{sec:symbols}

\begin{tabular}{p{1.8 cm} p{6.05cm}}
$f: \Re^d \mapsto \bC$ & Field of interest\\
$F: \Re^d \mapsto \bC$ & Fourier transform of the field of interest defined in (\ref{eqn:fourtran})\\
$\|v\|$ & Euclidean norm of the vector $v$\\
$\langle v, w\rangle$ & Euclidean inner product between vectors $v$ and $w$\\
$\delta(.)$& Dirac delta function\\
$\delta_{ij}$& Kronecker delta function evaluated at $(i,j)$\\
Uniformly discrete set& Any set $\Lambda \subset \Re^d$ satisfying $\inf\{\|x-y\|: x,y \in \Lambda, x \neq y\} > 0$\\
$\bmo$ & Collection of fields bandlimited to a given set $\Omega$ as defined in (\ref{eqn:bmodefn})\\
$\amo$ & Collection of all uniformly discrete sets $\Lambda \subset \Re^d$ with the property that any field $f \in \bmo$ can be reconstructed exactly from its values on $\Lambda$\\
$\Omega(s)$& Set $\Omega$ shifted by the vector $s$ as defined in (\ref{eqn:omegashiftdefn})\\
$\ell(P)$& Path density of trajectory set $P$ defined in (\ref{eqn:pddefn})\\
$B_a^d$& The $d$-dimensional spherical ball of radius $a$ centered at the origin in $\Re^d$: $B_a^d := \{y\in\Re^d: \|y\| \leq a\}$\\
$B_a^d(x)$& The $d$-dimensional spherical ball of radius $a$ centered at $x \in \Re^d$: $B_a^d(x) := \{y\in\Re^d: \|y-x\| \leq a\}$\\
$\nom$& Collection of all Nyquist trajectory sets for $\Omega$, i.e., trajectory sets that satisfy conditions~\Con{recon2} and~\Con{path3}\\
Uniform set in $\Re^2$& Trajectory set composed of equispaced parallel lines as defined in (\ref{eqn:unifsetdefn})\\
Uniform set in $\Re^d$& Trajectory set composed of a periodic configuration of parallel lines of the form in (\ref{eqn:linetrajsdD}) \\
\end{tabular}
\newpage
\begin{tabular}{p{1.8 cm} p{6.05cm}}
$\fmo$& Collection of all uniform sets that are in $\nom$\\
$\emo$& Collection of all trajectory sets in $\nom$ that can be expressed as a finite union of uniform sets\\
$\clW(\Omega)$& Width of a convex set $\Omega$ as defined in (\ref{eqn:widthdefn})\\
$\mu_\kappa(P)$& Manifold density of a set $P$ of $\kappa$-manifolds as defined in (\ref{eqn:manifdendefn})\\
$\nomk$& Collection of all Nyquist $\kappa$-manifold sets for $\Omega$, i.e., manifold sets satisfying condition~\Mon{recon1}\\
$\emoh$& Collection of all manifold sets in $\nomh$ that can be expressed as a finite union of uniform hyperplane sets
\end{tabular}

\subsection{Proof of Lemma \ref{lem:pdforregular}}\label{lemproof:pdforregular}
\begin{figure}
\centering
\subfigure[Total length of interest]{
\includegraphics[width=1.6in]
{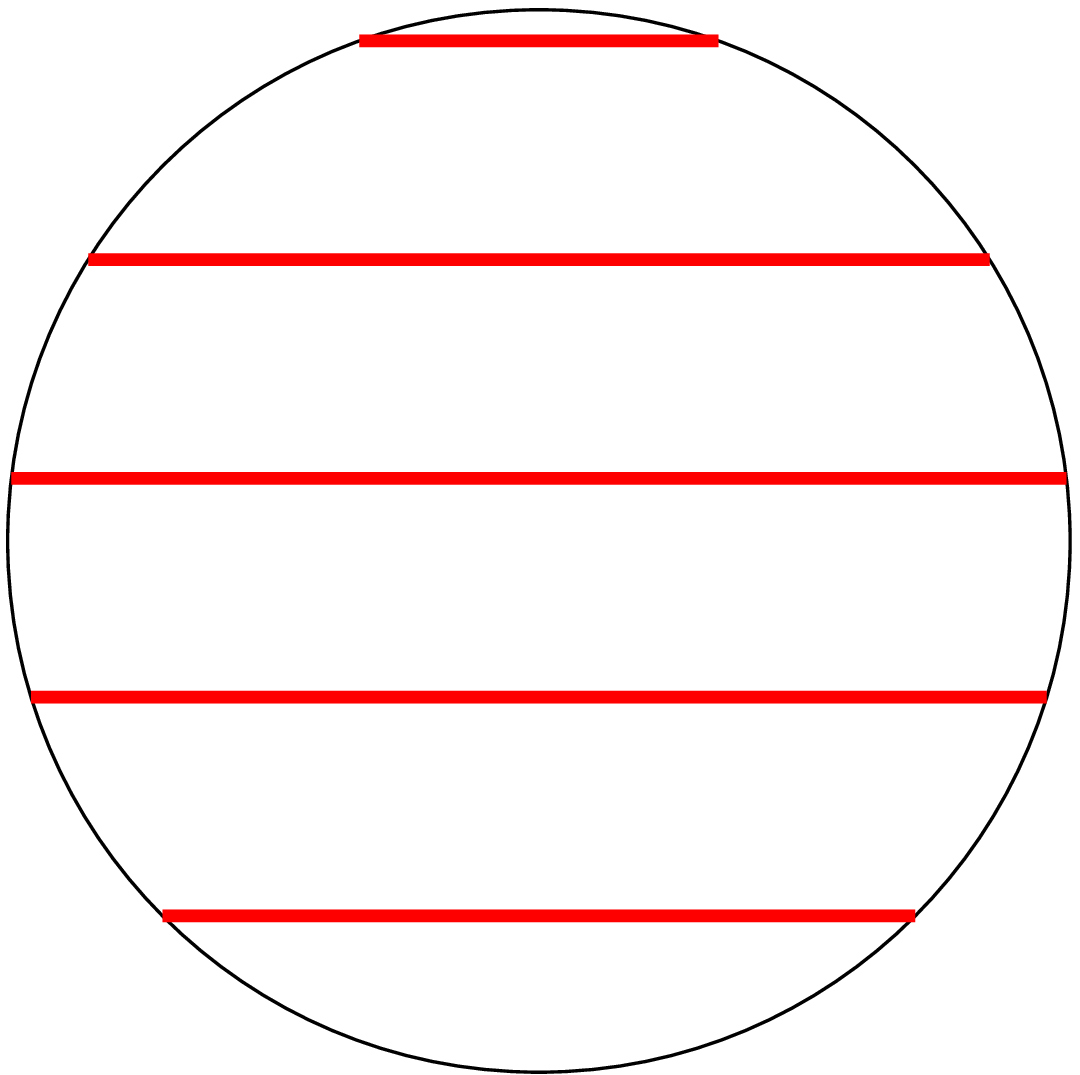}
\label{fig:pdbound}
}
\subfigure[Area representing lower bound in (\ref{eqn:pdproofineqs})]{
\includegraphics[width=1.6in]
{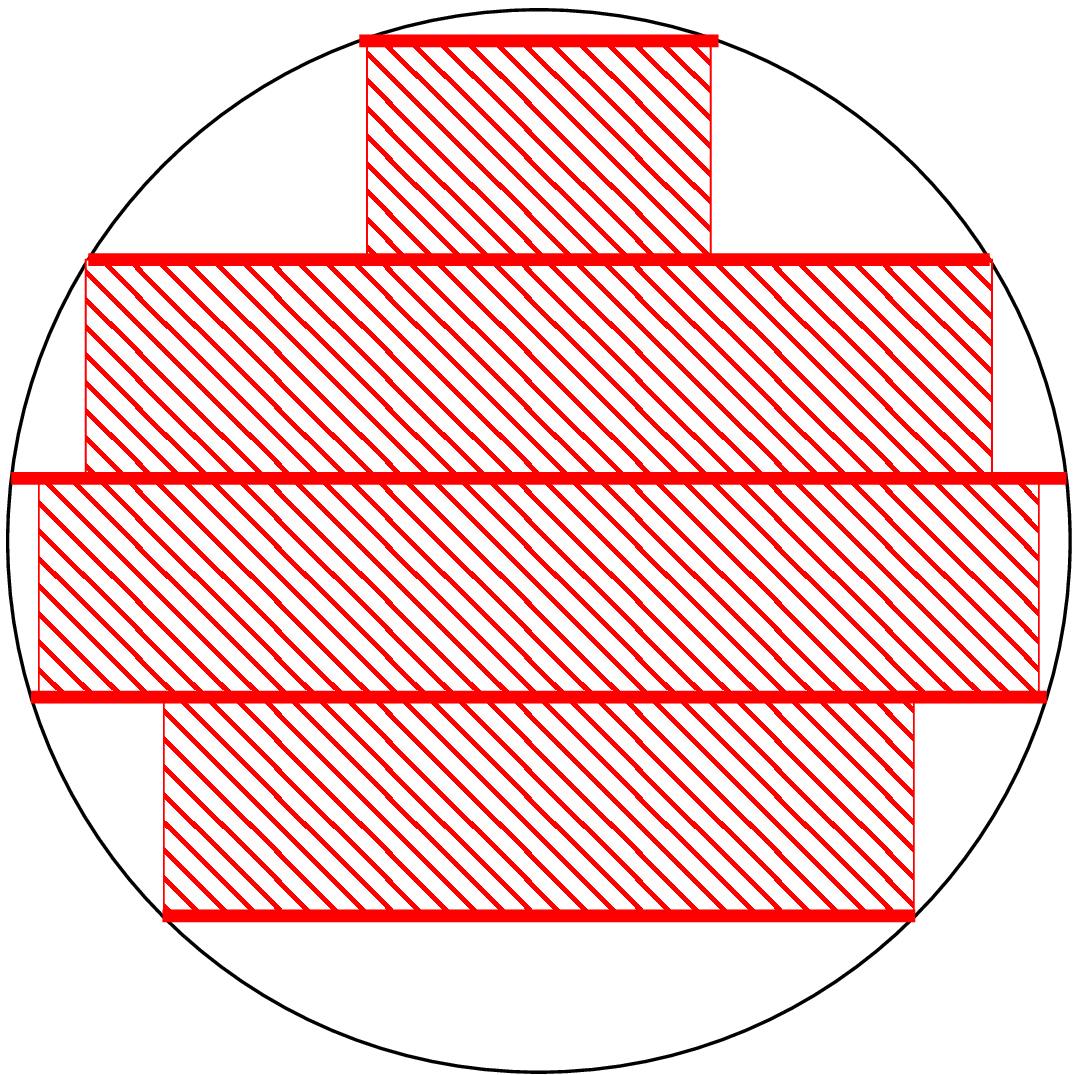}
\label{fig:pdbound1}
}
\subfigure[Area representing upper bound in (\ref{eqn:pdproofineqs})]{
\includegraphics[width=1.6in]
{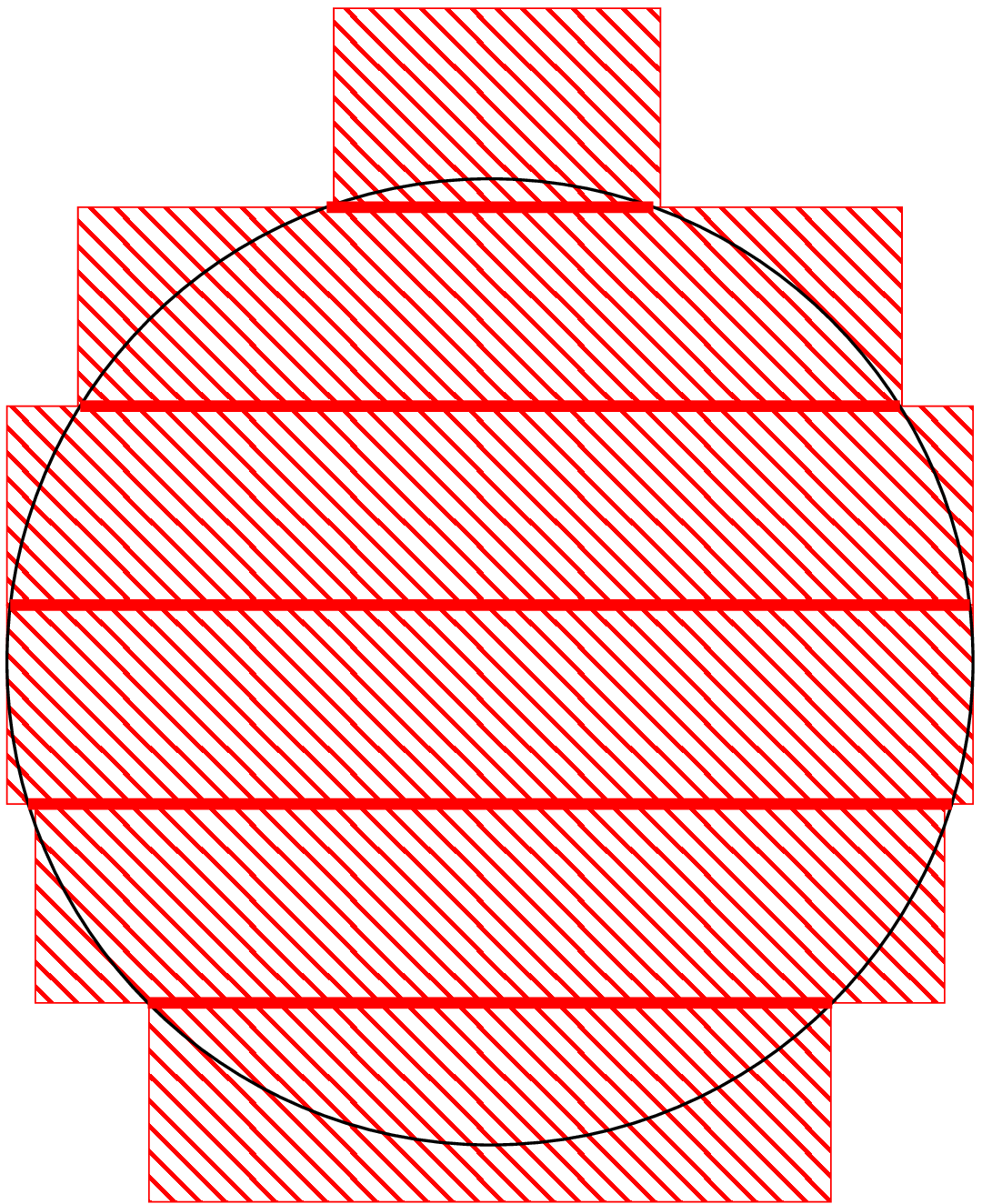}
\label{fig:pdbound2}
}
\caption[Computing the path density of a uniform set]{Computing the path density of a uniform set}
\end{figure}

Let $p_i$ be the straight line trajectory defined in (\ref{eqn:unifsetdefn}). Let $\ell_i$ denote the length of the portion of the line $p_i$ located within the disc $B_a^2(x)$. Let $[i_-, i_+]$ denote the range of indices $i$ such that $\ell_i > 0$. We are interested in the total length of the portions of the lines within the disc $B_a^2(x)$. This total length is as illustrated in Figure \ref{fig:pdbound} when the lines are horizontal. Let $\hat i = \argmax_i \ell_i$. Now we can approximate the area of the disc above and below by rectangular regions with sides $\ell_i \times \Delta$ to obtain the inequality:
\begin{equation}
\sum_{i=i_-}^{i^+} \ell_i \Delta - \ell_{\hat i}\Delta \leq \pi a^2 \leq  \sum_{i=i_-}^{i^+} \ell_i \Delta + 2 a \Delta.\label{eqn:pdproofineqs}
\end{equation}
The areas representing the lower and upper bounds are illustrated respectively in Figures \ref{fig:pdbound1} and \ref{fig:pdbound2}. Now $\clD^{P}(a,x) =\sum_{i=i_-}^{i^+} \ell_i$. Hence we have
\[
\frac{\clD^{P}(a,x) -  \ell_{\hat i}}{\pi a^2} \leq \frac{1}{\Delta} \leq \frac{\clD^{P}(a,x) + 2 a}{\pi a^2}
\]
and (\ref{eqn:regpartrajratio}) follows by taking limits. It is also immediate that line segments of length $\ell_i$ for $i \in [i_-, i_+]$ can all be connected by a continuous curve by connecting portions the ends of adjacent segments via a part of the circumference. The length of such a curve is no more than $\clD^{P}(a,x) + \pi a$ and hence $P$ satisfies condition~\Con{path3}.

\subsection{Proof of Proposition \ref{prop:neccondunion}}\label{propproof:neccondunion}
By the given condition it follows that $\clQ(-s) \subset \overset{\circ}{\Omega}$. Now consider the field
\begin{eqnarray*}
f_c(r) := c\exp(-{\sf i} \langle s, r \rangle) \prod_{i=1}^N\sin\left(\frac{\langle u_i, r- w_i \rangle}{2 }\right), r\in\Re^d.
\end{eqnarray*}
The Fourier transform $F_c$ of the field $f_c$ is supported on the set $\clQ(-s)$ in the Fourier domain. Although the fields $f_c$ are not in $L^2(\bC)$ they can be approximated by functions $g_c \in L^2(\bC)$ defined by
\[
g_c(r)=f_c(r)\prod_{i=1}^2\sinc(\frac{r_i\epsilon}{\pi})
\]
where $\sinc(x) := \frac{\sin (\pi x)}{\pi x}$ and $r_i$ denotes the $i$-th component of vector $r$. The Fourier transform of $g_c$ is supported on points close to $\clQ(-s)$. In particular, we have
\[
G_c(\omega) = 0 \mbox{ if } \omega \notin \cup_{t \in \clQ(-s)} R_\epsilon(t)
\]
where $R_\epsilon$ is the square $[-\epsilon, \epsilon]^2$
and $R_\epsilon(t)$ is the set $R_\epsilon$ shifted by $t$ following our convention from (\ref{eqn:omegashiftdefn}).

Since $\epsilon$ can be made arbitrarily small, it follows from the fact that $\clQ(-s) \subset \overset{\circ}{\Omega}$ that there exists an $\epsilon$ such that $G_c$ is supported on a subset of the set $\overset{\circ}{\Omega}$ in the Fourier domain. It can also be verified that for all $c \in \bC$, the functions $g_c$ vanish on the straight line trajectories in $P$. Hence, all fields $g_c$ are bandlimited to $\Omega$ but they cannot be reconstructed uniquely based on their values on the lines in $P$. Thus it follows that $P$ does not satisfy condition~\Con{recon2} and hence ${P} \notin \nom$.

\subsection{Proof of Theorem \ref{thm:gentrajsetnecsuff}}\label{thmproof:gentrajsetnecsuff}

We prove only (\ref{eqn:gentrajssufcondn}) since result of (\ref{eqn:gentrajsneccondn}) clearly follows from Proposition \ref{prop:neccondunion}. We need the following definition and lemma. The lemma is made more general than what is required for this proof because the general result is required later.

For any set $\clS \subset \bZ^N$, we say that $\clS$ is \emph{lattice-convex} if the following condition holds: if $a, b \in \clS$ then $\delta a + (1-\delta) b \in \clS$ for all $\delta \in (0,1)$ such that $\delta a + (1-\delta) b \in \bZ^N$. For the lattice $\bZ^N$ we define the unit-cell $\clU^N$ as the subset of points
\begin{eqnarray*}
\clU^N := \left\{\sum_{i=1}^N x_i e^i : x_i \in \{0,1\} \mbox{ for all } i \right\} \subset \bZ^N
\end{eqnarray*}
where $x_i$ denotes the $i$-th component of $x$, and $e^i$ denotes the unit-vector along the $i$-th principal axis. A translate of the unit-cell is any shifted version of the $\clU^N$ expressed in the form
$
\{x + z: x \in \clU^N\}
$
for some $z \in \bZ^N$.

\begin{lemma}\label{lem:latconvdecode}
Let $\sfN \subset \bZ^N$ be a compact lattice-convex set. Associated with each element $n \in \sfN$ is a value $v(n)$. These values satisfy the linear relations
\begin{eqnarray}
g_n^i = \sum_{m \in \sfM_{n,i}} \tau_{i,m} v(m) \quad 1\leq i\leq N  \label{eqn:linrelations}
\end{eqnarray}
where $\tau_{i,m} \in \bC \setminus \{0\}$ and $g_n^i$ and $\tau_{i,m}$ are known, and
\[
\sfM_{n,i} := \{m \in \sfN : m -n = k e^i, \mbox{ for some }k \in \bZ\}.
\]
Suppose that these equations are consistent, i.e., they admit at least one solution. Further suppose that $\sfN$ does not contain $\clU^N$ or its translates.


Then the values $\{v(n): n\in \sfN\}$ can be uniquely identified from the relations in (\ref{eqn:linrelations}).
\end{lemma}
\begin{IEEEproof}
Since the equations in (\ref{eqn:linrelations}) are known to be consistent, we just have to show that these equations admit a unique solution. This is equivalent to showing that the homogenous system obtained from those equations with the $g_n^i$ set to zero admits a unique solution. In other words we need to establish the following claim:

\medskip

\noindent \emph{Claim}: Under the conditions of the statement of Lemma \ref{lem:latconvdecode}, the equations
\begin{eqnarray}
0 = \sum_{m \in \sfM_{n,i}} \tau_{i,m} v(m) \quad 1\leq i\leq N  \label{eqn:linrelationshomo}
\end{eqnarray}
admit a unique solution given by $v(n) = 0$ for all $n \in \sfN$.

\medskip

We prove the claim using induction on $N$. Suppose $N=1$. In that case it is easy to see that if $\sfN$ does not contain $\clU^N$ or its translates, $\sfN$  is a singleton and hence the claim holds trivially. Now suppose that the claim holds for $N = N^0 - 1$. Now let $N = N^0$. Suppose $\sfN$ does not contain $\clU^{N^0}$ or its translates. Define
\[
\overline \nu := \max \{n_{N^0}: n = (n_1,n_2,\ldots,n_{N^0})^T \in \sfN\}
\]
and
\[
M(\sfN) := | \{n_{N^0}: n = (n_1,n_2,\ldots,n_{N^0})^T \in \sfN\}|
\]
where $n_{N^0}$ denotes the $N^0$-th component of vector $n$. Let
\[
\sfN^1 := \{n = (n_1,n_2,\ldots,n_{N^0})^T \in \sfN: n_{N^0} = \overline \nu\}.
\]
Now let $\sfN^2 \subseteq \sfN^1$ be defined as
\[
\sfN^2 := \{n \in \sfN^1: n - e^{N^0} \in \sfN\}.
\]
Clearly, by construction, for all $m \in \sfN^1 \setminus \sfN^2$, we have $\sfM_{m,N^0} = \{m\}$. Since $\tau_{N^0,m} \neq 0$ this implies that
\[
v(m) = \frac{g_m^{N^0}}{\tau_{N^0,m}}\mbox{ for } m \in \sfN^1 \setminus \sfN^2
\]
and hence all values $\{v(m): m \in \sfN^1 \setminus \sfN^2\}$ are necessarily equal to $0$.

It is easy to see that there is a set $\tilde \sfN^2 \subset \bZ^{N^0-1}$ such that one can define a bijection between elements $n \in \sfN^2$ and $\tilde n \in \tilde \sfN^2$ as
\[
\tilde n := (n_1,n_2,\ldots, n_{N^0 - 1})^T
\]
and $n = (n_1,n_2,\ldots, n_{N^0 - 1}, \overline \nu)^T = (\tilde n_1,\tilde n_2,\ldots, \tilde n_{N^0 - 1}, \overline \nu)^T$. It is also easy to see from this definition that $\tilde \sfN^2$ is lattice-convex, because $\sfN$ was itself lattice-convex. Furthermore, $\tilde \sfN^2$ does not contain $\clU^{N^0-1}$ or its translates, since $\sfN$ does not contain $\clU^{N^0}$ or its translates. To each element $\tilde n \in \tilde \sfN^2$ we associate a value $w(\tilde n) := v(n)$  where $n = (\tilde n_1,\tilde n_2,\ldots, \tilde n_{N^0 - 1}, \overline \nu)^T$. It is clear by the definition of $\tilde \sfN^2$ and $w(.)$ that the equations in (\ref{eqn:linrelationshomo}) corresponding to $n \in \sfN^2$ and $i \in \{1, 2, \ldots, N^0-1\}$  involve only the values $\{v(m): m \in \sfN^2\}$ and hence can be rewritten with $v(m)$ replaced by $w(\tilde m)$ where $\tilde m$ is the element of $\tilde \sfN^2$ corresponding to $m$. Thus it follows via the induction assumption for $N = N^0 - 1$ applied to $\tilde N^2$ that the values $\{w(\tilde n): \tilde n \in \tilde \sfN^2\}$, or equivalently the values $\{v(m): m \in \sfN^2\}$, are necessarily equal to $0$. Thus we have shown that $v(m) = 0$ for all $m \in \sfN^1$.

In order to complete the proof of the claim, we need to show that $v(m) = 0$ for all $m \in \sfN \setminus \sfN^1$. This follows by a straightforward induction on $M(\sfN)$. Since $\sfN^1 = \sfN$ for $M(\sfN) = 1$, we have established that the claim is true for all lattice-convex sets $\sfN \subset \bZ^{N^0}$ such that $M(\sfN)=1$. Now with $N$ still fixed at $N=N^0$ assume that the statement of the claim is true for all lattice-convex sets $\sfN \subset \bZ^{N^0}$ with $M(\sfN) = M^0-1$. Now consider any lattice-convex set $\sfN \subset \bZ^{N^0}$ with $M(\sfN) = M^0$. By the argument presented in the previous paragraph, we have $v(m) =0$ for all $m \in \sfN^1$ where $\sfN^1 := \{n \in \sfN: n_{N^0} = \overline \nu\}$ as before. It is straightforward to see that $\sfN \setminus \sfN^1$ is a lattice-convex subset of $\bZ^{N^0}$ with $M(\sfN \setminus \sfN^1) = M^0 -1$. Furthermore, since we have $v(m) = 0$ for all $m \in \sfN^1$ the variables $\{v(m): m \in \sfN^1\}$ can be removed from the linear relations (\ref{eqn:linrelationshomo}) to obtain linear relations involving only the variables $\{v(m): m \in \sfN \setminus \sfN^1\}$. By the induction assumption on $M = M^0 - 1$, it follows that $v(m) = 0$ must hold for all $m \in \sfN \setminus \sfN^1$. This completes the induction on $M(\sfN)$ and thus the induction on $N$. Therefore this completes the proof of the claim and hence the lemma.
\end{IEEEproof}



We now proceed to the proof of (\ref{eqn:gentrajssufcondn}). Suppose
\begin{equation}
\clQ \nsubseteq \Omega(s), \mbox{ for all } s\in \Re^2. \label{eqn:nsubsetcondnproof1}
\end{equation}
Let $F$ denote the two-dimensional Fourier transform of any field $f$ bandlimited to $\Omega$. Let $\omega \in \Omega$ be arbitrary. Consider the following set
\begin{equation}
\sfN := \{n \in \bZ^N: \omega + U n \in \Omega\} \label{eqn:clNdefn1}
\end{equation}
where $U$ is a $2 \times N$ matrix with $i$-th column given by $u_i$ defined in (\ref{eqn:uicondndefn}). It is easy to see via the convexity of $\Omega$ that $\sfN$ is a lattice-convex subset of $\bZ^N$. Furthermore, condition (\ref{eqn:nsubsetcondnproof1}) implies that translates of the unit-cell $\clU^N$ are not contained within $\sfN$. To each element $n$ of $\sfN$ we associate the value $v(n) = F(\omega + Un)$.
Now the expressions in (\ref{eqn:specrepsregparunion}) for the spectrum $F_s^i$ of the samples taken on points on the parallel lines in $P_i$ can be rewritten as
\begin{eqnarray}
\lefteqn{\exp({\sf i} \langle \omega, w_i \rangle)F_s^i(\omega)} \nonumber\\
&=& \sum_{j \in \bZ} \exp({\sf i} \langle \omega + j u_i,w_i\rangle) F(\omega + j u_i), \omega \in \Omega,\label{eqn:specrepsregparunionnew}
\end{eqnarray}
for all $i \in \{1,2,\ldots, N\}$. Evaluating these equations at the points $\{\omega + Un: n\in \sfN\}$ we
obtain consistent linear equations in $v(.)$ of the form (\ref{eqn:linrelations}) with
$g_n^i =  \exp({\sf i} \langle \omega + Un, w_i \rangle) F_s^i(\omega + Un)$.
Applying Lemma \ref{lem:latconvdecode} to the set $\sfN$ defined in (\ref{eqn:clNdefn1}) we conclude that for each $\omega \in \Omega$ the values of the Fourier transform $\{F(\omega + Un): n\in \sfN\}$ can be decoded using the values $\{F_s^i(\omega + Un): n\in \sfN, 1\leq i \leq N\}$. Since $\omega \in \Omega$ was arbitrary this proves that $F(\omega)$ can be recovered for all $\omega \in \Omega$. Thus we have verified that $P$ satisfies condition~\Con{recon2} whenever (\ref{eqn:nsubsetcondnproof1}) holds.

We now verify that $ P$ satisfies condition~\Con{path3}. For each $i$, it is clear that for any $x \in \Re^2$ the portions of the lines in the uniform set $P_i$ within $B_a^2(x)$ can be connected by a continuous path containing just these portions as well as portions of the circumference of the circle. Thus the length of such a curve is just $\clD^{P_i}(a,x) + \BigO(a)$. Hence $N$ such curves corresponding to each of the $\{P_i\}_1^N$ can be joined into a single curve with length $\sum_{i=1}^N\clD^{P_i}(a,x) + \BigO(a) = \clD^{ P}(a,x) + \littleo(a^2)$. Thus condition~\Con{path3} in the definition of $\nom$ also holds and hence $P \in \nom$. This completes the proof of (\ref{eqn:gentrajssufcondn}).

\subsection{Proof of Lemma \ref{lem:bwlincirc}}\label{lemproof:bwlincirc}
We have
\[
s(t) = \frac{1}{2\pi} \int_{\Re^2} F(\omega) \exp({\sf i} (a \omega_x \cos(\nu t)  + a \omega_y \sin(\nu t)) ) d \omega
\]
where $F$ is the Fourier transform of $f$. The Fourier transform of $s$ is given by
\begin{eqnarray}
S(\xi) 
&=&  \frac{1}{2\pi} \int_{\Re^2} F(\omega)K(\omega_x,\omega_y,\xi)d \omega\label{eqn:kerneleqn}
\end{eqnarray}
where
\begin{eqnarray}
\lefteqn{K(\omega_x,\omega_y,\xi)}\nonumber\\
&=& \int \exp({\sf i} (a \omega_x \cos(\nu t)  + a \omega_y \sin (\nu t) -\xi t))  dt \nonumber\\
&=&  \int \exp({\sf i}( A \sin(\nu t + B) -\xi t) ) dt \label{eqn:ABCintros}\\
&=& \int \sum_{n=-\infty}^{+\infty} J_n(A)   \exp({\sf i}( n(\nu t + B) -\xi t) ) dt\nonumber\\
&=& 2\pi \sum_{n=-\infty}^{+\infty} J_n(A) \delta(\xi - n \nu) \exp({\sf i} n B) \nonumber
\end{eqnarray}
where $J_n(.)$ denotes the Bessel function of order $n$ and in (\ref{eqn:ABCintros}) we use $A := a \|\omega\|$ and $B:= \tan^{-1}\frac{\omega_x}{\omega_y}$. Using the approximation that $J_n(x) \approx 0$ for $|n| > |x| + 1$ we get the desired result from (\ref{eqn:kerneleqn}) via the bandlimitedness of $F$.

\subsection{Proof of Lemma \ref{lem:pdspirals}}\label{lemproof:pdspirals}
From the geometry of the problem it is clear that the spiral trajectories are densest around the origin and hence the supremum in the definition of path density in (\ref{eqn:pddefn}) is achieved at the origin. Now let $x_i(t) := c t \cos (2 \pi (t - i/N))$ and $y_i(t) := c t \sin (2 \pi (t - i/N))$ denote the $x$ and $y$ coordinates of the trajectory $sp_i$. Then the total length of the trajectories located within a disc of radius $a$ centered at the origin is given by
\begin{eqnarray*}
\clD_{SP}(a,0) &=& \sum_{i=1}^N\int_0^{\frac{a}{c}}(\dot{x_i}(t)^2 + \dot{y_i}(t)^2)^\half dt\\
&=&  N\int_0^{\frac{a}{c}} \sqrt{c^2 + c^2 4\pi^2 t^2} dt\\
&=& \frac{cN}{2}\left[t \sqrt{t^2 4\pi^2 + 1} \right.\\
&& \qquad + \left.\frac{1}{2\pi} \log (t + \sqrt{t^2 + 1/(4 \pi^2)}) \right] \bigg|_0^{\frac{a}{c}}\\
&=& \frac{\pi a^2N}{c} + \littleo(a^2).
\end{eqnarray*}
The result follows.

\subsection{Proof of Theorem \ref{thm:suffcondsnD}}\label{thmproof:suffcondsnD}
Before we proceed we need the following lemmas.
\begin{lemma}\label{lem:conpath3linetrajsdD}
For any $d \geq 2$ let $P$ denote a uniform set in $\Re^d$ as defined in (\ref{eqn:linetrajsdD}). Then for any $u \leq a$ and any $r \in \Re^d$, there is a continuous curve of length $\clD^P(u,r) + \BigO(a^{d-1})$ that contains the portions of the straight line trajectories from $P$ that are located within $B^d_{u}(r)$. Here $\clD^P(u,r)$ is as defined in (\ref{eqn:clLdefn}).
\end{lemma}
\begin{IEEEproof}
We prove this claim by induction on $d$. For $d=2$ the uniform set $P$ comprises a set of parallel lines and the ball $B^2_{u}(r)$ is a circle for any $r \in \Re^2$. We assume without loss of generality that the lines are parallel to the $x$-axis. In this case it is easy to visualize a non-self-intersecting continuous curve that passes through the portions of these lines within $B^2_{u}(r)$.
A possible curve is one that starts with the portion of the line with the lowest ordinate and then proceeds to portions of lines with higher ordinates sequentially, using the portion of the circumference of the circle between adjacent portions as a connection. Hence the total length of such a curve would be no more than the $\clD^P(u,r) + 2\pi u$. Hence the statement of the lemma is proved for $d=2$.

Now assume that the statement of the lemma is true for $d = D-1$. Under this assumption we will now establish the lemma for $d = D$. Let $r \in \Re^D$ and $u < a$ be arbitrary. Consider a uniform set of the form (\ref{eqn:linetrajsdD}). Let $\clK := \{k \in \bZ: \exists m \in \bZ^{D-1}, t\in \Re \mbox{ such that } m_{D-1} = k \mbox{ and } p_m(t) \in B^D_{u}(r)\}$. Let $\hat p_k := \{p_m: m \in \bZ^{D-1}, m_{D-1} = k\}$. Clearly, $P =\{\hat p_k: k \in \bZ\}$.
Now, for each $k \in \clK$, let $I_k$ denote the intersection of $\hat p_k$ with $B^D_{u}(r)$ and let $\ell_k$ denote the total length of the line segments in $I_k$. Clearly, $\sum_{k \in \clK} \ell_k = \clD^P(u,r)$. By the definition of $\hat p_k$ it is clear that the points in $I_k$ lie on a hyperplane $\clH_k$ in $\Re^D$ defined by
\[
\clH_k = k v_{D-1} + \mbox{span}\{v_1,v_2,\ldots,v_{D-2}, v_D\}.
\]
Clearly the intersection $\clH_k \cap B^D_{u}(r)$ is congruent to a ball in $\Re^{D-1}$ of the form $B^{D-1}_{\hat u}$ where $\hat u \leq u$. Hence by the induction hypothesis, it follows that there is a continuous curve $C_k$ that covers $I_k$ with length $\overline \ell_k = \ell_k + O(a^{D-2})$. Since the longest distance between any two points in $B^{D}_{u}(r)$ is no more than $2a$, any two distinct curves $C_k$ and $C_j$ can be joined into a single continuous curve of length $\overline \ell_k + \overline \ell_j + O(a)$. Thus there is a single continuous curve of length
\[
\sum_{k \in \clK} \overline \ell_k  + (|\clK| - 1) O(a) = \clD^P(u,r) + |\clK| (O(a) + O(a^{D-2}))
\]
that covers $\cup_{k\in\clK} I_k$. Since $D \geq 3$, we have $O(a) + O(a^{D-2}) = O(a^{D-2})$. Moreover, by the definition of $\clK$ it is clear that $|\clK| = O(a)$. Thus the total length of the curve is indeed $\clD^P(u,r) + O(a^{D-1})$. Thus the lemma is proved by the principle of mathematical induction.
%
%
%
\end{IEEEproof}

\begin{lemma}\label{lem:noalias3Dgen}
Suppose $\Omega \subset \Re^d$ is a compact convex set with a point of symmetry at the origin. Then $\Omega \cap \Omega\left(\sum_{i=1}^{d-1}m_i u_i\right) \neq \emptyset$ for some $m \in \bZ^{d-1} \setminus \{0\}$ if and only if $\half \sum_{i=1}^{d-1}m_i u_i \in \Omega $.
\end{lemma}
\begin{IEEEproof}
The if part of the statement follows via the convexity and symmetry of $\Omega$. Now suppose
\[
s \in \Omega \cap \Omega\left(\sum_{i=1}^{d-1}m_i u_i\right) \mbox{ for some } m \in \bZ^{d-1} \setminus \{0\}.
\]
If $\tilde s:= s - \sum_{i=1}^{d-1}m_i u_i$, it follows by symmetry that $\{s,-s, \tilde s, -\tilde s\} \subset \Omega$. By convexity therefore $\half(s - \tilde s) = \half \sum_{i=1}^{d-1}m_i u_i \in \Omega$.
\end{IEEEproof}

\medskip

Now we proceed to the proof of Theorem \ref{thm:suffcondsnD}. We first prove the sufficiency of (\ref{eqn:suffcondndD}). Suppose $P$ satisfies (\ref{eqn:suffcondndD}). 
Lemma \ref{lem:conpath3linetrajsdD} shows that $P$ satisfies condition~\Con{path3}. Now let us consider condition~\Con{recon2}. Since the field can be measured at all points on the lines in $P$ it follows from the structure of $P$ in (\ref{eqn:linetrajsdD}) that we can sample the field at points of the form
\begin{equation}
\sum_{i=1}^{d-1}m_i v_i + m_d \epsilon v_d, \qquad m = (m_1,m_2,\ldots,m_d)^T \in \bZ^d \label{eqn:3Dlattice}
\end{equation}
where $\epsilon > 0$ is arbitrary. The points in (\ref{eqn:3Dlattice}) correspond to a lattice with finite density in $\Re^d$ generated by the vectors $\{w_1, w_2,\ldots, w_d\}$ where $w_i = v_i, 1\leq i\leq d-1$ and $w_d = \epsilon v_d$. It follows from \cite{petmid62} that
the Fourier transform of the sampled field, sampled at points on the lattice (\ref{eqn:3Dlattice}) is made up of repetitions of the Fourier transform on the reciprocal lattice of points of the form $\{\sum_{i=1}^{d}m_i u_i:m \in \bZ^{d}\}$ generated by the vectors $\{u_1,u_2,\ldots, u_d\}$ chosen to satisfy
\begin{equation}
\langle u_i,  w_j \rangle = 2\pi \delta_{ij} \label{eqn:Fdomlattice}
\end{equation}
where $\delta_{ij}$ denotes the Kronecker delta function. Clearly, the first $d-1$ vectors $\{u_i\}$ defined above are exactly as defined in the statement of the theorem. Now since $\epsilon$ and hence $\|w_d\|$ can be made arbitrarily small, it follows that $\|u_d\|$ can be made arbitrarily large. Hence for compact sets $\Omega$, the spectral repetitions in the direction $u_d$ can be ignored and hence perfect reconstruction is guaranteed provided spectral repetitions at the points
$\{\sum_{i=1}^{d-1}m_i u_i:m \in \bZ^{d-1}\}$ do not overlap. Or equivalently,
\begin{equation}
\Omega \cap \Omega\left(\sum_{i=1}^{d-1}m_i u_i\right) = \emptyset, \qquad m \in \bZ^{d-1}\setminus \{0\}. \label{eqn:perfcondn3D}
\end{equation}
Thus it follows via Lemma \ref{lem:noalias3Dgen} that under the assumptions of Theorem \ref{thm:suffcondsnD}, we have $P \in \nom$ provided
\begin{equation}
\half \sum_{i=1}^{d-1}m_i u_i \notin \Omega, \mbox{ for all } m \in \bZ^{d-1}\setminus \{0\}. \label{eqn:condnnoaliasnD}
\end{equation}

Now let us consider the necessary condition. Suppose $\exists m \in\bZ^{d-1} \setminus \{0\}$ such that $\half \sum_{i=1}^{d-1}m_i u_i \in \overset{\circ}{\Omega}$. Consider the field defined by
\[
f_\epsilon(r) = \sin(\half \sum_{i=1}^{d-1} m_i \langle r,  u_i\rangle)\prod_{i=1}^d \sinc(\frac{ \langle r, e^i \rangle \epsilon}{\pi}) , r\in\Re^d
\]
where $e^i$ denotes the unit vector along the $i$-th principal axis. The Fourier transform of $f_\epsilon$ is supported within a cubic box of width $2 \epsilon$ centered at the points $\{\half \sum_{i=1}^{d-1} m_i u_i, - \half \sum_{i=1}^{d-1} m_i u_i\}$. By the condition on $m$ it follows that  there exists some $\epsilon$ small enough such that $f_\epsilon \in \bmo$. Now, from the condition on $\{u_i\}$ given in the statement of the theorem, it follows that $f_\epsilon(p_m(t)) = 0$ for all $t \in \Re$ and for all $m\in \bZ^{d-1}$ where $p_m(t)$ is as defined in (\ref{eqn:trajpm}). Hence $f_\epsilon$ cannot be recovered from its samples on $p_m(t)$ since it is not distinguishable from the field $g \in \bmo$ that is identically $0$ - i.e., $g(r) = 0$ for all $r \in \Re^d$. Thus $P \notin \nom$.

\subsection{Outline of proof of Lemma \ref{lem:pdDdimns}}\label{lemproof:pdDdimns}
Lemma \ref{lem:pdDdimns} can be proved using the same approach as in the proof of Lemma \ref{lem:pdforregular}. Just as we approximated a circle with rectangles in the proof of Lemma \ref{lem:pdforregular} we will now approximate a $d$-spherical ball with parallelotopes. As before, let $B^d_a(x)$ denote a $d$-dimensional spherical ball of radius $a$ centered at $x \in \Re^d$. For $r \in \Re^d$ let $\beta_i(r)$ denote the coefficients in the basis expansion
$
r = \sum_{i=1}^d \beta_i(r) v_i.
$
Now for $m \in \bZ^{d-1}$ define
\[
S_m := \{r \in B^d_a(x): \beta_i(r) \in [m_i, m_i+1), 1\leq i\leq d-1\}.
\]
Let
\[
M_a := \{m \in \bZ^{d-1}: p_m(t) \in B^d_a(x) \mbox{ for some } t \in \Re\}
\]
with $p_m(t) = \sum_{i=1}^{d-1}m_i v_i + t v_d$ as defined in (\ref{eqn:trajpm}).
For each $m \in M_a$ let $\ell_m(a)$ denote the length of the line segment representing the intersection of the line $p_m(.)$ with $\clS^d_a$.
We then have
\[
\mbox{Vol}(B^d_a(x)) = \sum_{m \in M_a} \mbox{Vol}(S_m) + \littleo(a^{d-1}).
\]
Now we can approximate $S_m$ with the parallelotope generated by the vectors $\{v_1,v_2,\ldots,v_{d-1}, {\ell_m(a)}{}v_d\}$ so that $\mbox{Vol}(S_m) \approx {|\mbox{det}(G)|^\half}{}\ell_m(a)$. In fact it can be shown that
\[
\sum_{m \in M_a} \mbox{Vol}(S_m) = \sum_{m \in M_a} {|\mbox{det}(G)|^\half}{}\ell_m(a) + \littleo(a^{d-1}).
\]
Now $ \sum_{m \in M_a} \ell_m(a) = \clD^P(a,x)$ and hence the result follows.

\subsection{Proof of Theorem \ref{thm:regparopt}}\label{thmproof:regparopt}
From Theorem \ref{thm:gentrajsetnecsuff} it easily follows that $P^\epsilon \in \emo$. Suppose $P^\epsilon$ is expressed in the form $P_1$ as defined in (\ref{eqn:trajthetai2}). Then the vector $u_1$ defined in (\ref{eqn:uicondndefn}) corresponding to $ P_1 = P^\epsilon$ is given by
\[
u_1 = \left(1- \frac{\epsilon \clW(\Omega)}{2\pi}\right)^{-1} \hat u
\]
and hence by the definition of $\hat u$ and by (\ref{eqn:gentrajssufcondn}) we have $P^\epsilon \in \emo$ for all $\epsilon > 0$. Also from Lemma \ref{lem:pdforregular} we have
\[
\lim_{\epsilon \to 0}\ell(P^\epsilon) = \frac{\clW(\Omega)}{2\pi}.
\]

In order to complete the proof we will establish that for all $P \in \emo$, we have $\ell(P) \geq \frac{\clW(\Omega)}{2\pi}$. Suppose $P \in \emo$ is expressed in the form defined in (\ref{eqn:unionphat}). Using the notations and assumptions of Theorem \ref{thm:gentrajsetnecsuff} it follows that if $P \in \emo$ we need
\[
\clQ \nsubseteq \overset{\circ}\Omega(s) \mbox{ for all }s\in\Re^d
\]
where $\overset{\circ}\Omega(s)$ denotes the interior of set $\Omega(s)$. Together with the relation (\ref{eqn:widthdefn2}) satisfied by the width of a convex set, it follows that there should exist two points in $\clQ$ separated by a distance greater than $\clW(\Omega)$. This means that
\[
\|\sum_{i=1}^N {k_i} u_i\| \geq \clW(\Omega) \mbox{ for some } k_i \in \{-1,0,1\}, 1 \leq i \leq N.
\]
This means that the path density calculated using Lemma \ref{lem:pdforregular} satisfies
\begin{eqnarray*}
\ell(P) &=&\frac{1}{2\pi}\sum_{i=1}^N \|u_i\| \geq \frac{1}{2\pi}\sum_{i=1}^N \|k_i u_i\|\\
&\geq& \frac{1}{2\pi}\|\sum_{i=1}^N {k_i} u_i\| \geq \frac{\clW(\Omega)}{2\pi}.
\end{eqnarray*}

\subsection{Sketch of proof for Proposition \ref{prop:neccondunionnD}}\label{propproof:neccondunionnD}
The result follows via the exact same steps as those followed in proving Proposition \ref{prop:neccondunion} by considering the field
\begin{eqnarray*}
g_c(r) &=& c \exp(-{\sf i} \langle s, r \rangle) \prod_{i=1}^N\sin\left(\frac{\langle u_i, r- w_i \rangle}{2 }\right)\\
&&\qquad \qquad \qquad \qquad \prod_{i=1}^d\sinc(\frac{r_i\epsilon}{\pi}), \quad r\in \Re^d.\\
\end{eqnarray*}
It is easily verified that the field $g_c$ vanishes at points on the hyperplanes in $P$ for all values of $c$. Hence $g_c$ cannot be uniquely identified from its values on $P$. Moreover, for $\epsilon$ small enough the field $g_c$ has a Fourier transform supported on the set $\Omega$. Thus $P$ does not satisfy condition~\Mon{recon1} and hence $P \notin \nomh$.

\subsection{Proof of Theorem \ref{thm:ddimsgentrajsetnecsuffmanifolds}}\label{thmproof:ddimsgentrajsetnecsuffmanifolds}
We only prove (\ref{eqn:ddimsgentrajssufcondn}) since (\ref{eqn:ddimsgentrajsneccondn}) follows directly from Proposition \ref{prop:neccondunionnD}.

Let $P$ be the manifold set defined in (\ref{eqn:ddimsunionphat}) with $P_i$ defined in (\ref{eqn:ddimsmanifset}). Let $\{b^i_1, b^i_2, \ldots, b^i_{d-1}\} \subset \Re^{d}$ form an orthonormal set of vectors spanning the subspace $H_i$. Suppose that for each $i$ we are given samples on the parallel hyperplanes in $P_i$ at points of the form $\{w_i + j  \Delta_i h_i + \sum_{k=1}^{d-1} n_k \epsilon b^i_k: j, n_k \in \bZ \}$. Let $F_s^i$ denote the Fourier transform of the following impulse stream $f_s^i$ of samples from the $i$-th manifold set $P_i$ in (\ref{eqn:unionphat}):
\begin{eqnarray}
f_s^i(r) &=& \sum_{j \in \bZ, n\in \bZ^{d-1}} f(w_i + j  \Delta_i h_i + \sum_{k=1}^{d-1} \epsilon n_k b^i_k)\nonumber \\
&& \quad \qquad \delta(r - w_i -  j \Delta_i h_i - \sum_{k=1}^{d-1} \epsilon n_k b^i_k )\label{eqn:sampledfield}
\end{eqnarray}
where $\delta(.)$ represents the Dirac-delta function in $d$-dimensions. We have the following lemma.
\begin{lemma}\label{lem:sampspec1set}
Let $\Omega \subset \Re^d$ be a compact set. If $\epsilon$ is small enough then the spectrum $F_s^i$ of the sampled field in (\ref{eqn:sampledfield}) can be expressed in terms of the Fourier transform $F$ of the field as
\begin{equation}
F_s^i(\omega) = \sum_{j  \in \bZ } \exp({\sf i} \langle j  u_i, w_i \rangle) F(\omega + j u_i), \quad \omega \in \Omega. \label{eqn:samspecddims}
\end{equation}
where $u_i$ are as defined in the statement of Theorem \ref{thm:ddimsgentrajsetnecsuffmanifolds}.
\end{lemma}
\begin{IEEEproof}
As before, let $\{b^i_1, b^i_2, \ldots, b^i_{d-1}\} \subset \Re^{d}$ form an orthonormal set of vectors spanning the subspace $H_i$. Let $r_1 = \Delta_i h_i$ and $r_k = \epsilon b^i_{k-1}$ for $2 \leq k \leq d$. Thus the sampling points form a lattice defined by $\{r_k: 1\leq k\leq d\}$ shifted by $w_i$. It follows from \cite{petmid62} that the sampled spectrum of the impulse stream in (\ref{eqn:sampledfield}) can be expressed as

\begin{eqnarray}
F_s^i(\omega) = \sum_{n \in \bZ^{d}} \exp({\sf i}   \sum_{k=1}^{d} \langle n_k  s_{k}, w_i \rangle) F(\omega + \sum_{k=1}^{d}n_k s_{k}) \label{eqn:gsamp}
\end{eqnarray}
where $s_i$ satisfy $\langle s_j, r_k \rangle = 2\pi \delta_{jk}, 1\leq j,k \leq d$. This implies that $s_1 = u_i$. Also, making $\epsilon$ arbitrarily small ensures that $\|r_k\| \to 0$ for $k \geq 2$ whence $\|s_k\| \to \infty$ for $k \geq 2$. Hence since $\Omega$ is compact it follows that for $\epsilon$ small enough only terms with $n_i = 0$ for all $i \geq 2$ contribute to the summation in (\ref{eqn:gsamp}) for $\omega \in \Omega$. Replacing $n_1$ with $j$ leads to (\ref{eqn:samspecddims}).
\end{IEEEproof}

We now proceed to the proof of (\ref{eqn:ddimsgentrajssufcondn}). Suppose
\begin{equation}
\clQ \nsubseteq \Omega(s), \mbox{ for all } s\in \Re^d. \label{eqn:nsubsetcondnproof}
\end{equation}
We have to verify the condition~\Mon{recon1} holds. Let $F$ denote the two-dimensional Fourier transform of any field $f$ bandlimited to $\Omega$. Let $\omega \in \Omega$ be arbitrary. Consider the following set
\begin{equation}
\sfN := \{n \in \bZ^N: \omega + Un \in \Omega\} \label{eqn:clNdefn}
\end{equation}
where $U$ is a $d \times N$ matrix with $i$-th column given by the vector $u_i = \frac{2\pi h_i}{\Delta_i}$. It is easy to see via the convexity of $\Omega$ that $\sfN$ is a lattice-convex subset of $\bZ^N$. Furthermore, condition (\ref{eqn:nsubsetcondnproof}) implies that translates of the unit-cell $\clU^N$ are not contained within $\sfN$. To each element $n$ of $\sfN$ we associate the value $v(n) = F(\omega + Un)$. Now the conclusion (\ref{eqn:samspecddims}) of Lemma \ref{lem:sampspec1set} can be restated as
\begin{eqnarray*}
\lefteqn{\exp({\sf i} \langle  \omega, w_i \rangle) F_s^i(\omega)}\\
&=& \sum_{j  \in \bZ } \exp({\sf i} \langle  \omega + j  u_i, w_i \rangle) F(\omega + j u_i), \quad \omega \in \Omega.
\end{eqnarray*}
These equations
can be evaluated at the points $\{\omega + Un: n\in \sfN, 1\leq i \leq N\}$ to obtain consistent linear equations in $v(.)$ of the form (\ref{eqn:linrelations}) with 
$g_n^i = \exp({\sf i} \langle  \omega + Un, w_i \rangle) F_s^i(\omega + Un)$, where $F_s^i$ represents the spectrum of the samples taken on points on the manifolds in $P_i$. 
Applying Lemma \ref{lem:latconvdecode} to the set $\sfN$ defined in (\ref{eqn:clNdefn}) we conclude that for each $\omega \in \Omega$ the values of the Fourier transform $\{F(\omega + Un): n\in \sfN\}$ can be decoded using the values $\{F_s^i(\omega + Un): n\in \sfN, 1\leq i \leq N\}$. Since $\omega \in \Omega$ was arbitrary this proves that $F(\omega)$ can be recovered for all $\omega \in \Omega$ thus completing the proof of (\ref{eqn:ddimsgentrajssufcondn}).

\subsection{Proof of Theorem \ref{thm:ddimsregparopt}}\label{thmproof:ddimsregparopt}
From Theorem \ref{thm:ddimsgentrajsetnecsuffmanifolds} it easily follows that $P^\epsilon \in \emoh$. Suppose $P^\epsilon$ is expressed in the form $ P_1$ as defined in (\ref{eqn:ddimsmanifset}). Then the vector $u_1$ appearing in Theorem \ref{thm:ddimsgentrajsetnecsuffmanifolds} corresponding to $ P_1 = P^\epsilon$ is given by
\[
u_1 = \left(1- \frac{\epsilon \clW(\Omega)}{2\pi}\right)^{-1} \hat u
\]
and hence by the definition of $\hat u$ and by (\ref{eqn:ddimsgentrajssufcondn}) we have $P^\epsilon \in \emoh$ for all $\epsilon > 0$. Also from Lemma \ref{lem:ddimsmdunion} we have
\[
\lim_{\epsilon \to 0}\mu_\kappa(P^\epsilon) = \frac{\clW(\Omega)}{2\pi}.
\]

Now consider any $P \in \emoh$ expressed in the form of (\ref{eqn:ddimsunionphat}). Using the notations and assumptions of Theorem \ref{thm:ddimsgentrajsetnecsuffmanifolds} it follows that if $P \in \emoh$ we need
\[
\clQ \nsubseteq \overset{\circ}\Omega(s) \mbox{ for all }s\in\Re^d
\]
where $\overset{\circ}\Omega(s)$ denotes the interior of set $\Omega(s)$. Following the same steps as in the proof of Theorem \ref{thm:regparopt} we get $\mu_\kappa(P) \geq \frac{\clW(\Omega)}{2\pi}$.

\bibliographystyle{IEEEtran}
\bibliography{TomrefsIT}
\end{document}